\documentclass[pdflatex,sn-mathphys-num]{sn-jnl}% Math and Physical Sciences Numbered Reference Style
\usepackage{graphicx}%
\usepackage{amsmath,amssymb,amsfonts}%
\usepackage{amsthm}%
\usepackage{mathrsfs}%
\usepackage[title]{appendix}%
\usepackage{xcolor}%
\usepackage{textcomp}%
\usepackage{manyfoot}%
\usepackage{booktabs}%
\usepackage{listings}%
\usepackage{setspace}
\usepackage{amsmath}  %gestion des symboles math\'{e}matiques
\usepackage{amsfonts} %gestion des fontes math\'{e}matiques
\usepackage{amssymb}  %gestion des symboles math\'{e}matiques
\usepackage{bm}
\usepackage{dsfont}
\usepackage{tcolorbox}
\usepackage{float}
\usepackage{xurl}
\usepackage{hyperref}
\usepackage[title]{appendix}
\hypersetup{
    colorlinks = true,
    allcolors = [rgb]{0,0.29,0.6}
    }
\usepackage{pdflscape}
\usepackage{enumitem} 
\usepackage{epstopdf}
\usepackage[official]{eurosym}
\usepackage{MnSymbol,wasysym}
\usepackage{caption}
\usepackage{subcaption}
\usepackage{rotating}
\usepackage{booktabs}
\usepackage{tabularx}
\usepackage{threeparttable}
\usepackage{pdflscape}
\usepackage{adjustbox,lipsum}
\usepackage{algorithm,algpseudocode}
\usepackage{changepage}
\usepackage{multirow}
\usepackage{dcolumn}

\newtheorem{property}{Property}

\newtheorem{remark}{Remark}%

\ifx\proof\undefined
  \newenvironment{proof}[1][Proof]{\par\noindent\textbf{#1.}\ }{\hfill$\square$\par}
\fi
\theoremstyle{thmstyleone}%
\newtheorem{theorem}{Theorem}%  meant for continuous numbers
\newtheorem{proposition}[theorem]{Proposition}% 

\theoremstyle{thmstyletwo}%

\theoremstyle{thmstylethree}%

\begin{document}

\title{Peer-to-Peer Basis Risk Management for Renewable Production Parametric Insurance}


\author*[1]{\fnm{Fallou} \sur{Niakh}}\email{fallou.niakh@ensae.fr}

\author[2]{\fnm{Alicia} \sur{Bassière}}\email{alicia.bassiere@centralesupelec.fr}

\author[3]{\fnm{Michel} \sur{Denuit}}\email{michel.denuit@uclouvain.be}

\author[4]{\fnm{Christian} \sur{Robert}}\email{christian.robert@univ-lyon1.fr}

\affil*[1]{\orgname{Centre de Recherche en Économie et de Statistiques (CREST), CNRS, École Polytechnique, GENES, ENSAE Paris, Institut Polytechnique de Paris}, 
\orgaddress{\street{5 Avenue Henry Le Chatelier}, \city{Palaiseau}, \postcode{91120}, \country{France}}}

\affil[2]{\orgname{Laboratoire de Génie Industriel (LGI), CentraleSupélec}, 
\orgaddress{\street{3 Rue Joliot Curie}, \city{Gif-Sur-Yvette}, \postcode{91190}, \country{France}}}

\affil[3]{\orgname{Institut de statistique, biostatistique et sciences actuarielles - ISBA, Université Catholique de Louvain},
\orgaddress{\street{1 Place de l'université}, \city{Louvain}, \postcode{B-1348}, \country{Belgium}}}

\affil[4]{\orgname{Institut de Science Financière et d’Assurances (ISFA), Université Claude-Bernard Lyon 1},
\orgaddress{\street{50 Av. Tony Garnier}, \city{Lyon}, \postcode{69007}, \country{France}}}


\abstract{The financial viability of renewable energy projects is challenged by the variability and unpredictability of production due to weather fluctuations. This paper proposes a novel risk management framework combining parametric insurance and peer-to-peer (P2P) risk sharing to address production uncertainty in solar electricity generation. We first design a weather-based parametric insurance scheme to protect against forecast errors, recalibrated at the site level to mitigate geographical basis risk. To handle residual mismatches between insurance payouts and actual losses, we introduce a complementary P2P mechanism that redistributes the remaining basis risk among participants. The method leverages physically based simulation models to reconstruct day-ahead forecasts and realized productions, integrating climate data and solar farm characteristics. A second-order theoretical approximation links heterogeneous local models to a shared weather index, making risk sharing operationally feasible. In an empirical application to 50 German solar farms, our approach reduces the volatility of production losses by 55\%, demonstrating its potential to stabilize revenues and strengthen the resilience of renewable investments.}




\maketitle

\section{Introduction}

The transition to renewable energy is crucial for achieving a sustainable energy system in the face of climate change. Solar and wind energy are increasingly deployed among renewable sources, but their production remains highly variable and only partially predictable due to weather fluctuations. This uncertainty introduces financial risks, as deviations from expected generation levels, that can lead to unstable revenues and affect investment decisions. Current market structures, heavily reliant on selling production directly, are increasingly debated at the European level. To address these challenges, complementary mechanisms, such as revenue stabilization strategies or insurance frameworks, are needed to secure minimum returns and mitigate investment risks.

To tackle these issues, we propose a novel approach combining parametric insurance with a peer-to-peer (P2P) mechanism to manage unanticipated variability in renewable energy production. First, we design a classical parametric insurance scheme based on a weather-based reference index and quantify the basis risk—the mismatch between index-based payouts and actual losses. Then, we implement a complementary P2P insurance mechanism to redistribute this basis risk, reducing individual exposure to weather-related hazards.

The increasing reliance of renewable energy producers on market revenues introduces a significant source of uncertainty, which can deter investment decisions due to unpredictable cash flows. Power Purchase Agreements (PPA) were among the first solutions proposed to mitigate this uncertainty \citep{taheri2025physical}. Since then, several European countries have implemented Capacity Remuneration Mechanisms (CRMs) to address this, ensuring revenues are not solely tied to production, see e.g. \citet{kepplerRationalesCapacityRemuneration2017}. 

More recently, Contracts for Difference (CfDs) have emerged as a key instrument for stabilizing revenues, see e.g. \citet{onifadeHybridRenewableEnergy2016}. Under a CfD, renewable energy producers are guaranteed a fixed ``strike price" for their electricity. When market prices fall below this strike price, producers receive compensatory payments; conversely, when market prices exceed the strike price, producers return the surplus. This mechanism protects against price volatility, making renewable energy investments more predictable and attractive. However, while CfDs mitigate price risks, they fail to address production variability, leaving producers vulnerable to weather-related uncertainties and forecast errors. 
In addition to these market-based instruments, many European governments have introduced subsidies, such as feed-in tariffs and feed-in premiums, to stabilize renewable revenues. These policies provide guaranteed prices or additional premiums for renewable generation, ensuring predictable cash flows. However, they still do not eliminate the physical production risks.

As climate change intensifies weather-related variability in renewable energy production, the limitations of these existing mechanisms become increasingly apparent. This underscores the urgent need for complementary mechanisms, such as insurance, that address both price and production risks, ensuring the financial viability of renewable energy investments.

%% Contribution assurance paramétrique

Parametric insurance has emerged as a promising mechanism for mitigating financial risks associated with weather-related variability in renewable energy production. Unlike traditional insurance, which compensates for actual losses, parametric insurance triggers payouts based on predefined parameters, such as weather indices.
\citet{hanWeatherIndexInsurance2019} explore weather index insurance for wind energy, showing how such contracts protect producers from economic losses due to unfavorable wind conditions by linking payments to weather indices rather than direct financial losses. Similarly, \citet{liaoManagingVolatilityRisk2021} develop index-based insurance models using time-series forecasting historical wind speed data, building index insurance models to stabilize revenue and enhance investment security. Extending this approach to solar farms, \citet{liaoIndexbasedRenewableEnergy2022} design an index-based insurance product for photovoltaic plants in Taiwan, using long-term solar irradiation data and ARIMA models to calculate premiums and hedge against production shortfalls. The appeal of parametric insurance lies in its several advantages: it enables rapid payouts, relies on measurable third-party data (e.g., solar radiation or wind speed), and minimizes disputes by eliminating subjective claims assessments and human error. Moreover, by linking insurance triggers directly to measurable weather variables, insurers can more effectively hedge their exposure to weather-related losses in financial markets using weather derivatives. This enhances the stability of their portfolios and facilitates access to reinsurance and alternative risk transfer mechanisms, ultimately improving capital efficiency and enabling more sustainable coverage for renewable energy producers.

However, a key limitation of parametric insurance is basis risk—the mismatch between index-based payouts and actual losses when the chosen parameters fail to capture local damages. In particular, a component of this basis risk, known as geographical basis risk, arises when the weather index is computed at a single reference station whose conditions diverge from those at individual farms, undermining the insurance’s effectiveness in atypical or micro-climatic settings \citep{d2023minimizing}. The literature has proposed various solutions—e.g. probabilistic triggers that introduce uncertainty into the payout rule to better align compensation with realized losses \citep{figueiredoProbabilisticParadigmParametric2018}—yet these approaches still leave residual risks unaddressed. To mitigate geographic basis risk ex-ante, we therefore recalibrate the trigger for each site by fitting a simple farm-level model linking reference-index deviations to that farm’s historical production losses. This preliminary step smooths out spatial idiosyncrasies and significantly reduces basis risk.

% To address these limitations, we propose a novel hybrid model that combines parametric insurance with a P2P mechanism. 

Despite site-specific recalibration, some residual mismatches inevitably persist. To address these remaining gaps, we layer a peer-to-peer (P2P) basis risk sharing mechanism on top of our recalibrated parametric contracts. In this context of P2P risk-sharing mechanisms, the study by \citet{wang2025outage} provides valuable insights into the application of outage risk sharing in electric power systems. Their work highlights the benefits of decentralized risk-sharing frameworks in mitigating financial losses associated with power outages. A complementary line of research by \citet{moret2020heterogeneous} has already shown that pure P2P risk‐sharing, when coupled with simple financial contracts in a community‐based electricity market, can significantly attenuate participants’ risk aversion and restore fairness in cost allocation. This study strengthens our proposal by demonstrating how P2P mechanisms can be effectively implemented in the energy sector to enhance financial resilience. While \citet{wang2025outage} focus on outage risks, our approach extends the concept to revenue volatility in renewable energy production. The P2P layer complements parametric insurance by redistributing residual risks among participants, providing an additional buffer accounting for basis risk. This approach mitigates individual exposure to financial losses caused by forecast errors and weather variability, enhancing the robustness of the insurance framework. By allowing producers to pool their risks, our model helps smooth out compensation mismatches and provides a more effective hedge against production uncertainties.

P2P insurance, as implemented in our model, directly addresses the basis risk residual by redistributing financial impacts not covered by parametric insurance. Its main advantages include cost efficiency, as it eliminates overhead costs associated with traditional insurers; flexibility, as it can be tailored to meet the specific needs of renewable energy producers; and its cooperative structure, which strengthens risk-sharing among producers and aligns with the growing role of local energy initiatives in decentralized electricity markets. Our approach offers a practical and scalable solution for managing financial risks in renewable energy markets.

The theoretical foundations of risk pooling further strengthen our approach. A key aspect is to determine the risk-sharing rule that has to be accepted by all producers, a question addressed frequently \citep{clemente2023optimal}. In this paper, we work with a risk-sharing rule based on an approximation of the conditional mean risk-sharing rule proposed in \citet{denuit_convex_2012}. According to the conditional mean risk-sharing rule, each producer contributes the conditional expectation of the basis risk brought to the pool, given the total basis risk experienced by the entire pool (see \citet{denuitRiskSharingPure2021}). Some attractive properties of the conditional mean risk allocation have been studied in  \citet{denuit_size-biased_2019, denuit_large-loss_2020, denuit_risk-sharing_2022, jiao_axiomatic_2022}. As the number of participants in the pool grows, the conditional mean risk-sharing rule can be approximated by a linear risk-sharing rule. This principle is particularly relevant to our model, ensuring a simplified basis risk-sharing for large pools.

Building on these principles, we incorporate risk-sharing rules as an additional layer on top of parametric insurance. This integration not only mitigates instances where parametric payouts deviate from actual losses but also enhances the resilience and equity of the overall insurance system.

\paragraph{Main contributions} On the theoretical side, we derive a novel second-order approximation showing that each farm’s conditional loss—originally expressed through its own Generalized Linear Model (GLM)—can be closely approximated by a single common weather index plus a correction term. This result bridges heterogeneous local loss models and a global reference index, making parametric insurance both fair and tractable. To turn theory into practice, we develop a computational framework that estimates the necessary conditional moments via non-parametric kernel methods and enriches limited weather data through copula-based simulation.

On the empirical side, we apply our hybrid model to 50 solar farms in southern Germany. By combining physics-based day-ahead forecasts with high-resolution climate data, we compute each farm’s revenue shortfall, build the weather index at a reference site, and layer on a P2P basis risk sharing mechanism. In this real-world setting, our approach reduces loss volatility by 55\%, demonstrating its effectiveness in stabilizing returns and strengthening the financial case for renewable energy investments.

This paper is structured as follows: Section \ref{sectmodel} introduces the problem formulation and presents the theoretical model for insuring against revenue losses in renewable energy production. Section \ref{sectcasestudy} outlines the empirical strategy and describes the physical model employed to simulate revenue losses for renewable producers. Finally, Section \ref{sectresults} discusses the results obtained from the proposed insurance model. Additional details are provided in the appendices: \ref{app1} illustrates the data representation, \ref{app2} provides the proof of the approximation of the conditional expectation of the loss, \ref{app3} details the computation of the objective function considered in the paper, and \ref{app4} presents additional results.

%%%% A réécrire avec les résultats

\section{A combined parametric insurance and P2P basis risk sharing mechanism\label{sectmodel}}

This section introduces the combined parametric insurance and P2P basis risk sharing mechanism.

\subsection{Problem description}

We aim to recover the basis risk subsisting after implementing the parametric insurance.

\subsubsection{Individual loss and basis risk}

In renewable energy production, the random loss $X_i$ for producer $i$ is the financial loss associated with forecast uncertainties at a given date. The parametric insurance is triggered when an observed index $Z$ based on weather-related variables for a reference location reaches a predefined threshold $z_0$. Like the Consumer Price Index (CPI), which tracks changes in the cost of a standard basket of goods and services over time, the weather index $Z$ quantifies weather risks for renewable energy production through a composite of weather variables. An increase in the index reflects a rise in extreme weather events, leading to very low production, highlighting elevated risk levels for renewable energy systems. 

The threshold $z_0$ serves as a critical benchmark, beyond which the weather index $Z$ is considered sufficiently high to signify a production loss for renewable energy producers. The selection of $z_0$ is a pivotal decision for producers, as it directly impacts the balance between risk exposure and insurance costs. If $z_0$ is set lower, it reflects a higher level of risk aversion on the part of the producers. A lower threshold implies that parametric insurance is triggered more frequently, even for relatively moderate adverse weather conditions. While this provides more protection and reduces the likelihood of uncovered losses, it also results in a higher insurance premium as the insurer must account for the increased probability of payouts. Conversely, a higher $z_0$ indicates lower risk aversion, with producers opting to shoulder more potential losses. This reduces the insurance premium but increases the risk of significant uncovered losses if extreme weather events occur. 

Thus, the choice of $z_0$ should align with producers’ specific risk tolerance and financial resilience. By tailoring $z_0$ to producers’ level of risk aversion, the parametric insurance framework can provide an optimal balance between cost and coverage, ensuring that it meets the needs of diverse renewable energy stakeholders.

The parametric insurance contract compensates the producer $i$ with an amount defined by the expected loss given the weather indicator $Z$ for the reference location:
\begin{equation}
    m_i(z) = \mathbb{E}[X_i \mid Z = z], \quad \text{for } z > z_0.
    \label{fctmi}
\end{equation}
In return, the producer $i$ pays an insurance premium that also considers administrative costs, risk margins, and other expenses.

However, the compensation $m_i(Z)$ does not perfectly match the actual loss because the compensation depends on weather parameters rather than realized loss production data. This creates a \textit{basis risk}:

$$
\varepsilon_i = X_i - m_i(Z).
$$
We aim to find an additional mechanism to minimize the variance of this basis risk by risk pooling, when $Z>z_0$. We introduce a P2P risk-sharing scheme hereafter.

\subsubsection{Risk pooling and sharing}

Producers can form a risk pool to reduce basis risk, where all participants share the aggregated total basis risk $S_\varepsilon = \sum_{i=1}^n \varepsilon_i$ where $n$ is the number of participants. This pooled risk is redistributed ex-post, with each producer $i$ retaining the individual contribution $\delta_i$ defined by:
$$
\delta_i = \frac{\sigma_i^2(Z)}{\sum_{j=1}^n \sigma_j^2(Z)} S_\varepsilon,
$$
where $\sigma_i^2(z) = \text{Var}[X_i \mid Z = z],$ is the conditional variance of $X_i$  for producer $i$ given that the weather index $Z$ is equal to $z$ that exceeds the threshold $z_0$. This formula is not only a proportional allocation based on conditional variances but also serves as an approximation of the conditional expectation $\mathbb{E}[\varepsilon_i \mid S_\varepsilon, Z = z]$ when $n$ is large, as shown in \citet{denuitRiskSharingPure2021}. It ensures fairness and aligns each producer’s contribution with the variability of their losses. The derivation of this formula as an approximation of $\mathbb{E}[\varepsilon_i \mid S_\varepsilon, Z = z]$ assumes that the basis risks $\varepsilon_i$ are independent, conditional on $Z$. This conditional independence is crucial, as it requires the weather index $Z$ to act as a sufficiently informative linear score, capturing all potential conditional dependencies between individual losses. $Z$ should comprehensively reflect the shared weather-related risk factors influencing producers’ financial outcomes. By carefully designing $Z$, this assumption becomes a valid and practical simplification, ensuring the effectiveness and fairness of the risk-sharing mechanism.

Moreover, the variance of each producer's adjusted contribution $\delta_i$ is given by:

$$
\text{Var}[\delta_i \mid Z > z_0] = \mathbb{E}\left[\frac{\sigma_i^4(Z)}{\sum_{j=1}^n \sigma_j^2(Z)} \mid Z > z_0\right]
$$
and reflects the residual uncertainty after pooling.
This variance quantifies how weather-driven production variability influences the fairness and effectiveness of risk-sharing. The total risk is shared proportionally by pooling, reducing individual variability while aligning contributions with producers' inherent risks.

\subsection{Parameterizing the insurer’s index benchmark for effective risk sharing}
The aim of this section, and a central focus of this paper, is to parameterize the insurer’s index benchmark $Z$ as effectively as possible. This involves close collaboration with the insurer to design $Z$ to minimize the variance of individual producers’ contributions to the pooled basis risk. The design and the calibration of $Z$ are crucial, as the index must act as a representative benchmark that captures key weather-related risk factors and minimizes residual risks for the producers. 

In this paper, the construction of the index $Z$ draws inspiration from the North American Actuaries Climate Index (ACI) in the sense that both use a linear combination of climate variables (\citet{garrido_climate_2024}). While the methodology used to build our index differs from the ACI, this linear structure is standard across many actuarial climate indices, providing an intuitive way to summarize weather conditions relevant to risk assessment and insurance design.

\begin{remark}
The ACI incorporates six components: frequency of temperatures $> 90^{th}$ percentile ($T90$), frequency of temperatures $< 10^{th}$ percentile ($T10$), maximum rainfall per month in five consecutive days ($P$), annual maximum consecutive dry days ($D$), frequency of wind speed $> 90^{th}$ percentile ($W$) and change in sea level ($S$). Each monthly value is standardized with the mean and standard deviation of the monthly values of the corresponding month over the reference period. For example, for an observation for January, the standardisation is with the mean and standard deviation of the 30 months of January in the reference period. Then, the standardized anomalies of all six components are averaged to form the composite
index: 
\begin{equation}
    ACI = \frac{1}{6} \left(T90_{std} - T10_{std} + P_{std} + D_{std} + W_{std} + S_{std}\right)
\end{equation}
where the temperature lows $T10_{std}$ is subtracted instead of added; as the climate warms, the occurrence of extremely low temperatures decreases, and the temperature distribution curve
shifts to the right.
\end{remark}
It is important to note that the ACI assigns equal weight ($\frac{1}{6}$) to each climate variable. This approach emphasizes the relative evolution of the index over time rather than its absolute values. However, the ACI must be adjusted to align with the assessed risks. In our context, where the focus is on parametric insurance on weather variables influencing solar energy production, the absolute value of the index $Z$ becomes crucial. Since not all weather variables impact production equally—solar radiation being a stronger determinant than temperature—this weighting adjustment ensures a more accurate representation of risk exposure. Therefore, we propose a weighted sum that prioritizes weather variables relevant to energy production. Hence, the index $Z$ is constructed as a linear combination of $J$  weather covariates $Y_j$:

$$
Z = \sum_{j=1}^J a_j Y_j.
$$
The weights $\boldsymbol{a} = \left(a_1, \ldots, a_J\right)^{\prime}$ determine the relative importance of each covariate on the global index. The group of participants and the insurer aim to choose $\boldsymbol{a} = \left(a_1, \ldots, a_J\right)^{\prime}$ to minimize the sum of the expected individual variances:

$$
\sum_{i=1}^n\mathbb{E}\left[ \sigma_i^2(Z) \mid Z > z_0\right].
$$

%$$
%\text{Var}\left[\sum_{i=1}^n \delta_i \mid Z > %z_0\right]=\mathbb{E}\left[\sum_{i=1}^n \sigma_i^2(Z) %\mid Z > z_0\right].
%$$

% Or the sum of the variance of the ex-post basis risks

% $$
% \sum_{i=1}^n \text{Var}\left[\delta_i\right]=\mathbb{E}\left[\frac{\sum_{i=1}^n \sigma_i^4(Z)}{\sum_{i=1}^n \sigma_i^2(Z)}\right].
% $$

We assume that for a typical production day, the conditional distributions of $X_i$, given the vector of weather covariates %$\textcolor{blue}{\boldsymbol{Y} =
% \begin{pmatrix}
% Y_1 \\
% .\\
% .\\
% .\\
% Y_J
% \end{pmatrix}}$ 
$\boldsymbol{Y} = \left(Y_1, \ldots, Y_J\right)^{\prime}$ follow a Generalized Linear Model (GLM). GLMs provide an attractive framework for this study because of their ability to model parametrically $X_i$ as a function of covariates $\boldsymbol{Y}$ while considering the specific nature of the data. GLMs allow us to specify a link and variance functions, which establish a relationship between the mean and the variance of $X_i$, and the covariates guaranteeing interpretable predictions. These properties are essential for constructing the index $Z$ as an optimized linear combination of covariates, where the coefficients are based on the GLMs. Finally, their flexibility and robustness make it possible to model park specificities while interpreting the relationships between covariates and losses, essential for achieving the study's objectives. We consider the following setup characterized by:

$$
\begin{aligned}
\mathbb{E}\left[X_i \mid \boldsymbol{Y}\right] & = g_i^{-1}\left(a_{i0} + \boldsymbol{a}_i^{\prime} \boldsymbol{Y}\right), \\
\text{Var}\left[X_i \mid \boldsymbol{Y}\right] & = \phi_i V_i\left(\mathbb{E}\left[X_i \mid \boldsymbol{Y}\right]\right),
\end{aligned}
$$
where $a_{i0} \in \mathbb{R}$ is the intercept, $\boldsymbol{a}_i \in \mathbb{R}^J$ is the vector of coefficients, $\phi_i \in \mathbb{R}_{+}$ is the dispersion parameter, $g_i$ is the GLM link function, and $V_i(.)$ represents the variance function associated with the GLM. We allow the sensitivity to the covariate vector $\boldsymbol{Y}$ to differ between producers, as captured by the parameters $\boldsymbol{a}_i$. However, we assume that the heterogeneity of $\boldsymbol{a}_i$ between the producers is not too strong so that a chosen common parameter $\boldsymbol{a}$ approximates the individual coefficients $\boldsymbol{a}_i$ reasonably well.

We now detail the approximation of the following quantities of interest: $\mathbb{E}\left[X_i \mid Z=z\right]$ and $\text{Var}\left[X_i \mid Z=z\right]$. The major theoretical issue in this paper lies in the fact that the conditional expectation $\mathbb{E}[X_i \mid \boldsymbol{a}_i^{\prime} \boldsymbol{Y}]$ is known analytically (because it is deduced from an individual GLM), but not $\mathbb{E}[X_i \mid Z]$. However, it is possible to find an approximation which depends only on the density distribution function of $Z$ and moments of the distribution $\boldsymbol{Y}$ given $Z$. These moments, even if they do not have an analytical expression, can be obtained using a non-parametric estimation method based on simulations.

\subsubsection{Approximation of the conditional expectation $\mathbb{E}\left[X_i \mid Z=z\right]$} 

We consider the following decomposition of the producer $i$'s linear score:
$$
a_{i0} + \boldsymbol{a}_i^{\prime} \boldsymbol{Y} = a_{i0} + Z + \left\|\boldsymbol{a}_i - \boldsymbol{a}\right\| \boldsymbol{\alpha}^{\prime}_i \boldsymbol{Y},
$$
where $Z = \boldsymbol{a}^{\prime} \boldsymbol{Y}$, $\boldsymbol{\alpha}_i = \frac{\boldsymbol{a}_i - \boldsymbol{a}}{\|\boldsymbol{a}_i - \boldsymbol{a}\|}$, and $\left\|\boldsymbol{a}_i - \boldsymbol{a}\right\|$ is assumed to be small where $\left\|.\right\|$ is the Euclidean norm. 

\begin{proposition}[Second-order approximation of conditional expectation]
% Extending Lemma 2 from \citet{gourierouxSensitivityAnalysisValues2000} to this context, we apply expansion formula of $\mathbb{E}[X_i \mid \boldsymbol{a}_i^{\prime} \boldsymbol{Y}=z]$ given as:

Let us assume that $\boldsymbol{a}$ and $\boldsymbol{a}_i$ are not collinear vectors and that $\left\|\boldsymbol{a}_i - \boldsymbol{a}\right\|$ is small. There exist two functions $L_i $ and $F_i$ such that:
\begin{equation}
\mathbb{E}[X_i \mid \boldsymbol{a}_i^{\prime} \boldsymbol{Y}=z] = \mathbb{E}[X_i \mid Z = z] + \|\boldsymbol{a}_i - \boldsymbol{a}\| L_i(z,\boldsymbol{a}) 
+\frac{1}{2} \|\boldsymbol{a}_i - \boldsymbol{a}\|^2 F_i(z,\boldsymbol{a}) + o(\|\boldsymbol{a}_i - \boldsymbol{a}\|^2).
    \label{expansionformula}
\end{equation}
% The terms $L_i(\cdot, \cdot)$ and $F_i(\cdot, \cdot)$ capture first- and second-order deviations, respectively, and adjust for individual differences in the weight vectors $\boldsymbol{a}_i$.
The term $L_i(\cdot, \cdot)$ accounts for the first-order deviation and is expressed as:
\begin{equation}
    \begin{aligned}
L_i(z,\boldsymbol{a}) = & -\frac{\partial}{\partial z} \log f(z) \mathrm{Cov}[X_i, \boldsymbol{\alpha}_i^{\prime}\boldsymbol{Y} \mid Z = z] 
- \frac{\partial}{\partial z} \mathrm{Cov}[X_i, \boldsymbol{\alpha}_i^{\prime}\boldsymbol{Y} \mid Z = z] \\
& + \mathbb{E}[\boldsymbol{\alpha}_i^{\prime}\boldsymbol{Y} \mid Z = z] \frac{\partial}{\partial z} \mathbb{E}[X_i \mid Z = z].
\end{aligned}
\label{li}
\end{equation}
where $f$ is the marginal probability density function of $Z$.\newline
The term $F_i(\cdot, \cdot)$ captures the second-order deviation and is given by:
\begin{equation}
    \begin{aligned}
        F_{i}\left( z,\boldsymbol{a}\right) =&\left[ \frac{\partial ^{2}}{\partial z^{2}}\log f\left( z\right) +\left( 
\frac{\partial }{\partial z}\log f\left( z\right) \right) ^{2}\right] 
\mathrm{Cov}\left[ X_{i},(\mathbf{\alpha }_{i}^{^{\prime }}\boldsymbol{Y}%
)^{2}\boldsymbol{|}Z=z\right]  \\
&+2\frac{\partial }{\partial z}\log f\left( z\right) \left[ \frac{\partial 
}{\partial z}\mathrm{Cov}\left[ X_{i},(\mathbf{\alpha }_{i}^{^{\prime }}%
\boldsymbol{Y})^{2}\boldsymbol{|}Z=z\right] +\mathbb{E}[(\mathbf{\alpha }%
_{i}^{^{\prime }}\boldsymbol{Y})^{2}|\left. Z=z\right. ]\frac{\partial }{%
\partial z}\mathbb{E}[\left. X_{i}|Z=z\right. ]\right]  \\
&+\frac{\partial ^{2}}{\partial z^{2}}\mathrm{Cov}\left[ X_{i},(\mathbf{%
\alpha }_{i}^{^{\prime }}\boldsymbol{Y})^{2}\boldsymbol{|}Z=z\right] +2\frac{%
\partial }{\partial z}\mathbb{E}[(\mathbf{\alpha }_{i}^{^{\prime }}%
\boldsymbol{Y})^{2}|\left. Z=z\right. ]\frac{\partial }{\partial z}\mathbb{E}%
[\left. X_{i}|Z=z\right. ] \\
&+\mathbb{E}[(\mathbf{\alpha }_{i}^{^{\prime }}\boldsymbol{Y})^{2}|\left.
Z=z\right. ]\frac{\partial ^{2}}{\partial z^{2}}\mathbb{E}[\left.
X_{i}|Z=z\right. ] \\
&-2\left[ \frac{\partial }{\partial z}\log f\left( z\right) \mathbb{E}[%
\mathbf{\alpha }_{i}^{^{\prime }}\boldsymbol{Y}|Z=z]+\frac{\partial }{%
\partial z}\mathbb{E}[\mathbf{\alpha }_{i}^{^{\prime }}\boldsymbol{Y}|Z=z]%
\right] L_{i}\left( z,\boldsymbol{a}\right) .
\end{aligned}
\label{fi}
\end{equation}
\label{gour}
\end{proposition}
The proof of the approximation of the conditional expectation of the loss in Proposition \ref{gour} is given in Appendix \ref{app2}. This result formalizes the idea that, when the losses $X_i$ depend on a vector of covariates $\boldsymbol{Y}$, their conditional expectation given a linear score $\boldsymbol{a}_i^{\prime} \boldsymbol{Y}$ can be approximated by the conditional expectation given a common reference index $Z = \boldsymbol{a}^{\prime} \boldsymbol{Y}$, up to a second-order correction. This is crucial in our context because:
\begin{itemize}
    \item The theoretical conditional expectation $\mathbb{E}[X_i \mid Z = z]$ is required for basis risk sharing computations, but it is not directly estimable from the GLM.
    \item Each park $i$ has its own weight $\boldsymbol{a}_i$, leading to model heterogeneity. However, operational implementation of the index requires a common threshold based on a shared index $Z$.
\end{itemize}
This approximation thus bridges individual heterogeneity and a pooled reference structure, and is key to making the parametric index tractable for real-world insurance design.

To compute $\mathbb{E}[X_i \mid Z = z]$ in Proposition \ref{gour}, we develop a computational approach that estimates the relevant conditional expectations and densities using kernel-based methods. Because the approximation involves quantities that are not analytically tractable, this non-parametric strategy is essential. To address the challenges of limited data, we simulate additional weather data by calibrating the marginal distributions of the weather variables and modeling their dependencies with a vine copula \citep{thesisKernelMethodsVine}. This enriched dataset improves the stability and reliability of the conditional expectation estimation. Further implementation details are provided in Appendix \ref{app3} and support the empirical analysis in Section \ref{sectcasestudy}.
\subsubsection{Approximation of the conditional variance $\text{Var}\left[X_i \mid Z=z\right]$}

The conditional variance of $X_i$ given $Z = z$ is approximated by:
$$
\phi_i V_i\left(\mathbb{E}[X_i \mid Z = z]\right),
$$
and, using Equation \eqref{expansionformula} above, we write:
\begin{equation}
    \text{Var}[X_i \mid Z = z] \simeq \phi_i V_i\left(\mathbb{E}[X_i \mid \boldsymbol{a}_i^{\prime} \boldsymbol{Y}=z] - \|\boldsymbol{a}_i - \boldsymbol{a}\| L_i(z,\boldsymbol{a}) 
-\frac{1}{2} \|\boldsymbol{a}_i - \boldsymbol{a}\|^2 F_i(z,\boldsymbol{a})\right).
    \label{varapprox}
\end{equation}

\subsubsection{Final optimisation}

We aim to minimize the variance of the collective ex-post basis risks. We construct the following optimization problem, which seeks the optimal coefficient $\boldsymbol{a}$ (that is, the index $Z$) to minimize the sum of the expected individual variances:
\begin{equation}
    \sum_{i=1}^n \mathbb{E}\left[\text{Var}\left[X_i \mid  Z\right] \mid  Z > z_0\right],
    \label{problem}
\end{equation}
where $\text{Var}\left[X_i \mid  Z \right]$ is approximated using Equation \eqref{varapprox}.

\section{Case study: physics-based simulation and P2P basis risk sharing in southern Germany \label{sectcasestudy}}

We consider a parametric insurance contract that compensates solar producers in southern Germany for solar production losses due to weather variations. The producers further share their basis risk inherent in parametric insurance products to limit their financial losses. We apply the theoretical framework described above to real data to construct the index proposed in this study.

\subsection{Data collection}

To calibrate the insurance mechanism, we first quantify each producer's energy production loss using a physics-based model coupled with climate data. Since the financial impact of lost production depends on market conditions, assigning a monetary value to these losses is necessary. This valuation accounts for price variations over time, ensuring financial exposure reflects producers' economic risk. We establish a robust basis for defining the desired risk coverage by integrating energy production losses and their financial valuation.

\subsubsection{Power plants data}

We compiled a dataset of large-scale solar power plants in Germany using the German governmental database ; see, e.g., \citet{MaStR_Datenschutz}. This dataset was constructed by extracting and processing information on operational solar farms, focusing on installations exceeding 5 megawatt (MW). Such large-scale projects represent potential candidates for parametric insurance, as they primarily operate within electricity markets rather than for self-consumption.

\begin{figure}[h!]
        \centering
                \includegraphics[width=0.6\linewidth]{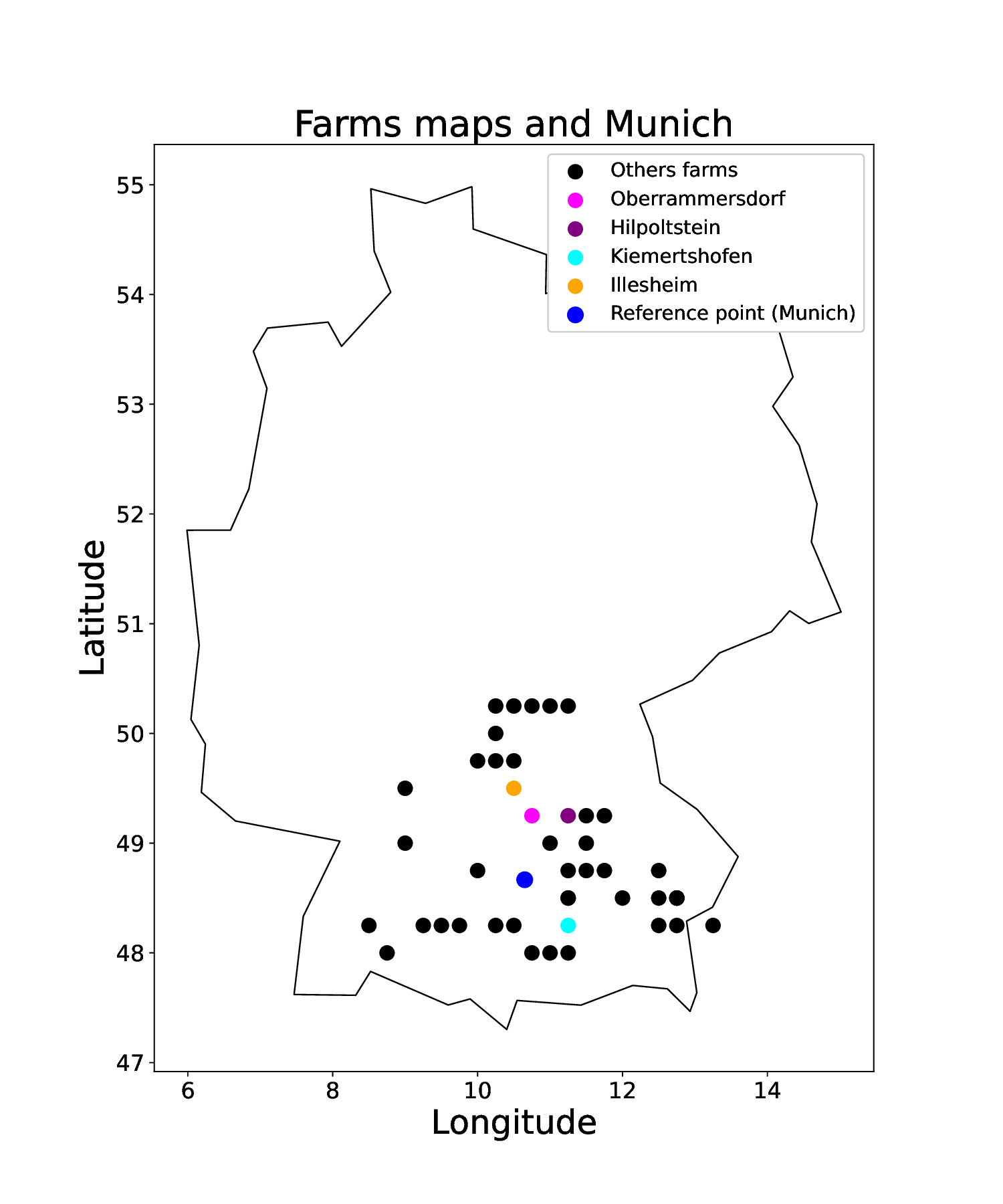}
                \caption{Geographical distribution of the 50 solar farms used in this study (black, magenta, cyan, purple, and orange points), along with the reference location in Munich (blue point). The four highlighted farms (magenta, cyan, purple, and orange) are frequently referenced throughout the paper as illustrative examples. These farms are spread across southern Germany, with Munich serving as the central weather index location for the parametric insurance model.}
                 \label{fig:illustrationadd3}
\end{figure}

The dataset comprises 50 power plants (as illustrated in Figure \ref{fig:illustrationadd3}) with key attributes such as installed capacity, commissioning date, geographical location, panel orientation, and economic support schemes (market-based mechanisms or feed-in tariffs, discussed in detail later). The 50 farms were selected for their geographical proximity and weather correlation with the reference location.

\subsubsection{Recovery of the energy production loss}

When German renewable energy producers participate in the day-ahead spot market, they must submit their expected production forecast by 12PM the following day, according to the rules of EPEX Spot. On this market platform, electricity is traded in Central Western Europe. However, because renewable electricity production depends on weather conditions, which cannot be predicted accurately, producers face inherent uncertainty in their forecasts. As a result, they can only estimate their future electricity generation rather than determine it with precision. This creates a trade-off between bidding the highest possible output to maximize revenues and avoiding an overestimation that could lead to costly imbalances, requiring repurchases on the intraday market at a higher price or, worst case, financial penalties.

To quantify these imbalances, we reconstruct both forecasted and realized conditions using meteorological variables from two datasets provided by the \href{https://www.ecmwf.int/en/forecasts/dataset/operational-archive}{European Centre for Medium-Range Weather Forecasts (ECMWF)}. The historical weather variables are obtained from the ERA5 reanalysis dataset \citep{hersbach2020era5}, which provides global coverage at a spatial resolution of 31 km with an hourly temporal resolution. This dataset, widely used in renewable energy research, spans 2012 to 2022 and includes near-surface air temperature (T2m) and surface solar radiation downwards (SSRD), which are essential for forecasting solar production.

Forecasted production was reconstructed using the ECMWF operational archive \href{https://www.ecmwf.int/en/forecasts/dataset/operational-archive}{(ECMWF archive data)}, which provides irradiance and temperature forecasts at a finer spatial resolution of approximately 10 km. This dataset was selected because it offers the highest available precision at an hourly temporal resolution, making it the most suitable choice for our analysis.

The forecast is initialized twice daily, at 12 PM and 12 AM. To ensure consistency with market rules, we used the forecast initialized at 12 PM, which provides predictions from midnight to 11 PM the following day. The expected production was then computed following the same methodology as realized production, using the PVlib model with forecasted SSRD and T2m inputs.

Since SSRD in the ECMWF archive is reported as a cumulative value over the day, we converted it into an hourly step-by-step format by computing the incremental difference between consecutive time steps. Additionally, as the dataset provides meteorological values on a discrete spatial grid, we applied bilinear interpolation to estimate temperature and irradiance at each power plant’s exact location.

For each given location, we model both forecasted and realized solar electricity production using the PVlib library \citep{holmgrenPvlibPythonPython2018} in Python, which simulates photovoltaic system performance based on climate variables. The energy production loss for each solar farm is then defined as the difference between its forecasted and actual production. The PVlib model relies on several key inputs, including SSRD, T2m, panel orientation, and panel tilt angle. The tilt angle, which represents the inclination of the solar panels relative to the horizontal plane, is set at 25° in our simulations. This value corresponds to the mean tilt across our dataset, where individual panel angles range from 20° to 30°. 

To recover energy output, we first compute Direct Normal Irradiance (DNI) and Diffuse Horizontal Irradiance (DHI) from SSRD using the Erbs model \citep{erbsEstimationDiffuseRadiation1982}. Since these values are not directly available, they must be derived using standard transposition models. The SSRD is then converted into Plane of Array (POA) irradiance, representing the amount of solar radiation incident on the tilted panel surface. The module temperature is estimated using a thermal model that accounts for both air temperature and POA irradiance. Finally, the PVWatts model \citep{dobosPVWattsVersion52014} is used to compute direct current (DC) power, which is then converted into alternating current (AC) power using an inverter model with default efficiency parameters.

By integrating forecasted and realized weather data with the technical characteristics of solar power plants, PVlib enables a precise reconstruction of expected and actual production. 

Finally, we computed the energy production loss in megawatt-hours (MWh) by subtracting the realized production from the forecasted production for each power plant. This metric quantifies the discrepancy between expected and actual generation, capturing the inherent uncertainty in renewable energy forecasting and its impact on market operations.

\subsubsection{Setting a financial value for lost production}

We now establish a financial value for lost production. Renewable energy producers typically face an arbitrage between two primary revenue sources: market-based income determined by the day-ahead market price and government subsidies under a feed-in tariff regime (a policy mechanism that guarantees a fixed rate for renewable energy producers over a set period). In some cases, producers combine both (partial feed-in), where they choose, on an hourly basis, whichever option --- day-ahead market price or feed-in tariff --- offers the highest return.

To characterize these revenue streams in practice, we distinguish three support schemes:
\begin{itemize}
    \item \textbf{Full Market Price}: Revenues come exclusively from selling electricity on the day-ahead market.
    \item \textbf{Partial Feed-in Tariff}: Producers compare, on an hourly basis, the day-ahead market price to the feed-in tariff and select the higher one.
    \item \textbf{Full Feed-in Tariff}: Revenues are fully determined by the feed-in tariff.
\end{itemize}

The German power plant database specifies which scheme applies to each facility. Day-ahead market prices were sourced from the ENTSO-E Transparency Platform~\citet{entsoe_prices}.

The feed-in tariff data are more challenging because only the minimum, maximum, and average solar subsidies are available. We use the average, noting that the variance in feed-in tariffs is lower than that observed for day-ahead market prices~\citet{BundesnetzagenturSolar}.

\subsection{Optimisation problem and resolution procedure} 

% To do :
% - Il ne faudrait parler que de la calibration (enlever la première phrase). Il faudrait quelques phrases en début de section expliquant la démarche qui va être mise en place et qui justifie les choix des GLMs, de la stationnarisation, etc.
% - Il est important de bien expliquer ce qu’est la perte : différence entre le revenue réalisé et le revenu anticipé (prédit). Est-ce que cela a été calculé heure par heure ? A la journée ? Etc.
% - Est-ce que cette perte est normalisée par la « taille » des fermes/parcs ? Il faut être plus précis sur le risque qui va être partagé entre les producteurs. Quel est le parc considéré dans la Figure 1 ? Quelle est l’unité de la variable en ordonnée ?
% - Même si la procédure de stationnarisation est renvoyée en annexe, il faut donner quelques explications dans le texte.
% - Sur la Figure 4, Direct Normal Irradiance, il persiste une forte hétéroscédasticité temporelle : il faudrait vérifier si cela bien été pris en compte.
% - Il faut préciser la fonction de lien et valider l’ajustement GLM empiriquement.
% Il faut bien expliquer comment on arrive au modèle Z>z_0

It is essential to clarify that the approach outlined earlier is based on a typical production day, assuming stationary weather and production conditions. However, both weather and production conditions vary throughout the year due to seasonal fluctuations. To address this, we aim to develop a scoring system with coefficients and thresholds that remain valid for any day of the year.

This requires first constructing an index within a stationary framework, where conditions are assumed to be constant. Once this stationary index is established, it can be adapted to a non-stationary model that accounts for the seasonal and temporal variations in weather and production. This ensures the scoring system is accurate and flexible for risk-sharing across all year periods.

Before delving into the detailed methodology, it is useful to outline the process. First, hourly meteorological and production data are collected from ECMWF, and relevant initial variables are defined. The data are then aggregated into a daily time series. PVlib allows for extracting the corresponding sunrise and sunset times for each day and farm location, allowing for the filtering of actual production periods. Next, the data are adjusted to meet stationarity assumptions. Following this, the function $m_i(z)$ in Equation \eqref{fctmi} is estimated to model the threshold $z_0$. The dataset is then filtered by keeping the days for which $z>z_0$. A Tweedie generalized linear model is applied to the filtered data to construct the scoring system. Finally, the optimal index is derived.
Below, we present the approach adopted to solve the above optimization problem on solar energy production. 

\subsubsection{Step 1: Data collection from ECMWF and initial variables}
The data used in this study cover the period from January 2012 to December 2022, providing a comprehensive view of weather conditions and solar production dynamics. We focus on the set of solar parks illustrated in Figure \ref{fig:illustrationadd3}.

For each solar park $i$ and each hour $h$ of a given day $d$, we collect both day-ahead forecasts (denoted with a hat symbol) and actual observations for key meteorological variables. These include the surface solar radiation downwards ($\widehat{SSRD}_{i,d,h}$ and $SSRD_{i,d,h}$), direct normal irradiance ($\widehat{DNI}_{i,d,h}$ and $DNI_{i,d,h}$), and temperature ($\widehat{T}_{i,d,h}$ and $T_{i,d,h}$). Additionally, weather data are collected for a designated reference station (indexed as 0). This reference station captures local weather conditions that reflect the circumstances of all participating producers. Hence, it allows us to construct a global index that can be applied uniformly across all producers. These variables are inputs for modeling the forecasted and realized solar energy outputs.

Solar production forecasts and the actual production of each park $i$ are obtained using the physical model provided in the PVlib package in Python:

% $$
% \begin{gathered}
% \widehat{P}_{i,d,h}=\mathcal{F}\left(\widehat{SSRD}_{i,d,h}, \widehat{DNI}_{i,d,h}, \widehat{T}_{i,d,h}\right) \\
% P_{i,d,h}=\mathcal{F}\left(SSRD_{i,d,h}, DNI_{i,d,h}, T_{i,d,h}\right)
% \end{gathered}
% $$

$$
\begin{aligned}
\widehat{P}_{i,d,h}=\mathcal{F}\left(\widehat{SSRD}_{i,d,h}, \widehat{DNI}_{i,d,h},  \widehat{T}_{i,d,h}\right), \quad i=1,\dots, n\\
P_{i,d,h}=\mathcal{F}\left(SSRD_{i,d,h}, DNI_{i,d,h}, T_{i,d,h}\right),  \quad i=1,\dots, n
\end{aligned}
$$
where $\mathcal{F}$ is the conversion function provided by the PVlib package in Python. \bigskip

We also define forecast errors for weather variables in the reference station and solar production loss of each park $i$, which are our variables of interest:

% $$
% \begin{gathered}
% \Delta P_{i,d,h}  = \widehat{P}_{i,d,h} - P_{i,d,h} \\
% \Delta SSRD_{0,d,h}  = \widehat{SSRD}_{0,d,h} - SSRD_{0,d,h}  \\
% \Delta DNI_{0,d,h}  = \widehat{DNI}_{0,d,h}  - DNI_{0,d,h}  \\
% \Delta T_{0,d,h}  = \widehat{T}_{0,d,h}  - T_{0,d,h} 
% \end{gathered}
% $$

$$
\begin{aligned}
\Delta P_{i,d,h} &= \widehat{P}_{i,d,h} - P_{i,d,h},  \quad i=1,\dots, n, \\
\Delta SSRD_{0,d,h} &= \widehat{SSRD}_{0,d,h} - SSRD_{0,d,h}, \\
\Delta DNI_{0,d,h} &= \widehat{DNI}_{0,d,h} - DNI_{0,d,h}, \\
\Delta T_{0,d,h} &= \widehat{T}_{0,d,h} - T_{0,d,h}.
\end{aligned}
$$

\subsubsection{Step 2: Aggregation to daily data}
% To convert hourly data into daily values, we aggregate the variables as follows:
% $$
% \begin{gathered}
% \Delta SSRD_{0,d} = \sum_{h = h_{\text{sunrise}}}^{h_{\text{sunset}}} \Delta SSRD_{0,d,h} \\ 
% \Delta DNI_{0,d}= \sum_{h = h_{\text{sunrise}}}^{h_{\text{sunset}}} \Delta DNI_{0,d,h}\\
% \quad \Delta T_{0,d}=\frac{1}{h_{\text{sunset}} - h_{\text{sunrise}} + 1} \sum_{h = h_{\text{sunrise}}}^{h_{\text{sunset}}} \Delta T_{0,d,h} \\ \quad \Delta P_{i,d}=\sum_{h = h_{\text{sunrise}}}^{h_{\text{sunset}}} \Delta P_{i,d,h}
% \end{gathered}
% $$

% Il manque une explication de pourquoi passer de hourly à daily 

We aggregate losses on a daily scale, recognizing the natural hourly seasonality of solar production. Since solar generation follows a predictable diurnal cycle, assessing losses at an hourly level would be impractical, as it would require accounting for systematic variations throughout the day. Moreover, structuring an insurance contract for each hour would be unrealistic. The daily scale provides a more practical and interpretable measure of forecasting discrepancies. However, this choice does not impose any fundamental constraint, and alternative time steps could be considered without loss of generality.

We filter the data to include only daytime hours, which vary by day and location. Let $h_{\text{sr},d}$ and $h_{\text{ss},d}$ represent the sunrise and sunset hours on day $d$, respectively. The daily aggregated variables are then computed as follows:

$$
\begin{aligned}
\Delta SSRD_{0,d} &= \sum_{h = h_{\text{sr},d}}^{h_{\text{ss},d}} \Delta SSRD_{0,d,h}, \\ 
\Delta DNI_{0,d} &= \sum_{h = h_{\text{sr},d}}^{h_{\text{ss},d}} \Delta DNI_{0,d,h}, \\
\Delta T_{0,d} &= \frac{1}{h_{\text{ss},d} - h_{\text{sr},d} + 1} \sum_{h = h_{\text{sr},d}}^{h_{\text{ss},d}} \Delta T_{0,d,h}, \\
\Delta P_{i,d} &= \sum_{h = h_{\text{sr},d}}^{h_{\text{ss},d}} \Delta P_{i,d,h}.
\end{aligned}
$$

To obtain daily values for the reference variables, we aggregate over daylight hours while considering the physical nature of each variable. For SSRD and DNI, accumulation is the appropriate approach, as these quantities represent energy fluxes that naturally sum over time. In contrast, temperature is not cumulative but rather a state variable. Therefore, we compute its daily average to better reflect the thermal conditions affecting solar panel efficiency. In the rest of the paper, the variable $\Delta T_{0,d}$ will no longer be considered, as it has no significant impact on individual park losses $\Delta P_{i,d}$.

Figure \ref{fig:illustration1} presents the daily production loss $\Delta P_{i,d}$ for the solar farm ``Kiemertsh\-ofen'' (the cyan point in Figure \ref{fig:illustrationadd3}). Figures \ref{fig:illustration2} and \ref{fig:illustration4} illustrate the time series $\Delta SSRD_{0,d}$ and $\Delta DNI_{0,d}$ for the reference point Munich represented in red in Figure \ref{fig:illustrationadd3}. Monthly heteroscedasticity is observed in these time series, and to mitigate this effect — as well as potential prediction bias in the model from the PVlib Python package — deterministic monthly stationarization is applied by centering and scaling the data.

\subsubsection{Step 3: Stationarization of data}

The data is grouped by month $m$ and year $y$ for each time series, and the corresponding monthly statistics are computed. The monthly average is given by  

$$
\mu_{m, y}^{\Delta X} = \frac{1}{N_{m, y}} \sum_{d \in \mathcal{D}_{m,y}} \Delta X_{d},
$$  
while the monthly standard deviation is computed as  

$$
\sigma_{m, y}^{\Delta X} = \sqrt{\frac{1}{N_{m, y}-1} \sum_{d \in \mathcal{D}_{m,y}} \left(\Delta X_{d} - \mu_{m, y}^{\Delta X} \right)^2}.
$$  
Here, $\mathcal{D}_{m,y}$ denotes the set of days in month $m$ of year $y$, and $N_{m,y}$ represents the number of days in this set. These statistics allow for the normalization of each time series, ensuring that deviations are evaluated relative to typical monthly conditions.\newline
Each time series is then standardized to obtain stationary variables:
% $$
% X_{i,d} = \frac{\Delta P_{i,d} - \mu_{m,y}^{\Delta P}}{\sigma_{m,y}^{\Delta P}}, \quad 
% Y^{(1)}_{0,d} = \frac{\Delta SSRD_{0,d} - \mu_{m,y}^{\Delta SSRD}}{\sigma_{m,y}^{\Delta SSRD}}, \quad 
% Y^{(2)}_{0,d} = \frac{\Delta DNI_{0,d} - \mu_{m,y}^{\Delta DNI}}{\sigma_{m,y}^{\Delta DNI}}, \quad \forall d \in \mathcal{D}_{m,y}.
% $$

$$
X_{i,d} = \frac{\Delta P_{i,d} - \mu_{m,y}^{\Delta P}}{\sigma_{m,y}^{\Delta P}}, \quad
Y^{(1)}_{0,d} = \frac{\Delta SSRD_{0,d} - \mu_{m,y}^{\Delta SSRD}}{\sigma_{m,y}^{\Delta SSRD}}, 
$$
$$
Y^{(2)}_{0,d} = \frac{\Delta DNI_{0,d} - \mu_{m,y}^{\Delta DNI}}{\sigma_{m,y}^{\Delta DNI}}, 
\quad \forall d \in \mathcal{D}_{m,y}.
$$

In Figures \ref{fig:illustrationf}, \ref{fig:illustration5} and \ref{fig:illustration7}, we represent the stationarization of the production loss $X_{i,d}$ of the solar park ``Kiemertshofen" and the covariates $Y^{(1)}_{0,d}$ and $Y^{(2)}_{0,d}$.

\subsubsection*{Step 4: Estimation of the functions $m_i(.)$ via Gaussian GLM}

In this step, we estimate the functions $m_i(.)$ in Equation \eqref{fctmi}, which will later be used to approximate the threshold $z_0$ for triggering the index. These functions describe the relationship between production losses and weather forecast errors at the reference location. To estimate them, we fit a Gaussian GLM for each solar park $i$:
 
$$
    \mathbb{E}\left[X_{i,d} \mid 
 Y^{(1)}_{0,d}, Y^{(2)}_{0,d}\right]  = \tilde{a}_{i,1} Y^{(1)}_{0,d} + \tilde{a}_{i,2} Y^{(2)}_{0,d}.
$$

We present in Figure \ref{fig:illustration8} the boxplot of the fitted Gaussian GLM coefficients by covariate. All covariates are statistically significant for every solar park at the $5\%$ level. As expected, the signs of the coefficients remain consistent across all parks, and their values exhibit slight variation between sites. Figure \ref{fig:illustrationadd9} illustrates the scatter plot of the realized losses against the Gaussian GLM predictions for the solar parks ``Oberrammersdorf", ``Hilpoltstein", ``Kiemertshofen", and ``Illesheim". These figures reveal a linear relationship between the losses and the weather variables. The Gaussian GLM provides a suitable initial approach when applied to the full dataset, ensuring a well-defined baseline for coefficient estimation. However, as we later introduce a thresholding step that restricts the dataset, this selection process induces asymmetry in the distribution of residuals. To account for this, we adopt the Tweedie GLM, which is better suited to handle skewed distributions and potential zero-inflation in the data. Furthermore, both $Y^{(1)}_{0,d}$ and $Y^{(2)}_{0,d}$ positively impact the financial losses of the parks, which is consistent with expectations, as an underestimation of solar irradiance leads to more significant production losses. The time series of the prediction and realized losses are represented in Figure \ref{fig:illustration9}. 

\bigskip

We now construct a first linear score that will prefigure the weather index with the coefficients $\tilde{a}_{i,j}$ as follows. We start by averaging coefficients across all parks:
\begin{equation}
    \bar{\tilde{a}}_j = \frac{1}{n} \sum_{i=1}^{n} \tilde{a}_{i,j}, \quad j \in \{1,2\}.
\end{equation}
The first linear score is then constructed as:  

\begin{equation}
    \tilde{Z}_d = \bar{\tilde{a}}_1 \times Y^{(1)}_{0,d} + \bar{\tilde{a}}_2 \times Y^{(2)}_{0,d}.
\end{equation}

Figure \ref{fig:illustration10} represents the daily times series $\tilde{Z}_d$. As this linear score is constructed from weather forecast errors, it is expected to exhibit a strong positive correlation with solar production losses. Figure \ref{fig:illustration11} illustrates this relation by representing the scatter plot of production losses with respect to $\tilde{Z}_d$.

\subsubsection{Step 5: Thresholding and data filtering}

To focus on significant production losses, we define a threshold based on the 80\% quantile of $\tilde{Z}_d$:  

\begin{equation}
    \widehat{z_0} = Q_{80\%}(\tilde{Z}_d).
\end{equation}
We keep only days where $\tilde{Z}_d > \widehat{z_0}$, defining the filtered dataset:
$$\mathcal{D}_f = \{ d \mid \tilde{Z}_d > \widehat{z_0} \}.$$

The true threshold $z_0$ is computed once the index is entirely constructed. However, in this case study, the approximation of the threshold obtained $\widehat{z_0}$ is close to the final true value, making this value a reliable early indicator. This threshold ensures that only the most impactful $20\%$ of forecast errors are considered for insurance coverage, representing the upper fifth of potential losses. The choice of the 80\% quantile reflects a trade-off between two key objectives: ensuring a sufficient number of observations for training the GLM while maintaining a realistic insurance design. In practice, insurance contracts rarely aim for complete risk aversion, and covering only extreme losses (e.g., beyond the 95\% quantile) would reduce the number of relevant data points for statistical modeling. The $80\%$ threshold is thus a reasonable compromise aligned with industry practices, balancing statistical robustness and practical applicability.

\subsubsection{Step 6: Tweedie GLM on filtered data}

We now fit a Tweedie Generalized Linear Model on the filtered data $\mathcal{D}_f$ that is for days $d$ such that $\tilde{Z}_d > \widehat{z_0}$. The Tweedie GLM is particularly suited for this task, as it can model right-skewed loss distributions while accommodating overdispersion, a common feature in financial loss data. Unlike a standard Gaussian GLM, which assumes normally distributed residuals, the Tweedie GLM captures the heavy-tailed nature of solar production losses, ensuring a more accurate representation of extreme events.

$$
\begin{aligned}
            \mathbb{E}\left[X_{i,d} \mid 
 \boldsymbol{Y}_{0,d}\right] & = \left( a_{i1} \times Y^{(1)}_{0,d}+ a_{i2}  \times Y^{(2)}_{0,d}\right)^{p_i}, \\
            \text{Var}\left[X_{i,d} \mid 
 \boldsymbol{Y}_{0,d}\right] & = \phi_i \left(\mathbb{E}\left[X_{i,d} \mid 
 \boldsymbol{Y}_{0,d}\right]\right)^{q_i}
            \end{aligned}
$$
where $\boldsymbol{Y}_{0,d} = \left(Y^{(1)}_{0,d}, Y^{(2)}_{0,d}\right)^{\prime}.$

We select a couple of hyperparameters $(p_i, q_i)$ by a grid search method that minimizes the deviances of the GLM Tweedie of solar park $i$. In this study, we search for $p_i \in [0.5, 2]$ and $q_i \in [0, 2]$. 

Figure \ref{fig:illustration12} presents the boxplot of the fitted Tweedie GLM coefficients for each covariate, while Figure \ref{fig:illustration13} illustrates the model's predictive performance.

\subsubsection{Step 7: Optimal index}
The objective now is to find an optimal weight $\boldsymbol{a}= \left(a_{1}, a_{2}\right)^{\prime}$ and, by consequence, an optimal weather index $Z_d=\boldsymbol{a}^{\prime} \boldsymbol{Y}_{0,d}$ that minimizes the variance of the after-sharing basis risk.\bigskip

We adopt the following steps:

\begin{enumerate}
    \item Assumption: the heterogeneity of the $\boldsymbol{a}_i$ is not too strong.

    \item Basis risk sharing
    
    The selected optimal index $Z_d$ is used to estimate the basis risks of each solar producer $i$ and for each day $d \in \mathcal{D}_f$:
    \begin{equation}
        \varepsilon_{i,d}= X_{i,d} - m_{i,d}(Z_d).
    \end{equation}
    
    Producers then share the aggregated total basis risk:
    
          $$S_{\varepsilon, d} = \sum_{i=1}^n \varepsilon_{i,d}.$$
    The after-sharing basis risk for producer $i$ is given by:
          $$
            \delta_{i,d} = \frac{\sigma_{i,d}^2(Z_d)}{\sum_{j=1}^n \sigma_{j,d}^2(Z_d)} \times S_{\varepsilon, d}
          $$
    where $\sigma_{i,d}^2(Z_d)$ is given in Equation \eqref{varapprox}.

We select $Z_d$ to minimize the sum of the expected variances of the individual basis risks:

$$\sum^{n}_{i=1} \mathbb{E}\left[ \sigma_{i,d}^2(Z_d)\mid Z_d > \widehat{z_0}\right].$$
The remaining steps below propose an approximation of $\sigma_{i,d}^2(Z_d)$ as a function of the chosen common weight $\boldsymbol{a}$. 

    \item Approximation of the conditional expectation $\mathbb{E}\left[X_{i,d} \mid Z_d=z\right]$, $z>\widehat{z_0}$ using Equation \eqref{expansionformula}

\item Approximation of the conditional variance $\text{Var}\left[X_{i,d} \mid Z_d=z\right]$, $z>\widehat{z_0}$ \newline

The conditional variance of $X_{i,d}$ given $Z_d = z$ is expressed as:
$$
\text{Var}[X_{i,d} \mid Z_d = z] = \phi_i \left(\mathbb{E}[X_{i,d} \mid Z_d = z]\right)^{q_i},
$$
and is approximated using Equation \eqref{varapprox}.
\item Final optimisation

We solve Problem \eqref{problem} to find the optimal coefficient $\boldsymbol{a}$. The detailed methodology used to compute the objective function on real data is presented in Appendix \ref{app3}. The optimization is solved with Basin Hopping in Python's Scipy package.\bigskip

The optimal coefficient obtained with $\mathcal{D}_f$ is 
\begin{equation}
  \boldsymbol{a}=(0.47, 0.16)^{\prime},  
\end{equation}
which is acceptable because it is not too far from the $\boldsymbol{a}_i$.\newline
We deduce the following optimal index for each day $d$ in the observation period:
$$Z_d=a_1 \times Y^{(1)}_{0,d} + a_2 \times Y^{(2)}_{0,d}.$$ The corresponding time series $Z_d$ is represented in Figure \ref{fig:illustration14}.

\begin{figure}[h!]
        \centering
                \includegraphics[width=0.9\linewidth]{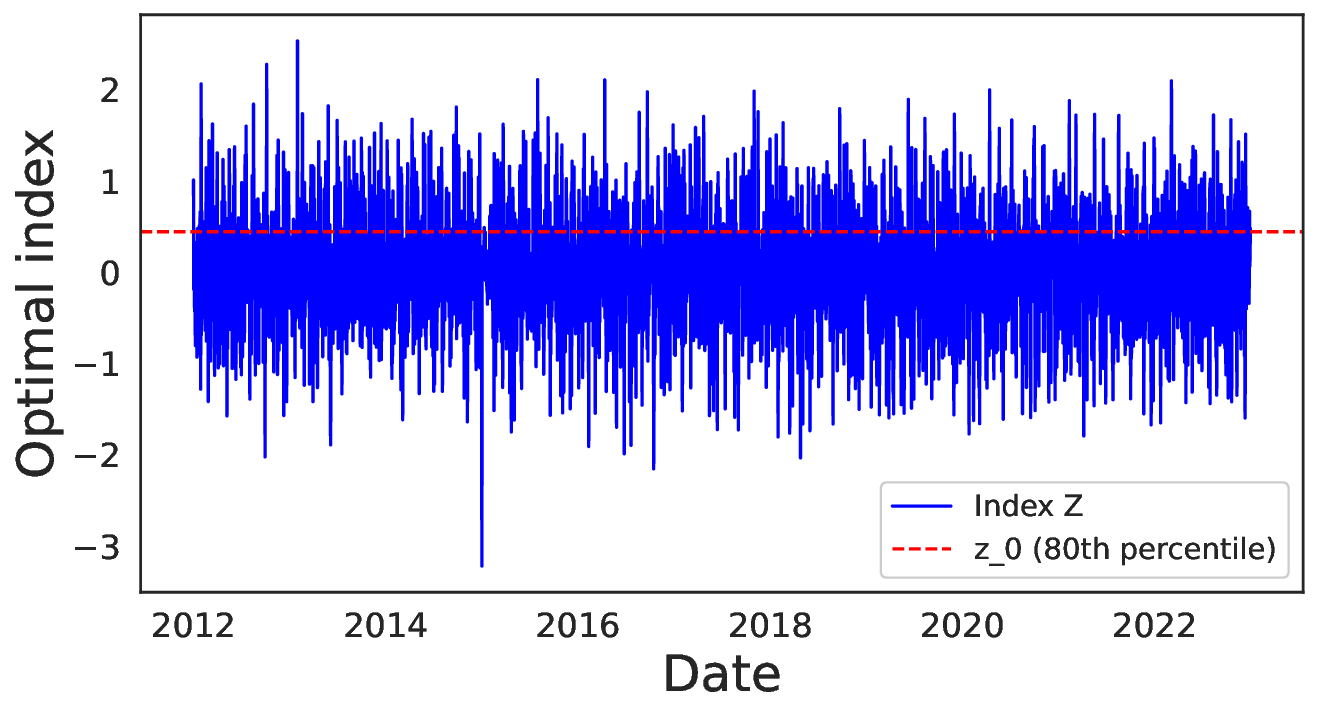}
                \caption{Index $Z$.}
                 \label{fig:illustration14}
\end{figure}

\item Choice of the threshold $z_0$ 

Now that the index has been obtained, we can determine the true threshold $z_0$ as follows:
$$z_0=Q_{80\%}(Z_d),$$
and is illustrated by the horizontal red line in Figure
 \ref{fig:illustration14}.

\end{enumerate}

\subsubsection{Method for moving from stationary to real basis risk}
The quantities $\varepsilon_{i,d}$ and $\delta_{i,d}$ represent the stationary before and after sharing basis risks. In practice, we need to translate these into their real-world counterparts, denoted as $\varepsilon^{*}_{i,d}$ and $\delta^{*}_{i,d}$, which account for the original data scale and variability.
\bigskip

We adopt the following steps:

\begin{enumerate}
    \item The real before-sharing basis risk $\varepsilon^{*}_{i,d}$ is given by:
    $$\varepsilon^{*}_{i,d}=\sigma_{m, y}^{\Delta P_{i,d}} \times \varepsilon_{i,d}.$$
    
    \item We calculate the aggregate real basis risk as follows: $$S^{*}_{\varepsilon, d}=\sum^{n}_{i=1}\varepsilon^{*}_{i,d}.$$
    
    \item The real after-sharing basis risk $\delta^{*}_{i,d}$ is given by:
        
        $$
        \delta^{*}_{i,d}=\frac{\sigma_{i,d}^2(Z_d)(\sigma_{m, y}^{\Delta P_{i,d}})^2}{\sum^{n}_{j=1}\sigma_{j,d}^2(Z_d)(\sigma_{m, y}^{\Delta P_{j,d}})^2} \times S^{*}_{\varepsilon, d} \quad \text{for} \ d \in \mathcal{D}_f  \cap \mathcal{D}_{m,y}.
        $$
\end{enumerate}

\subsection{Results}

% To do:
% Rajouter les résultats ici
% Mettre les graphiques B15 et B16
% Il faut une interprétation des résultats avec des chiffres clés au dela de la performance du modèle (par exemple combien est l'indemnité avant et après l'assurance)
% Est-ce que 100\% des fermes ont "gagné" par rapport à la situation initiale ? Sinon combien ?

\begin{figure}[h!]
\begin{subfigure}{.5\textwidth}
  \centering
  \includegraphics[width=.9\linewidth]{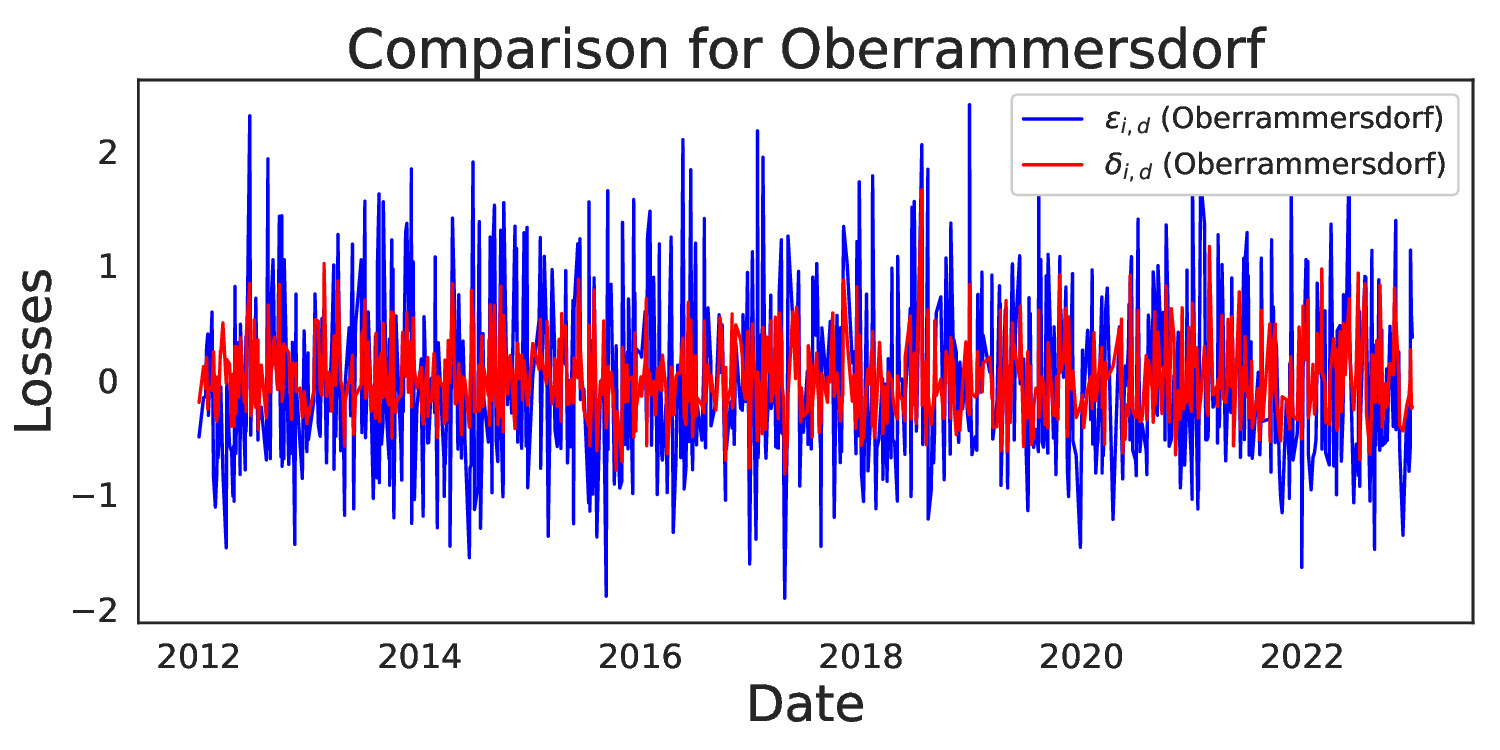}
  \caption{}
  \label{fig:sfig13}
\end{subfigure}%
\begin{subfigure}{.5\textwidth}
  \centering
  \includegraphics[width=.9\linewidth]{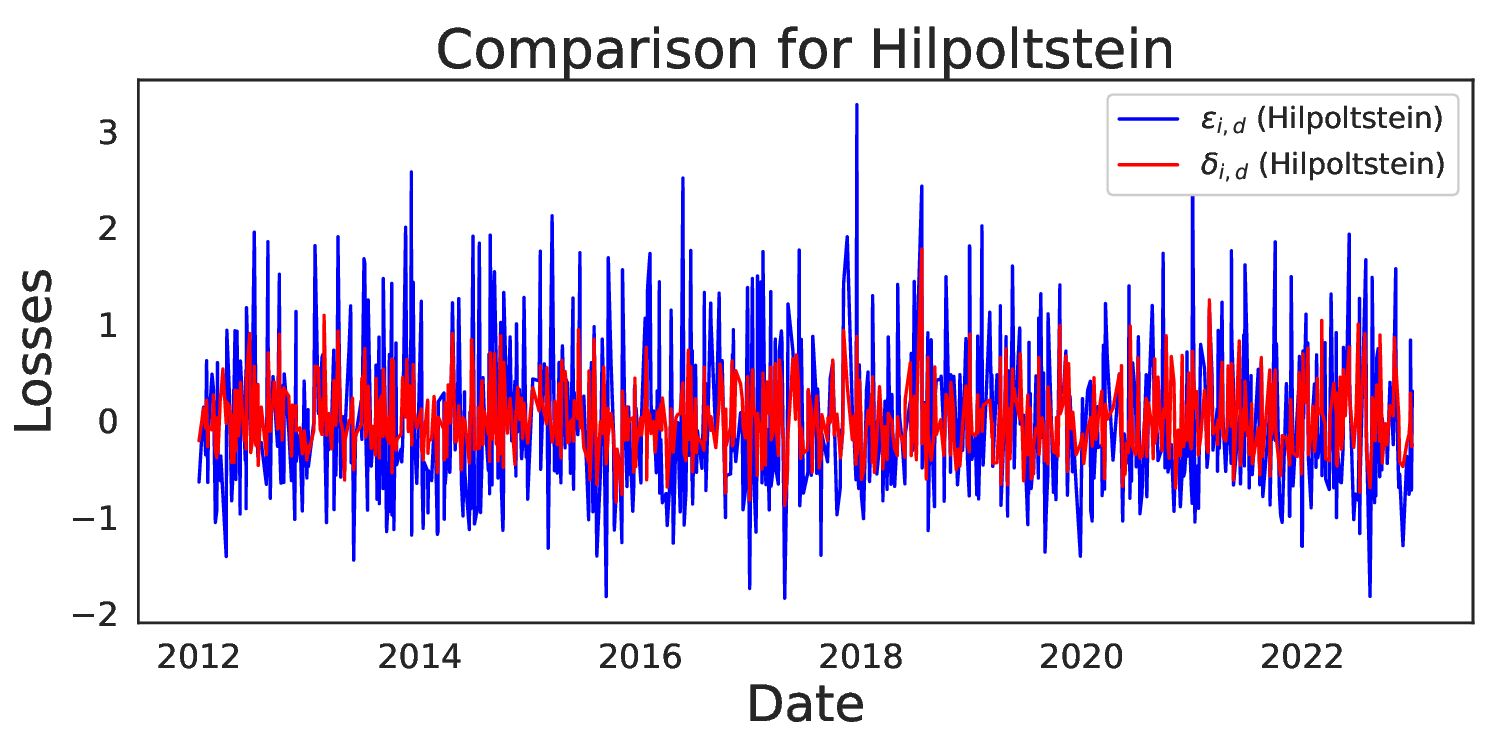}
  \caption{}
  \label{fig:sfig14}
\end{subfigure}
\begin{subfigure}{.5\textwidth}
  \centering
  \includegraphics[width=.9\linewidth]{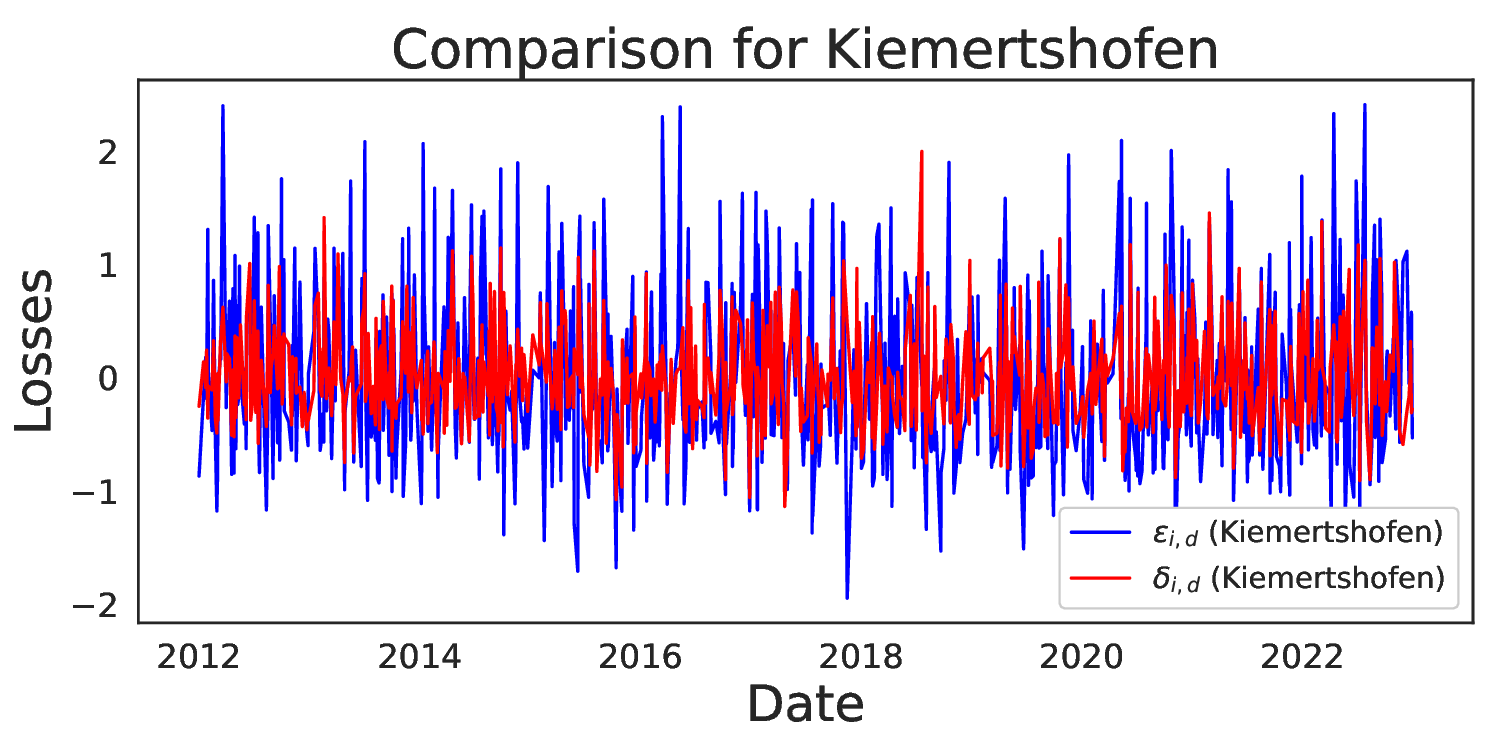}
  \caption{}
  \label{fig:sfig15}
\end{subfigure}
\begin{subfigure}{.5\textwidth}
  \centering
  \includegraphics[width=.9\linewidth]{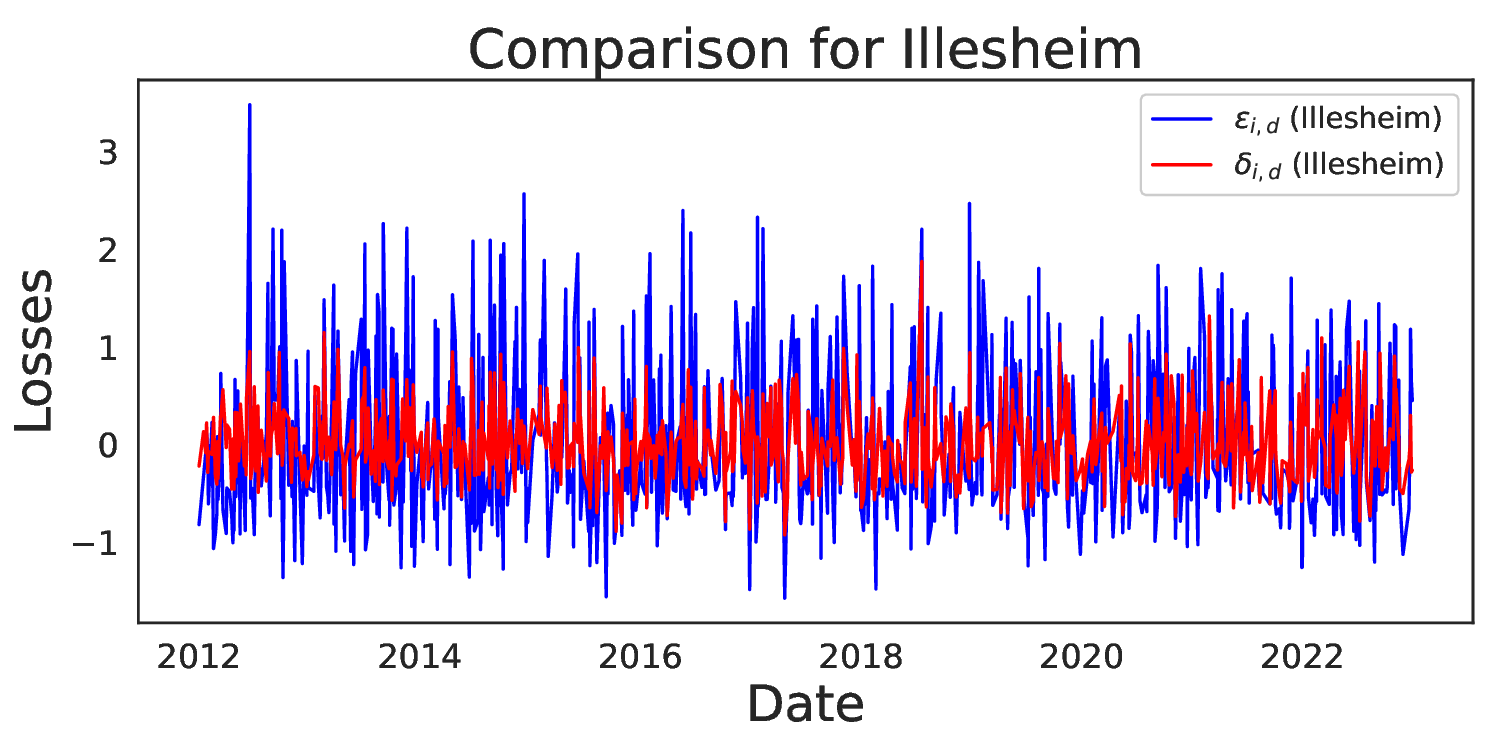}
  \caption{}
  \label{fig:sfig16}
\end{subfigure}
\caption{Stationary after and before sharing basis risk $\varepsilon_{{i,d}}$ and $\delta_{{i,d}}$.}
\label{fig:illustration15}
\end{figure}

\begin{figure}[h!]
\begin{subfigure}{.5\textwidth}
  \centering
  \includegraphics[width=.9\linewidth]{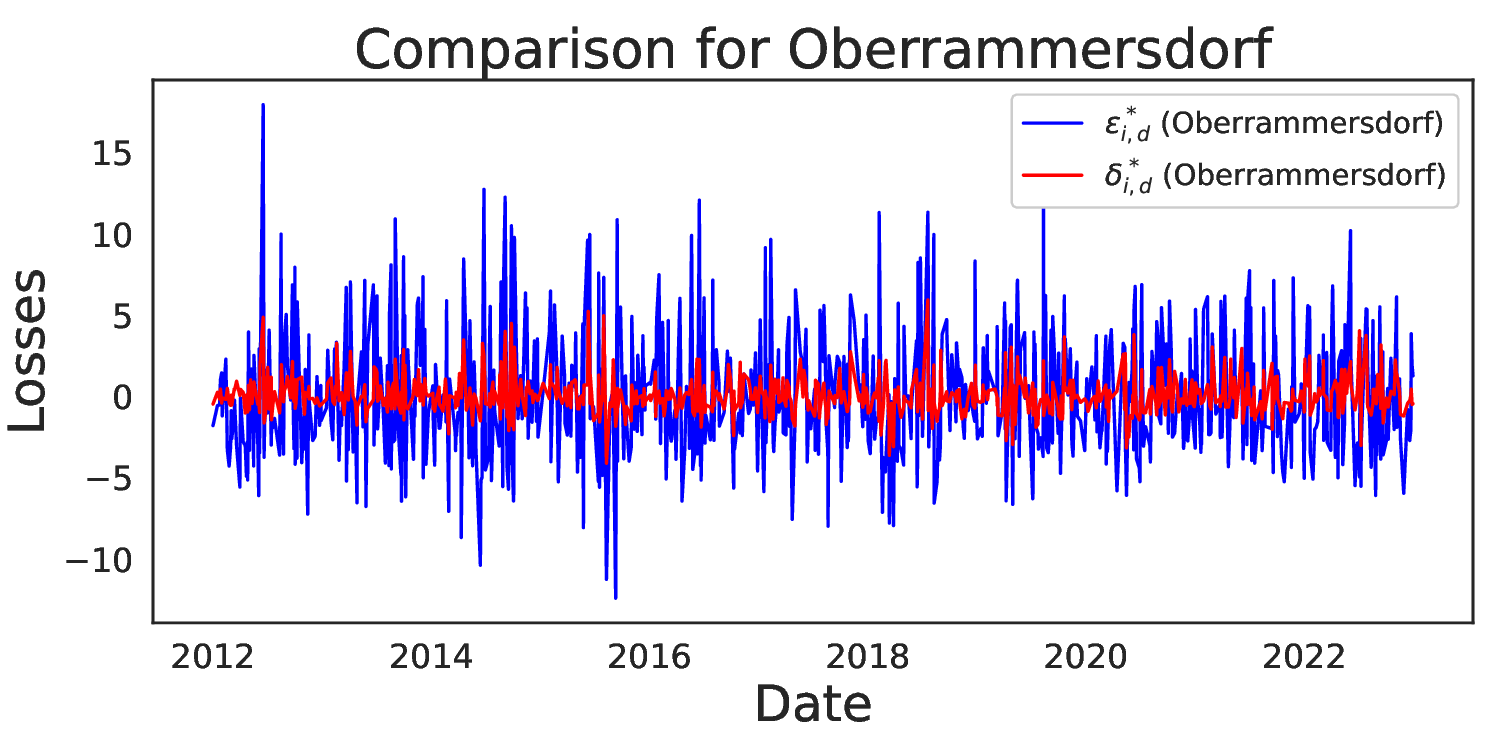}
  \caption{}
  \label{fig:sfig17}
\end{subfigure}%
\begin{subfigure}{.5\textwidth}
  \centering
  \includegraphics[width=.9\linewidth]{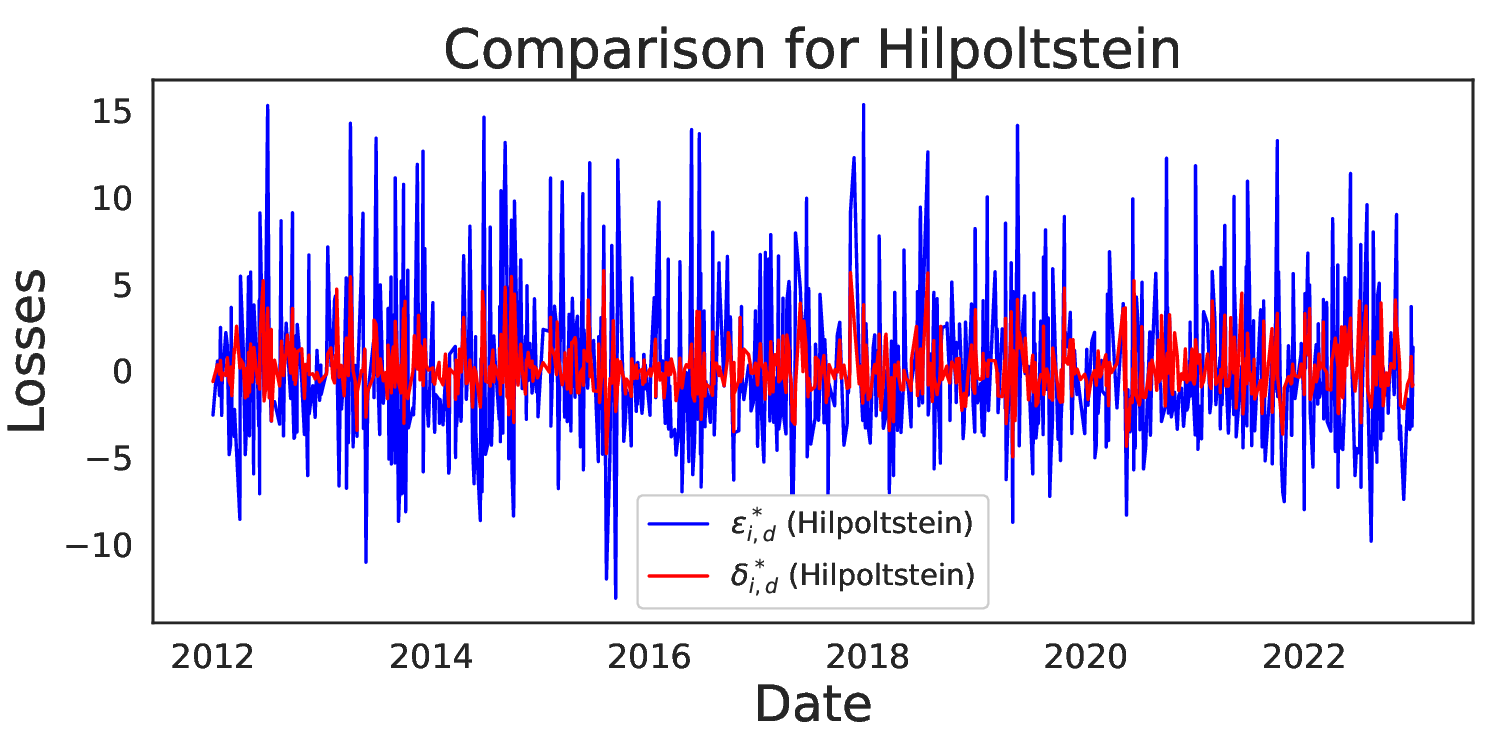}
  \caption{}
  \label{fig:sfig18}
\end{subfigure}
\begin{subfigure}{.5\textwidth}
  \centering
  \includegraphics[width=.9\linewidth]{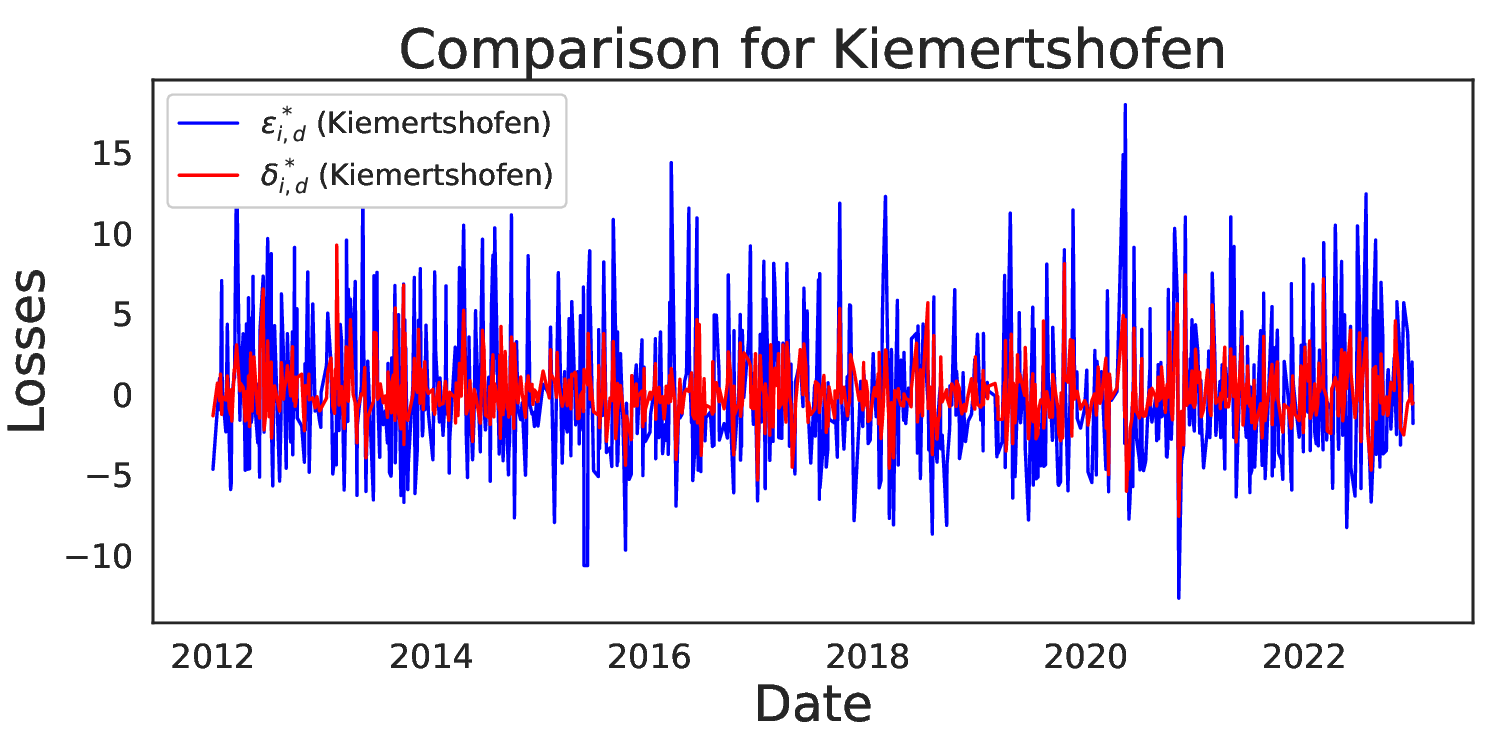}
  \caption{}
  \label{fig:sfig19}
\end{subfigure}
\begin{subfigure}{.5\textwidth}
  \centering
  \includegraphics[width=.9\linewidth]{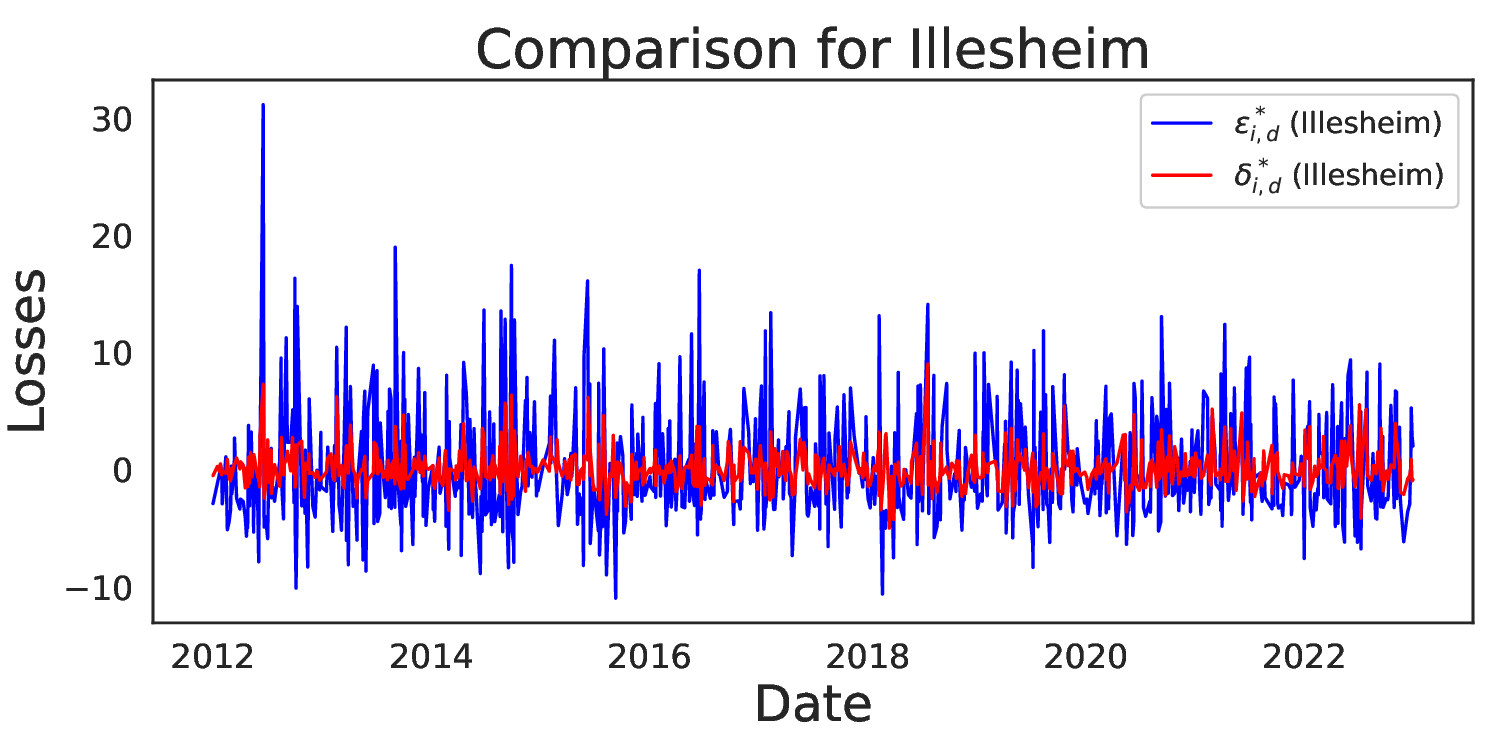}
  \caption{}
  \label{fig:sfig20}
\end{subfigure}
\caption{Real-world counterparts after and before sharing basis risk $\varepsilon^*_{{i,d}}$ and $\delta^*_{{i,d}}$.}
\label{fig:illustration16}
\end{figure}

\begin{figure}[h!]
        \centering
                \includegraphics[width=0.7\linewidth]{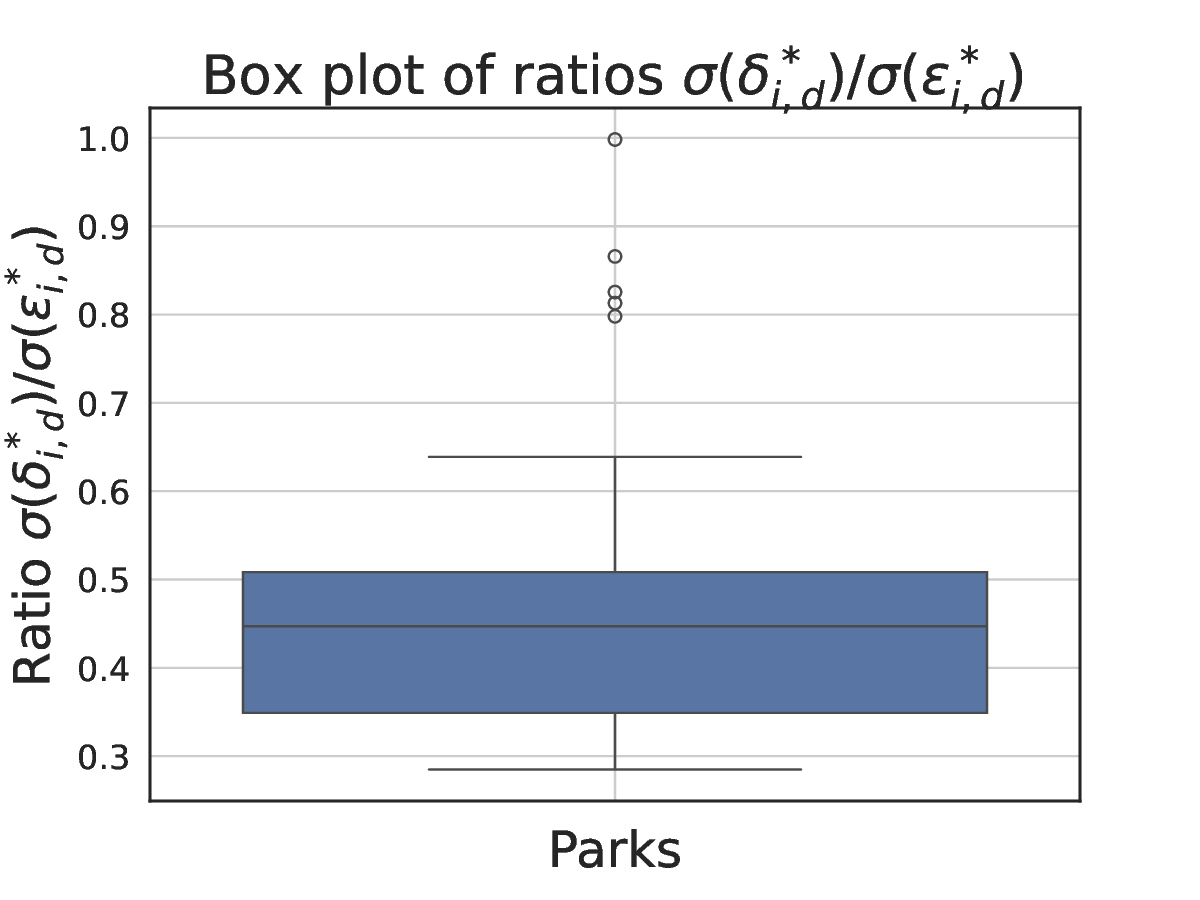}
                \caption{Boxplot of the ratio $\frac{\sigma(\delta^{*}_{i,d})}{\sigma(\varepsilon^{*}_{i,d})}$ of all parks.}
                 \label{fig:illustration17}
\end{figure}

% Moins de sous plot car on ne voit pas bien
% Imprimer les graphiques en haute résolution (.eps ou .pdf)
We compare in Figure \ref{fig:illustration15}, the evolution of the basis risk $\varepsilon_{i,d}$ and the shared basis risk $\delta_{i,d}$. We can see a significant reduction in the variability of the basis risk after risk sharing.
Their real-world counterparts $\varepsilon^{*}_{i,d}$ and $\delta^{*}_{i,d}$ are illustrated in Figure \ref{fig:illustration16}. We further quantify the percentage reduction of the riskiness of the basis risk by representing in Figure \ref{fig:illustration17} the Boxplot of the ratio $\frac{\sigma(\delta^{*}_{i,d})}{\sigma(\varepsilon^{*}_{i,d})}$ for all parks. One can see that this ratio is, on average, equal to $45\%$, which means that the volatility of the basis risk is reduced by $55\%$ after risk sharing. Moreover, Figure \ref{fig:illustrationa3} illustrates the spatial distribution of the ratio $\frac{\sigma(\delta^{*}_{i,d})}{\sigma(\varepsilon^{*}_{i,d})}$ across different parks, highlighting variations in risk reduction effectiveness. Parks with lower ratios (in green) experience greater volatility reduction, indicating that the risk-sharing mechanism mitigates financial uncertainty in these locations. Conversely, higher ratios (in red) suggest that certain parks contribute disproportionately to overall risk, limiting diversification benefits. We also include the location of the reference point to provide a geographical anchor for interpreting these variations. One can note that no clear spatial pattern emerges between its location and the observed variations in risk ratios. 

Now, we investigate the variability in volatility reduction across different solar parks (as observed in Figure \ref{fig:illustration17} and Figure \ref{fig:illustrationa3}) to better understand the factors driving heterogeneity in basis risk improvements. The objective is to explore whether characteristics such as park capacity and distance from the reference location influence the effectiveness of risk-sharing mechanisms.

\subsubsection{Volatility reduction and park capacity}
The top panel of Figure \ref{fig:illustration17add1} illustrates the relationship between volatility reduction and the capacity of each solar park. A clear negative correlation emerges: smaller solar parks benefit more from volatility reduction, while larger parks experience comparatively smaller gains, highlighting a size effect in the risk-sharing mechanism. Specifically, larger parks, which contribute more to the total production, tend to dominate the aggregate risk. As a result, their basis risk occupies a larger proportion of the total risk pool, which limits the potential for diversification.

\subsection{Volatility reduction and distance from reference location}
\begin{figure}[h!]
        \centering
                \includegraphics[width=0.7\linewidth]{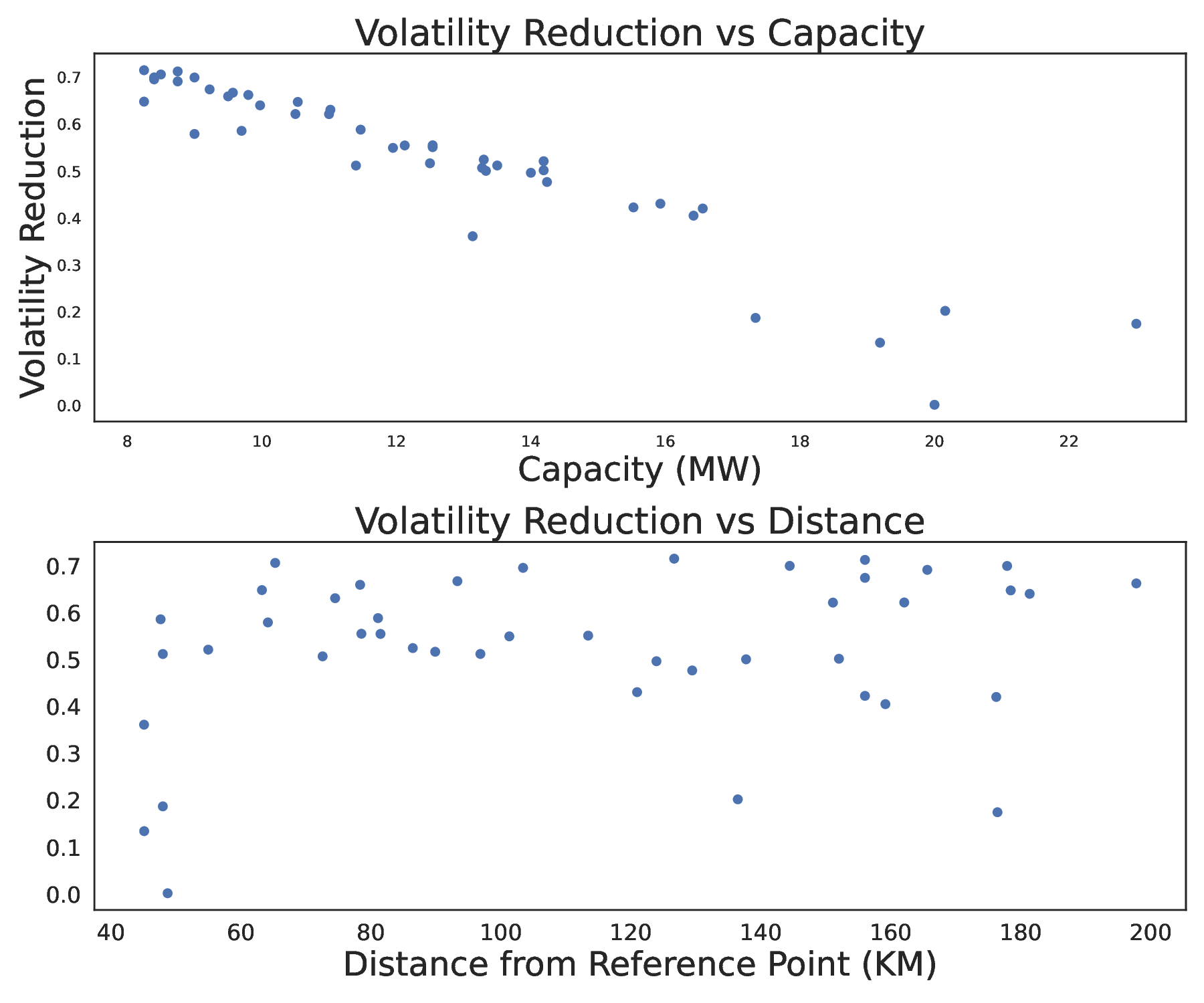}
                \caption{Relationship between volatility reduction vs capacity and distance.}
                 \label{fig:illustration17add1}
\end{figure}

The bottom panel of Figure \ref{fig:illustration17add1} shows the relationship between volatility reduction and the distance of each park from the reference point. Unlike the capacity relationship, no clear trend emerges here. Volatility reduction appears relatively scattered, with parks at varying distances experiencing high and low reductions. This suggests that the impact of distance is not purely linear and may depend on the spatial correlation structure of weather patterns. Parks far apart might face independent weather events, which could increase the benefits of risk-sharing, but the absence of a strong pattern indicates that distance alone is not the primary driver of volatility reduction.

\section{Discussion \label{sectresults}}

The selection of the 80\% quantile as a threshold appeared to be a reasonable trade-off between ensuring sufficient data for model training and maintaining a realistic level of risk aversion. However, this choice remains flexible and can be adjusted depending on the availability of reference data. With a larger dataset, it may be relevant to recalibrate the model using a higher threshold to focus on more extreme losses.

Since the choice of the reference point is crucial for the robustness of the weather index, a sensitivity analysis was conducted using alternative reference locations, corresponding to different geographical subdivisions of Germany. This evaluation confirmed the relevance of the selected reference location, as it provided satisfactory predictive performance.

Among the available methods for estimating solar production, we opted for a physical simulation approach rather than purely statistical models. This choice is motivated by the increasing frequency of rare and extreme weather events due to climate change. Physical models offer two key advantages: (i) they provide a precise and reliable relationship between weather indices and actual production, thereby enhancing the robustness of the weather index; (ii) they allow for scenario-based simulations, making it possible to assess the impact of climate variability on new solar installations even in the absence of historical production data.

The calculation of financial losses presents significant complexity, as it is likely to be individualized for each producer in practice. A major limitation of our approach is the absence of direct validation of estimated losses against observed financial losses in real-world settings. However, our methodology incorporates the most precise available data to minimize estimation bias. In particular, a producer relying on more distant forecasts would likely face greater financial losses due to the inherent uncertainty in weather predictions.

Moreover, our framework is designed to be adaptable. It could be directly applied by a consortium of producers with access to actual financial loss data. Regarding market timing, while a significant share of producers still receive full compensation through state subsidies, our methodology accounts for this by incorporating subsidy data where applicable instead of using market prices. Additionally, the need for a fine-scale evaluation of financial losses is less critical for producers not exposed to market fluctuations. As a result, our model tends to underestimate the losses of such producers due to its higher level of precision.

Ultimately, this modeling approach is aligned with a long-term perspective in which subsidies are expected to be gradually phased out, and producers are anticipated to fully participate in the market.

A key consideration is the comparative advantage of our model over Contracts-for-Differences (CfD) in managing financial risk. The first advantage of our approach is its immediate applicability, even in the presence of subsidies, whereas CfDs become more relevant once producers are fully exposed to market conditions. The second advantage is its superior protection against climate-related risks, which can be correlated with high market prices. In practice, during an extreme weather event, a CfD contract may prove detrimental to the producer: low solar production coincides with high electricity prices, requiring the producer to compensate for the price difference. Conversely, our approach compensates producers for reduced generation, explicitly accounting for the opportunity cost of high market prices.

However, our model does not offer protection against periods of excess renewable production, which tend to drive market prices downward—a scenario addressed by CfDs. Therefore, rather than being competing instruments, our approach and CfDs should be considered complementary risk management tools.

The proposed hybrid model, combining ex-ante parametric insurance with ex-post basis risk sharing, is both innovative and easy to implement. It relies on the cooperation of the insurer, who facilitates the purchase of a collective parametric contract for a pool of solar producers. Beyond managing parametric payouts, the insurer coordinates the ex-post sharing of basis risk among producers. This additional layer of risk pooling is mutually beneficial: producers gain from enhanced diversification of basis risk, reducing individual volatility, while the insurer strengthens its value proposition, potentially boosting group contract sales. This collaborative structure promotes collective resilience, aligning the interests of all parties involved.

\section{Conclusion}

In this paper, we proposed a novel P2P risk-sharing mechanism combined with a parametric insurance model based on a weather index. Our results show that P2P mechanisms can effectively mitigate the residual basis risk associated with common insurance indices. To illustrate this, we conducted a case study on German solar farms, analyzing the risk of production losses due to unexpected weather fluctuations. Since climate change is expected to increase the frequency of such fluctuations, stabilizing producers' revenues is crucial to sustaining investments in renewable energy.

Our model addresses this challenge by smoothing farm revenues through insurance against unexpected production drops while fostering a collaborative approach by pooling basis risks across farms. Our analysis demonstrates that risk sharing significantly reduces basis risk variability, with an average volatility reduction of 55\%. By comparing basis risk before and after risk sharing, we observe a substantial decrease in uncertainty. Additionally, smaller solar parks benefit more from this mitigation, while larger parks—dominating total production—experience comparatively smaller gains. The absence of a clear spatial pattern suggests that factors beyond geographical proximity influence the effectiveness of risk sharing. We also highlight the importance of methodological choices in constructing a robust weather index. Specifically, selecting an 80\% quantile threshold and carefully determining the reference location enhance the model’s predictive performance. Furthermore, opting for a physical simulation approach, rather than purely statistical models, improves robustness, particularly in accounting for extreme weather events.

Finally, our proposed hybrid model, which integrates ex-ante parametric insurance with ex-post basis risk sharing, offers a complementary risk management tool to traditional instruments like Contracts-for-Differences (CfDs). Unlike CfDs, our approach is immediately applicable, even in subsidized markets, and provides superior protection against climate-related risks. While CfDs help mitigate risks linked to price fluctuations, they do not directly address climate variability, making them complementary to our model. By promoting risk diversification and collective resilience, our framework supports a more stable and sustainable energy market.

Future research could refine the selection of the reference location by incorporating advanced statistical techniques such as optimal transport and Wasserstein barycentre. This would enhance the robustness of the weather index by identifying the most representative location for risk assessment. Additionally, our approach could be extended to other renewable energy sources, such as wind or hydroelectric power, focusing on longer-term forecasting and risk anticipation. 

Beyond the energy sector, the flexibility and modularity of our framework make it relevant for other industries exposed to weather-related uncertainties. In agriculture, for instance, the issue of basis risk has been long acknowledged as a key barrier to the adoption of index-based insurance. One promising avenue involves the development of collective index insurance schemes, where a group of farmers jointly purchases a parametric insurance contract and shares basis risk through informal or contractual within-group transfers. These mechanisms allow the group to pay a single collective premium, receive a single indemnity, and internally reallocate payouts based on individual discrepancies between actual losses and index-based triggers. Such pooling strategies, as studied by \citet{traerup2012informal} and \citet{santos2021dynamics}, offer a compelling parallel to our peer-to-peer compensation mechanism.

\section*{Acknowledgements}

Alicia Bassière's work has benefited from the support of the \textit{Agence Nationale de la Recherche} PowDev project under grant 22-PETA-0016.


\bibliography{P2P_risk_renewable}% common bib file

\pagebreak

%\setcounter{page}{1}
%\renewcommand{\theproperty}%{\Alph{section}.\arabic{property}}

%\begin{center}
%\textbf{\Large Appendix}
%\end{center}

% This appendix provides additional details on the methodologies and analyses conducted in this study. Appendix \ref{app1} outlines the numerical resolution methods, including value function estimation using machine learning techniques and simulation-based approaches. Appendix \ref{appendix-calib} describes the calibration of demand parameters and renewable capacity factors, alongside scenario visualizations used to support the main results.
\begin{appendices}

\section{Data representation \label{app1}}

\begin{figure}[H]
        \centering
                \includegraphics[width=0.9\linewidth]{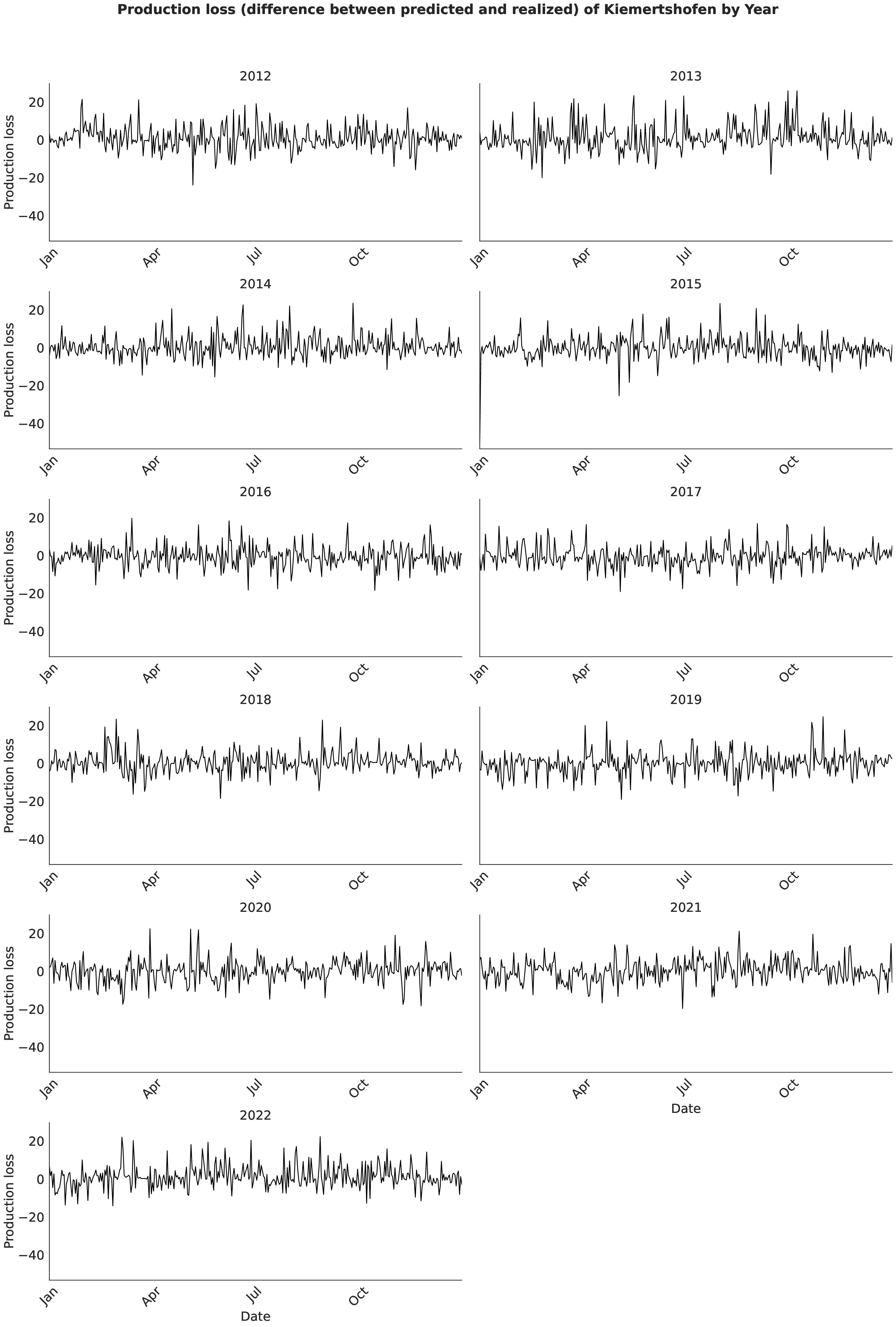}
                \caption{Production loss (difference between predicted and realized) of Kiemertshofen.}
                 \label{fig:illustration1}
\end{figure}

\begin{figure}[H]
        \centering
                \includegraphics[width=0.9\linewidth]{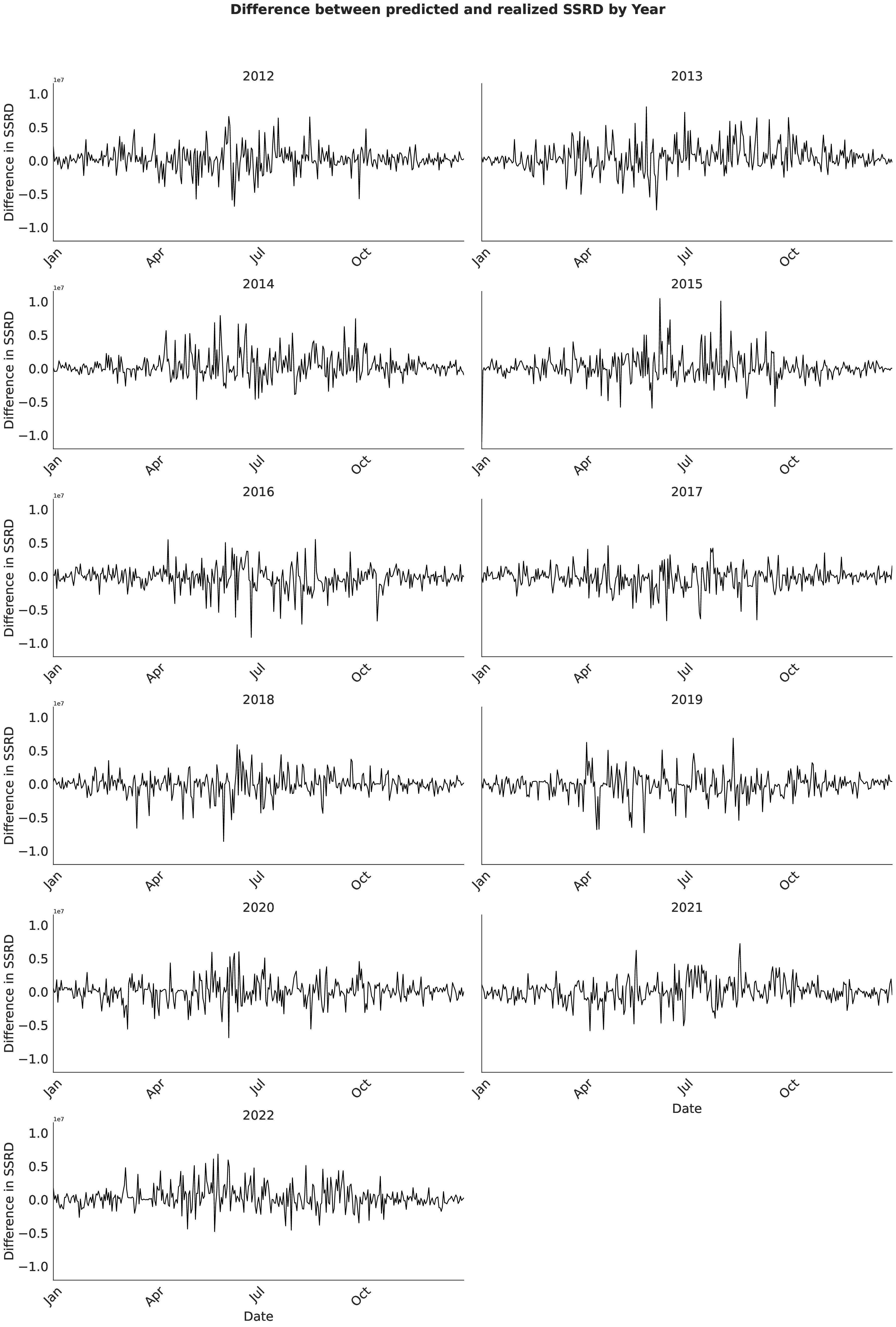}
                \caption{Difference between predicted and realized SSRD.}
                 \label{fig:illustration2}
\end{figure}

\begin{figure}[H]
        \centering
                \includegraphics[width=0.9\linewidth]{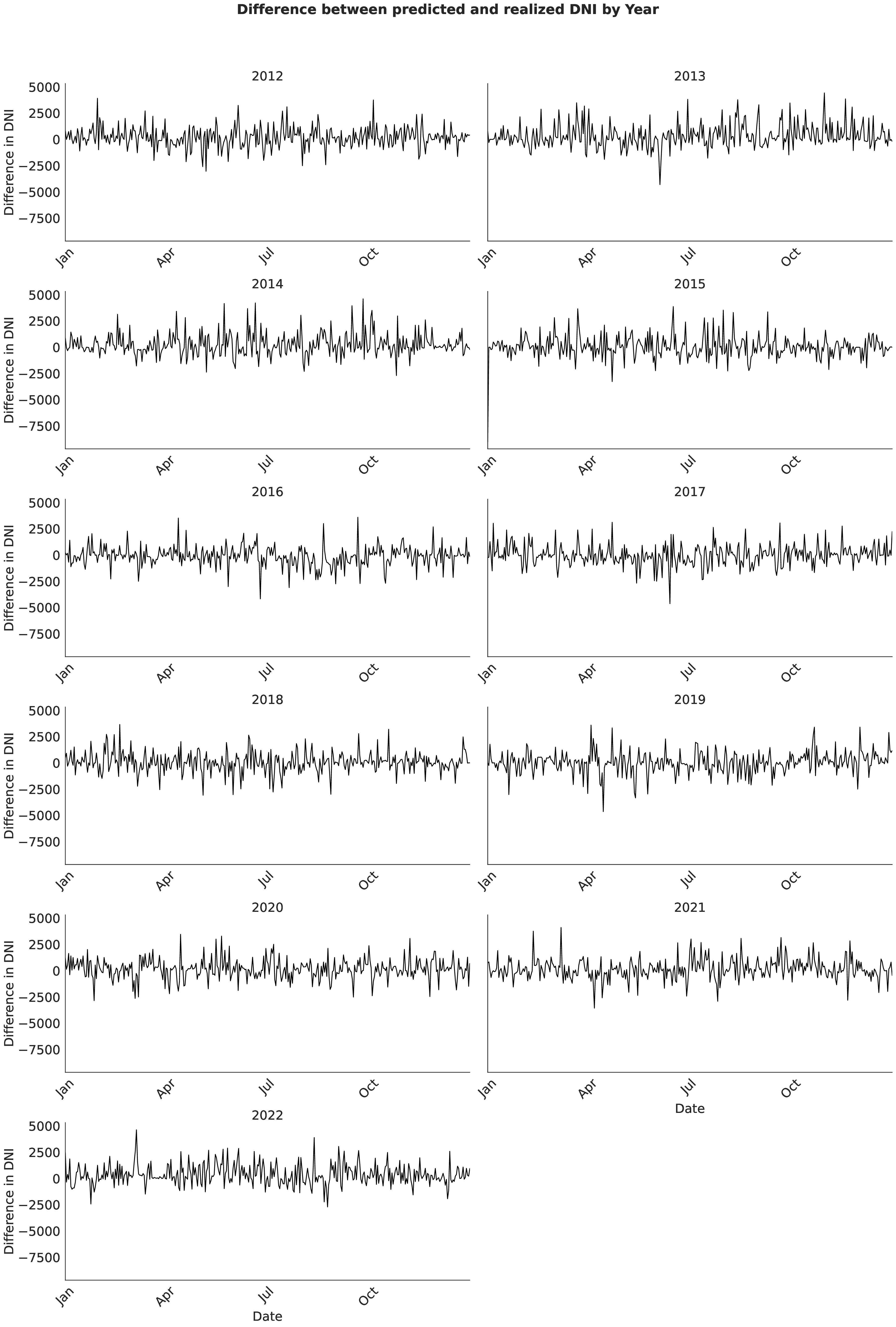}
                \caption{Difference between predicted and realized DNI.}
                 \label{fig:illustration4}
\end{figure}

\begin{figure}[H]
        \centering
                \includegraphics[width=0.9\linewidth]{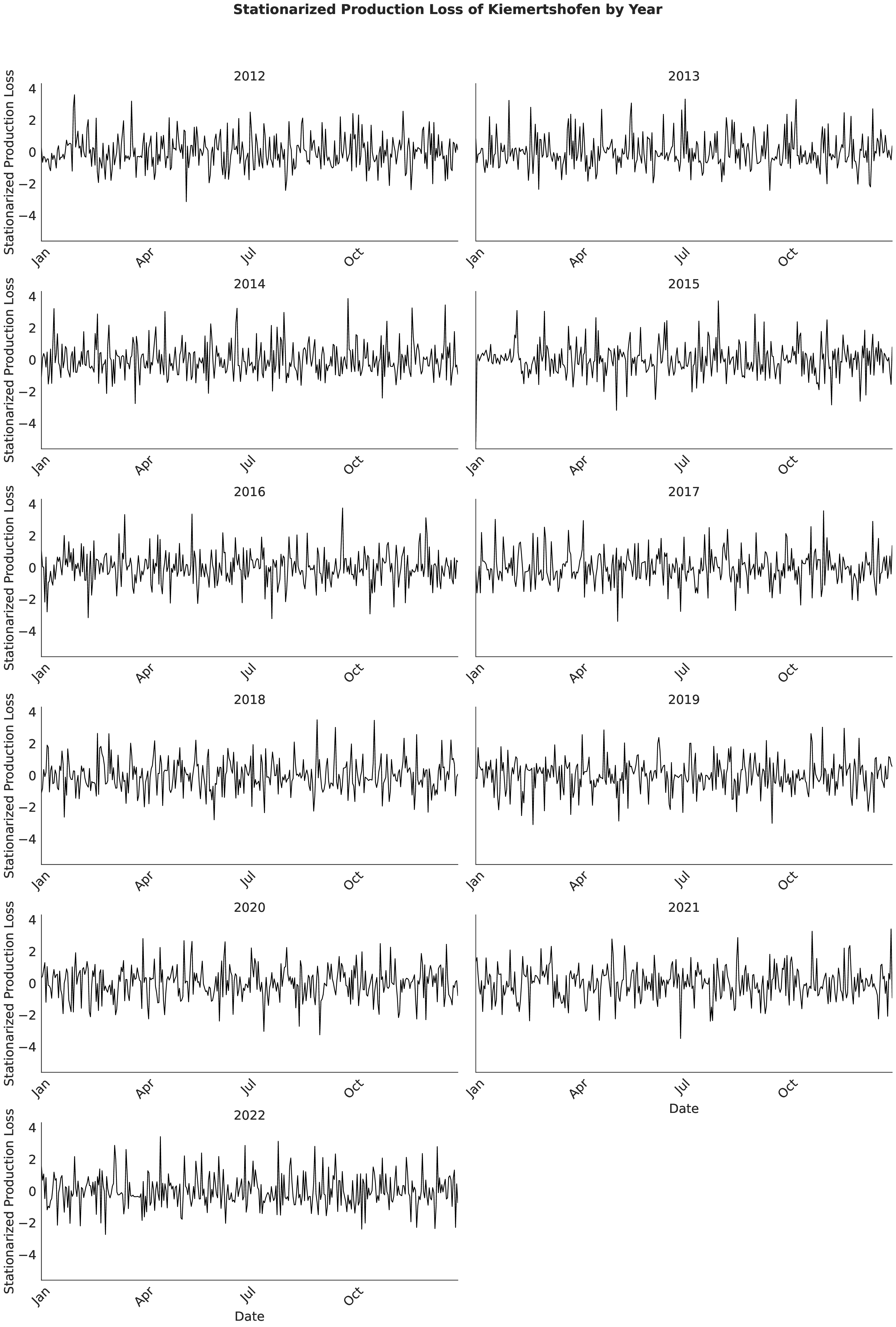}
                \caption{Stationnarized production loss of Kiemertshofen.}
                 \label{fig:illustrationf}
\end{figure}

\begin{figure}[H]
        \centering
                \includegraphics[width=0.9\linewidth]{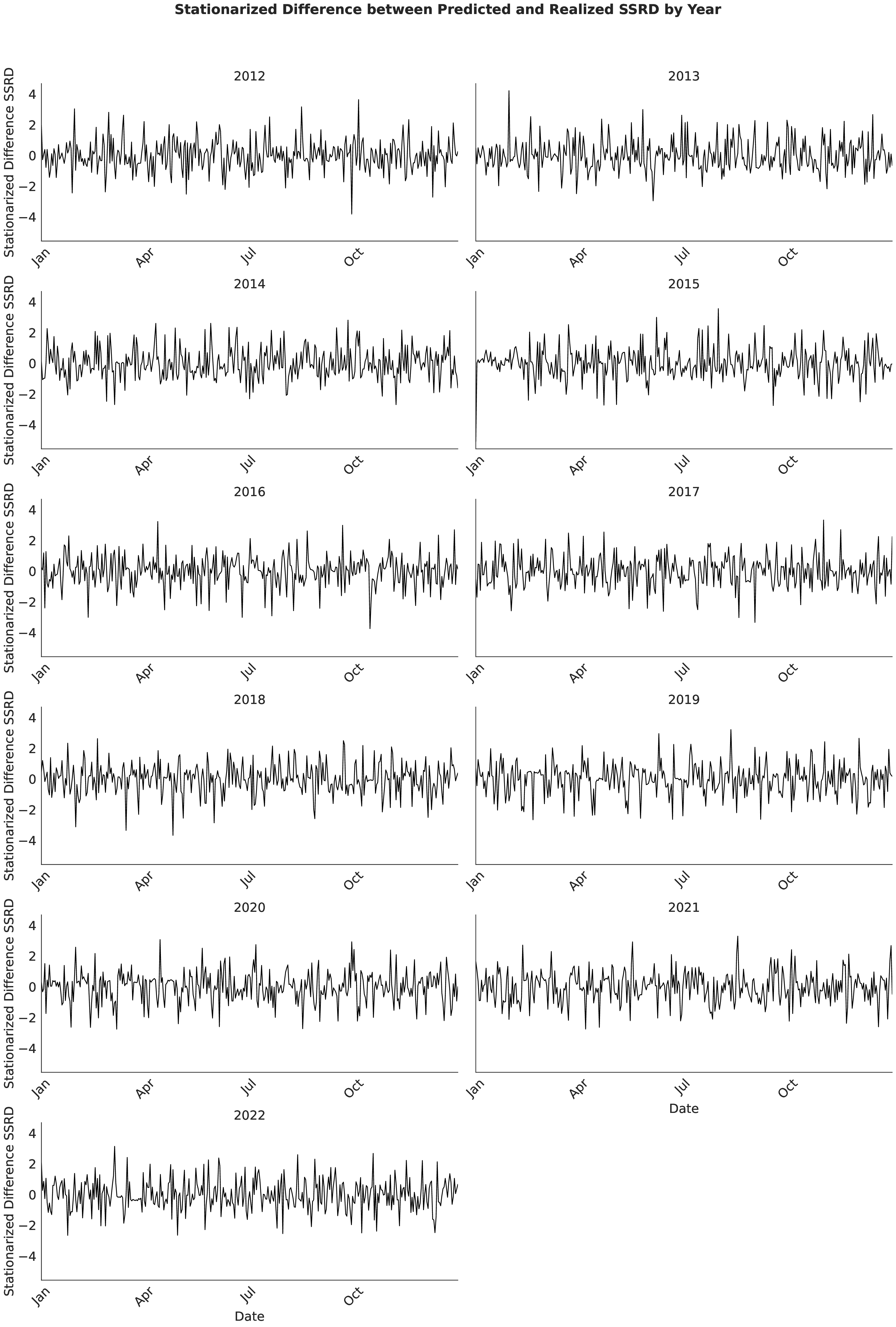}
                \caption{Stationarized of the difference between predicted and realized SSRD.}
                 \label{fig:illustration5}
\end{figure}

\begin{figure}[H]
        \centering
                \includegraphics[width=0.9\linewidth]{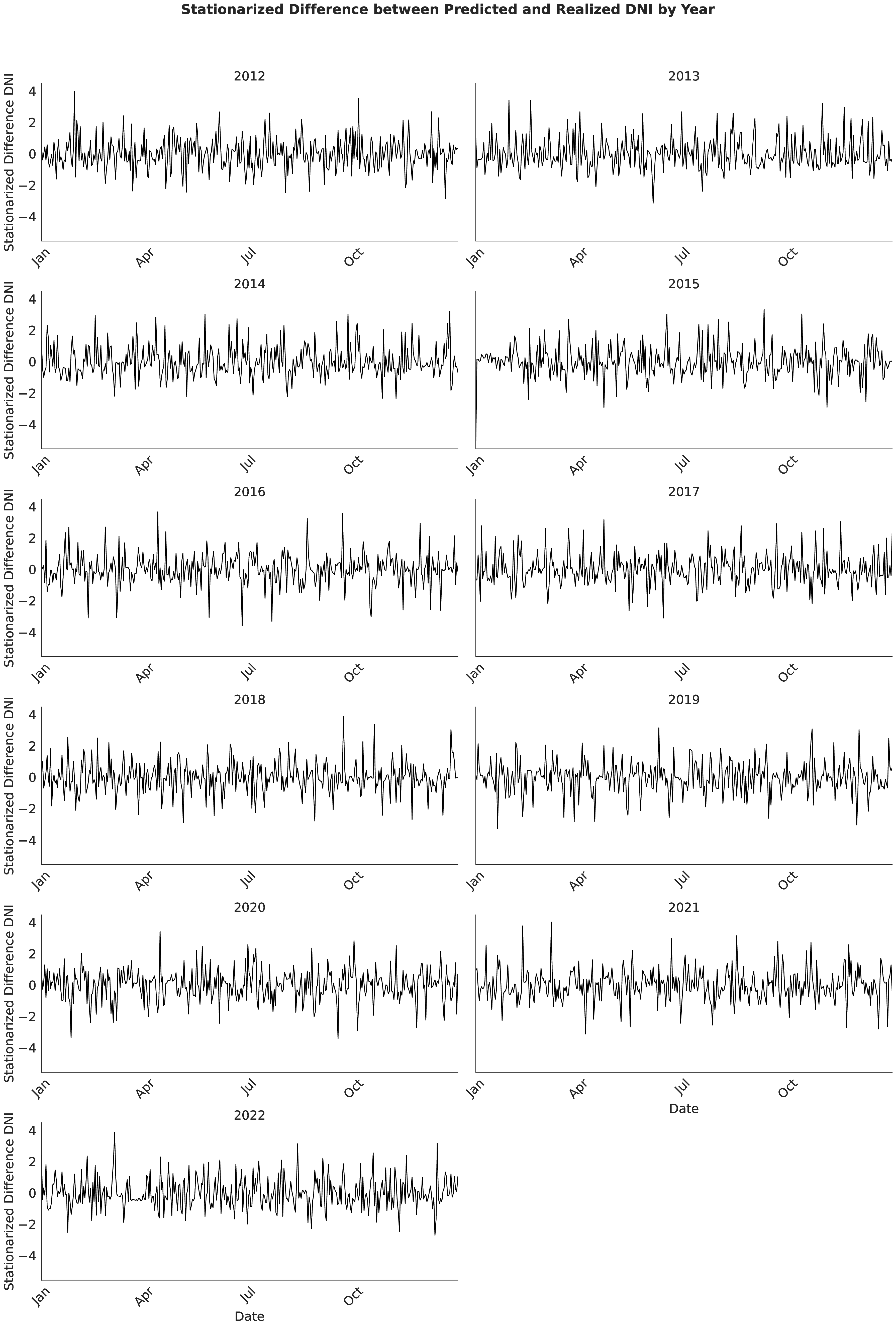}
                \caption{Stationnarized of the difference between predicted and realized DNI.}
                 \label{fig:illustration7}
\end{figure}

\clearpage

\section{Approximation of the conditional expectation of the loss\label{app2}}

We begin by recalling that if $\boldsymbol{a}_i \neq \boldsymbol{a}$, then the individual 
linear score $\boldsymbol{a}_i' \mathbf{Y}$ can be rewritten in terms of $Z = \boldsymbol{a}'\mathbf{Y}$. 

\paragraph{Rewriting the Linear Score}
Specifically,
$$
  \boldsymbol{a}_i' \mathbf{Y} 
  \;=\; Z \;+\; \tau_i\,\boldsymbol{\alpha}_i' \mathbf{Y},
$$
where 
$$
   Z \;=\; \boldsymbol{a}' \mathbf{Y}, 
   \quad 
   \tau_i \;=\; \bigl\|\boldsymbol{a}_i - \boldsymbol{a}\bigr\|,
   \quad
   \boldsymbol{\alpha}_i' \;=\; \frac{(\boldsymbol{a}_i - \boldsymbol{a})'}{\|\boldsymbol{a}_i - \boldsymbol{a}\|}.
$$
We introduce the following property: 

\begin{property}
\label{Property_proof}
Let us assume that $\boldsymbol{a}$ and $\boldsymbol{a}_i$ are not collinear vectors. Let $g$ be a real-valued measurable function and $f_i(x,y,z)$ be the joint p.d.f.\ of $(X_i,\,\boldsymbol{\alpha}_i' \mathbf{Y},\,Z)$. Then
\begin{adjustwidth}{-1cm}{0cm}
\begin{eqnarray*}
\frac{\partial }{\partial z}\mathrm{Cov}\left[ X_{i},g(\boldsymbol{\alpha }%
_{i}^{\prime }\boldsymbol{Y})\boldsymbol{|}Z=z\right]
&=&\mathrm{Cov}\left[ X_{i},g(\boldsymbol{\alpha }_{i}^{\prime }\boldsymbol{Y})\frac{\partial }{\partial z}\log f_{i}(X_{i},\boldsymbol{\alpha}_{i}^{^{\prime }}\boldsymbol{Y}|z)\boldsymbol{|}Z=z\right]\\
&& \hspace{2em} - \mathbb{E}[g(%
\boldsymbol{\alpha }_{i}^{\prime }\boldsymbol{Y})|\left. Z=z\right. ]\frac{%
\partial }{\partial z}\mathbb{E}[\left. X_{i}|Z=z\right. ]
\end{eqnarray*}%
and%
\begin{eqnarray*}
\frac{\partial ^{2}}{\partial z^{2}}\mathrm{Cov}\left[ X_{i},g(\boldsymbol{%
\alpha }_{i}^{\prime }\boldsymbol{Y})\boldsymbol{|}Z=z\right] &=& \mathrm{Cov}\left[ X_{i},g(\boldsymbol{\alpha}_{i}^{\prime }\boldsymbol{Y})\frac{\partial ^{2}}{\partial z^{2}}\log f_{i}(X_{i},\boldsymbol{\alpha}_{i}^{^{\prime}}\boldsymbol{Y}|z)\boldsymbol{|}Z=z\right] \\
&& \hspace{2em} + \mathrm{Cov}%
\left[ X_{i},g(\boldsymbol{\alpha }_{i}^{\prime }\boldsymbol{Y})\left( \frac{%
\partial }{\partial z}\log f_{i}(X_{i},\boldsymbol{\alpha}_{i}^{^{\prime }}%
\boldsymbol{Y}|z)\right) ^{2}\boldsymbol{|}Z=z\right]\\
&& \hspace{2em} - 2\frac{\partial }{\partial z}\mathbb{E}[g(\boldsymbol{\alpha }%
_{i}^{\prime }\boldsymbol{Y})|\left. Z=z\right. ]\frac{\partial }{\partial z}%
\mathbb{E}[\left. X_{i}|Z=z\right. ]\\
&& \hspace{2em} - \mathbb{E}[g(\boldsymbol{\alpha }%
_{i}^{\prime }\boldsymbol{Y})|\left. Z=z\right. ]\frac{\partial ^{2}}{%
\partial z^{2}}\mathbb{E}[\left. X_{i}|Z=z\right. ].
\end{eqnarray*}
\end{adjustwidth}
\end{property}

\begin{proof}
The first derivative of the conditional variance may be written in the following way:
\begin{eqnarray*}
\frac{\partial }{\partial z}\mathrm{Cov}\left[ X_{i},g(\boldsymbol{\alpha }%
_{i}^{\prime }\boldsymbol{Y})\boldsymbol{|}Z=z\right] 
&=& \frac{\partial }{\partial z}\left[ \mathbb{E}\left[ X_{i}g(\boldsymbol{\alpha }_{i}^{\prime }\boldsymbol{Y})\boldsymbol{|}Z=z\right] -\mathbb{E}%
\left[ X_{i}\boldsymbol{|}Z=z\right] \mathbb{E}\left[ g(\boldsymbol{\alpha }%
_{i}^{\prime }\boldsymbol{Y})\boldsymbol{|}Z=z\right] \right]  \\
&=&\mathbb{E}\left[ X_{i}g(\boldsymbol{\alpha }_{i}^{\prime }\boldsymbol{Y})%
\frac{\partial }{\partial z}\log f_{i}(X_{i},\boldsymbol{\alpha}_{i}^{^{\prime
}}\boldsymbol{Y}|z)\boldsymbol{|}Z=z\right] \\
&& \hspace{2em}-\mathbb{E}[\left.
X_{i}|Z=z\right. ]\mathbb{E}\left[ g(\boldsymbol{\alpha }_{i}^{\prime }%
\boldsymbol{Y})\frac{\partial }{\partial z}\log f_{i}(X_{i},\boldsymbol{\alpha}%
_{i}^{^{\prime }}\boldsymbol{Y}|z)\boldsymbol{|}Z=z\right] \\
&& \hspace{2em} -\mathbb{E}[g(\boldsymbol{\alpha }_{i}^{\prime }\boldsymbol{Y})|\left.
Z=z\right. ]\frac{\partial }{\partial z}\mathbb{E}[\left. X_{i}|Z=z\right. ]
\\
&=& \mathrm{Cov}\left[ X_{i},g(\boldsymbol{\alpha }_{i}^{\prime }\boldsymbol{Y%
})\frac{\partial }{\partial z}\log f_{i}(X_{i},\boldsymbol{\alpha}%
_{i}^{^{\prime }}\boldsymbol{Y}|z)\boldsymbol{|}Z=z\right] \\
&& \hspace{2em}-\mathbb{E}[g(%
\boldsymbol{\alpha }_{i}^{\prime }\boldsymbol{Y})|\left. Z=z\right. ]\frac{%
\partial }{\partial z}\mathbb{E}[\left. X_{i}|Z=z\right. ].
\end{eqnarray*}
The second derivative of the conditional variance may be written in the following way:

\begin{adjustwidth}{-1cm}{0cm}
\begin{eqnarray*}
\frac{\partial ^{2}}{\partial z^{2}}\mathrm{Cov}\left[ X_{i},g(\boldsymbol{%
\alpha }_{i}^{\prime }\boldsymbol{Y}) \,\middle|\, Z=z\right]
&=& \mathbb{E}\left[ X_{i}g(\boldsymbol{\alpha }_{i}^{\prime }\boldsymbol{Y})%
\frac{\partial ^{2}}{\partial z^{2}}\log f_{i}(X_{i},\boldsymbol{\alpha}%
_{i}^{\prime }\boldsymbol{Y} \,|\, z) \,\middle|\, Z=z\right]  \\
&& \hspace{2em} + \mathbb{E}\left[
X_{i}g(\boldsymbol{\alpha }_{i}^{\prime }\boldsymbol{Y})\left( \frac{%
\partial }{\partial z}\log f_{i}(X_{i},\boldsymbol{\alpha}_{i}^{\prime }%
\boldsymbol{Y} \,|\, z)\right) ^{2} \,\middle|\, Z=z\right]  \\
&& \hspace{2em} - \mathbb{E}\left[ g(\boldsymbol{\alpha }_{i}^{\prime }\boldsymbol{Y})%
\frac{\partial }{\partial z}\log f_{i}(X_{i},\boldsymbol{\alpha}_{i}^{\prime }%
\boldsymbol{Y} \,|\, z) \,\middle|\, Z=z\right] 
\cdot \frac{\partial }{\partial z}\mathbb{E}\left[ X_{i} \,|\, Z=z\right]  \\
&& \hspace{2em} - \mathbb{E}\left[ X_{i} \,|\, Z=z\right] \cdot \Bigg(
\mathbb{E}\left[ g(\boldsymbol{\alpha }_{i}^{\prime }\boldsymbol{Y})%
\frac{\partial ^{2}}{\partial z^{2}}\log f_{i}(X_{i},\boldsymbol{\alpha}%
_{i}^{\prime }\boldsymbol{Y} \,|\, z) \,\middle|\, Z=z\right]  \\
&& \hspace{4em} +\mathbb{E}\left[ g(\boldsymbol{\alpha }_{i}^{\prime }%
\boldsymbol{Y})\left( \frac{\partial }{\partial z}\log f_{i}(X_{i},\boldsymbol{\alpha}%
_{i}^{\prime }\boldsymbol{Y} \,|\, z)\right) ^{2} \,\middle|\, Z=z\right] 
\Bigg)  \\
&& \hspace{2em} - \frac{\partial }{\partial z} \mathbb{E}\left[ g(\boldsymbol{\alpha }_{i}^{\prime }%
\boldsymbol{Y}) \,|\, Z=z \right] \cdot \frac{\partial }{\partial z} \mathbb{E}\left[ X_{i} \,|\, Z=z \right]  \\
&& \hspace{2em} - \mathbb{E}\left[ g(\boldsymbol{\alpha }_{i}^{\prime }%
\boldsymbol{Y}) \,|\, Z=z \right] \cdot \frac{\partial ^{2}}{\partial z^{2}} \mathbb{E}\left[ X_{i} \,|\, Z=z \right] \\
&=& \mathrm{Cov}\left[ X_{i},g(\boldsymbol{\alpha }_{i}^{\prime }\boldsymbol{Y})%
\frac{\partial ^{2}}{\partial z^{2}}\log f_{i}(X_{i},\boldsymbol{\alpha}%
_{i}^{\prime }\boldsymbol{Y} \,|\, z) \,\middle|\, Z=z \right]  \\
&& \hspace{2em} + \mathrm{Cov}\left[ X_{i},g(\boldsymbol{\alpha }_{i}^{\prime }\boldsymbol{Y})%
\left( \frac{\partial }{\partial z}\log f_{i}(X_{i},\boldsymbol{\alpha}_{i}^{\prime }%
\boldsymbol{Y} \,|\, z) \right)^2 \,\middle|\, Z=z \right]  \\
&& \hspace{2em} - 2 \cdot \frac{\partial }{\partial z} \mathbb{E}\left[ g(\boldsymbol{\alpha }%
_{i}^{\prime }\boldsymbol{Y}) \,|\, Z=z \right] \cdot \frac{\partial }{\partial z} \mathbb{E}\left[ X_{i} \,|\, Z=z \right]  \\
&& \hspace{2em} - \mathbb{E}\left[ g(\boldsymbol{\alpha }_{i}^{\prime }%
\boldsymbol{Y}) \,|\, Z=z \right] \cdot \frac{\partial ^{2}}{\partial z^{2}} \mathbb{E}\left[ X_{i} \,|\, Z=z \right].
\end{eqnarray*}
\end{adjustwidth}
\end{proof}

\paragraph{Approximate \texorpdfstring{$\mathbb{E}[X_i \mid Z=z]$}{E[Xi|Z=z]}}

Following Lemma~2 in \cite{gourierouxSensitivityAnalysisValues2000}, we want 
to derive an approximation for 
$$
  \mathbb{E}[X_i \;\mid\; \boldsymbol{a}_i' \mathbf{Y} = z].
$$

Let $f_i(x,y,z)$ be the joint p.d.f.\ of $(X_i,\,\boldsymbol{\alpha}_i' \mathbf{Y},\,Z)$. 
Then
$$
  \mathbb{E}[X_i \;\mid\; \boldsymbol{a}_i' \mathbf{Y} = z]
  \;=\;
  \frac{ \displaystyle \int\!\!\int x \, f_i\bigl(x,y,\,z - \tau_i\,y\bigr)
                \,\mathrm{d}x\,\mathrm{d}y}
       { \displaystyle \int\!\!\int f_i\bigl(x,y,\,z - \tau_i\,y\bigr)
                \,\mathrm{d}x\,\mathrm{d}y}.
$$

\paragraph{Small-\texorpdfstring{$\tau_i$}{tau\_i} Expansion}
We next assume $\tau_i$ is small and expand the integrals in a second-order Taylor approximation around $z$. This yields:

\begin{adjustwidth}{-1.5cm}{2cm}
\begin{equation}
  \frac{ \displaystyle \int\!\!\int x\,f_i(x,y,z)\,\mathrm{d}x\,\mathrm{d}y
          - \tau_i \int\!\!\int x\,y\,\frac{\partial}{\partial z}f_i(x,y,z)\,\mathrm{d}x\,\mathrm{d}y
          + \tfrac{\tau_i^2}{2} \int\!\!\int x\,y^2\,\frac{\partial^2}{\partial z^2}f_i(x,y,z)\,\mathrm{d}x\,\mathrm{d}y}
       { \displaystyle \int\!\!\int f_i(x,y,z)\,\mathrm{d}x\,\mathrm{d}y
          - \tau_i \int\!\!\int y\,\frac{\partial}{\partial z}f_i(x,y,z)\,\mathrm{d}x\,\mathrm{d}y
          + \tfrac{\tau_i^2}{2} \int\!\!\int y^2\,\frac{\partial^2}{\partial z^2}f_i(x,y,z)\,\mathrm{d}x\,\mathrm{d}y } + o(\tau_i^2)
\label{eq:big-ratio}
\nonumber
\end{equation}
\end{adjustwidth}

\begin{center}
   Equation \eqref{eq:big-ratio}.
\end{center}
Since
$$
  \frac{ \displaystyle \int\!\!\int x\,f_i(x,y,z)\,\mathrm{d}x\,\mathrm{d}y }
        { \displaystyle \int\!\!\int f_i(x,y,z)\,\mathrm{d}x\,\mathrm{d}y }
  \;=\;
  \mathbb{E}\bigl[X_i \mid Z=z\bigr],
$$
We can rewrite \eqref{eq:big-ratio} in terms of the functions $A_i(z)$, $B_i(z)$, $C_i(z)$, $D_i(z)$ defined as follows.

\begin{eqnarray*}
A_i(z)
&=&
  \frac{ \int\!\!\int x\,y\,\frac{\partial}{\partial z}f_i(x,y,z)\,\mathrm{d}x\,\mathrm{d}y }
       { \int\!\!\int f_i(x,y,z)\,\mathrm{d}x\,\mathrm{d}y }
\\
&=&
  \mathbb{E}\Bigl[
    X_i\,(\boldsymbol{\alpha}_i' \mathbf{Y})\,\frac{\partial}{\partial z}\log f_i(X_i,\boldsymbol{\alpha}_i'\mathbf{Y},z)
    \,\bigm|\, Z=z
  \Bigr]
\\[1.5ex]
B_i(z)
&=&
  \frac{ \int\!\!\int x\,y^2\,\frac{\partial^2}{\partial z^2}f_i(x,y,z)\,\mathrm{d}x\,\mathrm{d}y }
       { \int\!\!\int f_i(x,y,z)\,\mathrm{d}x\,\mathrm{d}y }
\\
&=&
  \mathbb{E}\Bigl[ 
    X_i\,(\boldsymbol{\alpha}_i' \mathbf{Y})^2\,\frac{\partial^2}{\partial z^2}\log f_i(X_i,\boldsymbol{\alpha}_i'\mathbf{Y},z)
    \,\bigm|\, Z=z
  \Bigr]
\\
&&
  +\;
  \mathbb{E}\Bigl[
    X_i\,(\boldsymbol{\alpha}_i' \mathbf{Y})^2\,\Bigl(\frac{\partial}{\partial z}\log f_i(X_i,\boldsymbol{\alpha}_i'\mathbf{Y},z)\Bigr)^2
    \,\bigm|\, Z=z
  \Bigr]
\\[1.5ex]
C_i(z)
&=&
  \frac{ \int\!\!\int y\,\frac{\partial}{\partial z}f_i(x,y,z)\,\mathrm{d}x\,\mathrm{d}y }
       { \int\!\!\int f_i(x,y,z)\,\mathrm{d}x\,\mathrm{d}y }
\\
&=&
  \mathbb{E}\Bigl[
    (\boldsymbol{\alpha}_i' \mathbf{Y})\,\frac{\partial}{\partial z}\log f_i(X_i,\boldsymbol{\alpha}_i'\mathbf{Y},z)
    \,\bigm|\, Z=z
  \Bigr]
\\[1.5ex]
D_i(z)
&=&
  \frac{ \int\!\!\int y^2\,\frac{\partial^2}{\partial z^2}f_i(x,y,z)\,\mathrm{d}x\,\mathrm{d}y }
       { \int\!\!\int f_i(x,y,z)\,\mathrm{d}x\,\mathrm{d}y }
\\
&=&
  \mathbb{E}\Bigl[
    (\boldsymbol{\alpha}_i' \mathbf{Y})^2\,\frac{\partial^2}{\partial z^2}\log f_i(X_i,\boldsymbol{\alpha}_i'\mathbf{Y},z)
    \,\bigm|\, Z=z
  \Bigr]
\\
&&
  +\;
  \mathbb{E}\Bigl[
    (\boldsymbol{\alpha}_i' \mathbf{Y})^2\,\Bigl(\frac{\partial}{\partial z}\log f_i(X_i,\boldsymbol{\alpha}_i'\mathbf{Y},z)\Bigr)^2
    \,\bigm|\, Z=z
  \Bigr]
\end{eqnarray*}
Substituting these definitions back and expanding numerator and denominator in powers of $\tau_i$, we can rewrite \eqref{eq:big-ratio} as (up to $o(\tau_i^2)$ terms):

\begin{adjustwidth}{-1cm}{0cm}  % Push left margin -1.5cm just for this block
\begin{eqnarray*}
\frac{\int \int x\,f_{i}(x,y,z-\tau_i y)\,\mathrm{d}x\,\mathrm{d}y}
     {\int \int f_{i}(x,y,z-\tau_i y)\,\mathrm{d}x\,\mathrm{d}y}
&=&
  \frac{ \mathbb{E}[X_i|Z=z] - \tau_i A_i(z) + \left( \tau_i^2/2 \right) B_i(z) }
       { 1 - \tau_i C_i(z) + \left( \tau_i^2/2 \right) D_i(z) }
  + o(\tau_i^2)
\\[1.5ex]
&=&
  \mathbb{E}[X_i|Z=z] - \tau_i A_i(z) + \left( \tau_i^2/2 \right) B_i(z)
\\
&& \hspace{2em}
   \times \left( 1 + \tau_i C_i(z) + \tau_i^2 \left( C_i^2(z) - \tfrac{1}{2} D_i(z) \right) \right)
   + o(\tau_i^2)
\\[1.5ex]
&=&
  \mathbb{E}[X_i|Z=z] - \tau_i A_i(z) + \left( \tau_i^2/2 \right) B_i(z)
\\
&& \hspace{2em}
   + \tau_i C_i(z)\, \mathbb{E}[X_i|Z=z]
   - \tau_i^2 A_i(z)\, C_i(z)
\\
&& \hspace{2em}
   + \tau_i^2 \left( C_i^2(z) - \tfrac{1}{2} D_i(z) \right)\, \mathbb{E}[X_i|Z=z]
   + o(\tau_i^2)
\\[1.5ex]
&=&
  \mathbb{E}[X_i|Z=z] - \tau_i \left[ A_i(z) - C_i(z)\, \mathbb{E}[X_i|Z=z] \right]
\\
&& \hspace{2em}
   + \tau_i^2 \left[ \tfrac{1}{2} B_i(z) - A_i(z)\, C_i(z)
   + \left( C_i^2(z) - \tfrac{1}{2} D_i(z) \right)\, \mathbb{E}[X_i|Z=z] \right]
\\
&& \hspace{2em}
   + o(\tau_i^2)
\\[1.5ex]
&=&
  \mathbb{E}[X_i|Z=z]
  + \tau_i L_i(z,\boldsymbol{a})
  + \tfrac{1}{2} \tau_i^2 F_i(z,\boldsymbol{a})
  + o(\tau_i^2)
\end{eqnarray*}
\end{adjustwidth}
where 

\begin{eqnarray*}
L_{i}\left( z,\boldsymbol{a}\right)  &=&-\left[ A_{i}\left( z\right)
-C_{i}\left( z\right) \mathbb{E}[X_{i}|\left. Z=z\right. ]\right]  \\
F_{i}\left( z,\boldsymbol{a}\right)  &=&\left[ B_{i}\left( z\right)
-2A_{i}\left( z\right) C_{i}\left( z\right) +\left( 2C_{i}^{2}\left(
z\right) -D_{i}\left( z\right) \right) \mathbb{E}[X_{i}|\left. Z=z\right. ]%
\right]  \\
&=&\left[ B_{i}\left( z\right) -D_{i}\left( z\right) \mathbb{E}[X_{i}|\left.
Z=z\right. ]\right] -2C_{i}\left( z\right) L_{i}\left( z,\boldsymbol{a}%
\right) .
\end{eqnarray*}
Finally, we can express $L_i$ and $F_i$ in covariance terms involving the partial 
derivatives of $f_i$. We have

\begin{eqnarray*}
-L_{i}\left( z,\boldsymbol{a}\right)  
&=& A_{i}\left( z\right) - C_{i}\left( z\right) \mathbb{E}[X_{i}|Z=z]
\\[1.5ex]
&=& \mathbb{E}\left[ X_{i}\,\boldsymbol{\alpha}_{i}^{\prime }\boldsymbol{Y}\,\frac{%
\partial }{\partial z}\log f_{i}(X_{i},\boldsymbol{\alpha}_{i}^{\prime }%
\boldsymbol{Y},z) \,\big|\, Z=z \right]
\\
&& \hspace{2em} - \mathbb{E}\left[ \boldsymbol{\alpha}_{i}^{\prime }\boldsymbol{Y}\,\frac{\partial 
}{\partial z}\log f_{i}(X_{i},\boldsymbol{\alpha}_{i}^{\prime }\boldsymbol{Y},z) \,\big|\, Z=z \right]
\mathbb{E}[X_{i}|Z=z]
\\[1.5ex]
&=& \mathrm{Cov}\left[ X_{i},\boldsymbol{\alpha }_{i}^{\prime }\boldsymbol{Y}
\frac{\partial }{\partial z}\log f_{i}(X_{i},\boldsymbol{\alpha }_{i}^{\prime }
\boldsymbol{Y},z) \,\big|\, Z=z \right]
\\[1.5ex]
&=& \frac{\partial }{\partial z}\log f(z)\,\mathrm{Cov}\left[
X_{i},\boldsymbol{\alpha }_{i}^{\prime }\boldsymbol{Y} \,\big|\, Z=z \right]
\\
&& \hspace{2em} + \mathrm{Cov}\left[ X_{i},\boldsymbol{\alpha }_{i}^{\prime }\boldsymbol{Y}\,\frac{%
\partial }{\partial z}\log f_{i}(X_{i},\boldsymbol{\alpha }_{i}^{\prime }
\boldsymbol{Y} \,\big|\, z)\,\big|\, Z=z \right]
\end{eqnarray*}
where $f$ is the p.d.f. of $Z$, since $f\boldsymbol{_{i}}(x,y|z)=f_{i}\left(
x,y,z\right) /f(z)$. Moreover, using Property $\left( \ref{Property_proof}%
\right) $\ with $g\left( y\right) =y$, 

\begin{eqnarray*}
\frac{\partial }{\partial z}\mathrm{Cov}\left[ X_{i},\boldsymbol{\alpha }%
_{i}^{\prime }\boldsymbol{Y|}Z=z\right]  &=&\mathrm{Cov}\left[ X_{i},%
\boldsymbol{\alpha }_{i}^{\prime }\boldsymbol{Y}\frac{\partial }{\partial z}%
\log f_{i}(X_{i},\boldsymbol{\alpha}_{i}^{^{\prime }}\boldsymbol{Y}|z)%
\boldsymbol{|}Z=z\right]  \\
&& \hspace{2em} -\mathbb{E}[\boldsymbol{\alpha}_{i}^{\prime }\boldsymbol{Y}|\left.
Z=z\right. ]\frac{\partial }{\partial z}\mathbb{E}[\left. X_{i}|Z=z\right. ]
\end{eqnarray*}
and it follows that

\begin{eqnarray*}
L_{i}\left( z,\boldsymbol{a}\right)  &=&-\frac{\partial }{\partial z}\log
f\left( z\right) \mathrm{Cov}\left[ X_{i},\boldsymbol{\alpha }_{i}^{\prime }%
\boldsymbol{Y|}Z=z\right] -\frac{\partial }{\partial z}\mathrm{Cov}\left[
X_{i},\boldsymbol{\alpha }_{i}^{\prime }\boldsymbol{Y|}Z=z\right]  \\
&& \hspace{2em} +\mathbb{E}[\boldsymbol{\alpha }_{i}^{\prime }\boldsymbol{Y}|\left.
Z=z\right. ]\frac{\partial }{\partial z}\mathbb{E}[\left. X_{i}|Z=z\right. ].
\end{eqnarray*}
Second, we have

\begin{eqnarray*}
C_{i}\left( z\right)  &=&\mathbb{E}[\boldsymbol{\alpha}_{i}^{^{\prime }}%
\boldsymbol{Y}\frac{\partial }{\partial z}\log f_{i}(X_{i},\boldsymbol{\alpha}%
_{i}^{^{\prime }}\boldsymbol{Y},z)|Z=z] \\
&=&\frac{\partial }{\partial z}\log f\left( z\right) \mathbb{E}[\mathbf{%
\alpha }_{i}^{^{\prime }}\boldsymbol{Y}|Z=z]+\mathbb{E}[\boldsymbol{\alpha}%
_{i}^{^{\prime }}\boldsymbol{Y}\frac{\partial }{\partial z}\log f_{i}(X_{i},%
\boldsymbol{\alpha}_{i}^{^{\prime }}\boldsymbol{Y}|z)|Z=z] \\
&=&\frac{\partial }{\partial z}\log f\left( z\right) \mathbb{E}[\mathbf{%
\alpha }_{i}^{^{\prime }}\boldsymbol{Y}|Z=z]+\frac{\partial }{\partial z}%
\mathbb{E}[\boldsymbol{\alpha}_{i}^{^{\prime }}\boldsymbol{Y}|Z=z].
\end{eqnarray*}
A similar approach applies to $F_i(z,\boldsymbol{a})$, which ultimately involves the second derivatives $\tfrac{\partial^2}{\partial z^2}\log f_i(\dots)$ and mixed covariance terms. Specifically,
\begin{eqnarray*}
B_i(z) - D_i(z)\,\mathbb{E}[X_i\mid Z=z]
  &=& \mathrm{Cov}\Bigl[X_i,\;(\boldsymbol{\alpha}_i' \mathbf{Y})^2\,\tfrac{\partial^2}{\partial z^2}\log f_i(\dots)
    \;\bigm|\;Z=z
  \Bigr] \\
&& \hspace{2em} + \mathrm{Cov}\Bigl[
    X_i,\;(\boldsymbol{\alpha}_i' \mathbf{Y})^2\,\bigl(\tfrac{\partial}{\partial z}\log f_i(\dots)\bigr)^2
    \;\bigm|\;Z=z
  \Bigr]
\end{eqnarray*}
Further details appear in the property-based expansions below, ultimately giving us:
\begin{adjustwidth}{-1cm}{0cm}
\begin{eqnarray*}
F_{i}\left( z,\boldsymbol{a}\right) &=& \left[ \frac{\partial ^{2}}{\partial z^{2}}\log f\left( z\right) + \left(\frac{\partial }{\partial z}\log f\left( z\right) \right) ^{2}\right] 
\mathrm{Cov}\left[ X_{i},(\boldsymbol{\alpha}_{i}^{^{\prime }}\boldsymbol{Y})^{2}\boldsymbol{|}Z=z\right]  \\
&& \hspace{2em} + 2\frac{\partial }{\partial z}\log f\left( z\right) \left[ 
    \frac{\partial }{\partial z}\mathrm{Cov}\left[ X_{i},(\boldsymbol{\alpha}_{i}^{^{\prime }}\boldsymbol{Y})^{2}\boldsymbol{|}Z=z\right] 
    + \mathbb{E}[(\boldsymbol{\alpha}_{i}^{^{\prime }}\boldsymbol{Y})^{2}|\left. Z=z\right. ]\frac{\partial }{%
\partial z}\mathbb{E}[\left. X_{i}|Z=z\right. ]
\right]  \\
&& \hspace{2em} + \frac{\partial ^{2}}{\partial z^{2}}\mathrm{Cov}\left[ X_{i},(\mathbf{%
\alpha }_{i}^{^{\prime }}\boldsymbol{Y})^{2}\boldsymbol{|}Z=z\right] +2\frac{%
\partial }{\partial z}\mathbb{E}[(\boldsymbol{\alpha}_{i}^{^{\prime }}%
\boldsymbol{Y})^{2}|\left. Z=z\right. ]\frac{\partial }{\partial z}\mathbb{E}%
[\left. X_{i}|Z=z\right. ] \\
&& \hspace{2em} + \mathbb{E}[(\boldsymbol{\alpha}_{i}^{^{\prime }}\boldsymbol{Y})^{2}|\left.
Z=z\right. ]\frac{\partial ^{2}}{\partial z^{2}}\mathbb{E}[\left.
X_{i}|Z=z\right. ] \\
&& \hspace{2em} -2\left[ \frac{\partial }{\partial z}\log f\left( z\right) \mathbb{E}[%
\boldsymbol{\alpha}_{i}^{^{\prime }}\boldsymbol{Y}|Z=z]+\frac{\partial }{%
\partial z}\mathbb{E}[\boldsymbol{\alpha}_{i}^{^{\prime }}\boldsymbol{Y}|Z=z]%
\right] L_{i}\left( z,\boldsymbol{a}\right) .
\end{eqnarray*}
\end{adjustwidth}
We thus arrive at a fully expanded approximation for 
$\mathbb{E}[X_i \mid \boldsymbol{a}_i' \mathbf{Y} = z]$ up to second order in $\tau_i$, completing the derivation.

\section{Detail for the computation of the objective function \label{app3}}

In the following subsections, we detail all the steps to compute the objective function for a given weight vector $\boldsymbol{a}$. 

\subsection{Developing the covariance function}
To compute the covariance functions that appear in the term $L_i(\cdot,\cdot)$ in Equation \eqref{li} and the term $F_i(\cdot,\cdot)$ in Equation\eqref{fi}, we use their development in expectation form:
$$
\operatorname{Cov}\left[X_i, \boldsymbol{Y} \mid Z\right]=\mathbb{E}\left[X_i \boldsymbol{Y} \mid Z\right]-\mathbb{E}\left[X_i \mid Z\right] \mathbb{E}[\boldsymbol{Y} \mid Z]
$$

with

$$
\begin{aligned}
\mathbb{E}\left[X_i \boldsymbol{Y} \mid Z\right] & =\mathbb{E}\left[\boldsymbol{Y}\mathbb{E}\left[X_i \mid \boldsymbol{Y}, Z\right] \mid Z\right] \\
& =\mathbb{E}\left[\boldsymbol{Y}\mathbb{E}\left[X_i \mid \boldsymbol{Y}\right] \mid Z\right] \\
& =\mathbb{E}\left[\mathbf{Y} g_i^{-1}\left(a_{i 0}+\boldsymbol{a}_i^{\prime} \boldsymbol{Y}\right) \mid Z\right] \\
\mathbb{E}\left[X_i \mid Z\right] & =\mathbb{E}\left[g_i^{-1}\left(a_{i 0}+\boldsymbol{a}_i^{\prime} \boldsymbol{Y}\right) \mid Z\right]
\end{aligned}
$$
\subsection{Computation of density and derivatives}
In this part, we present the methodology adopted to compute the density $f$ of $Z$, and the derivatives $\frac{\partial}{\partial z} \log f(z)$ and $\frac{\partial ^{2}}{\partial z^{2}}\log f\left( z\right)$ which appears in the terms $L_i(\cdot,\cdot)$ \eqref{li} and $F _i(\cdot,\cdot)$ \eqref{fi}\newline
We use the kernel simulation-based method to compute these quantities.
\begin{itemize}
\item Density estimation using the kernel method
$$
f(z)\approx\frac{1}{Lh}\sum_{l=1}^L K\left(\frac{z-z_l}{h}\right)
$$
where $L$ is the number of simulations, $K$ is the kernel method, and $h$ is the bandwidth parameter.
\item Estimation of derivatives using the kernel simulation-based method:
$$
f^\prime(z)\approx\frac{1}{L h^2}\sum_{l=1}^L K^\prime\left(\frac{z-z_l}{h}\right), \quad f^{\prime\prime}(z)\approx\frac{1}{L h^3}\sum_{l=1}^L K^{\prime\prime}\left(\frac{z-z_l}{h}\right).
$$
\newline
We deduce :
$$\frac{\partial}{\partial z} \log f(z)=\frac{f^\prime(z)}{f(z)}, \quad \frac{\partial ^{2}}{\partial z^{2}}\log f\left( z\right)=\frac{f^{\prime\prime}(z)f(z)-(f^\prime(z))^2}{(f(z))^2}.$$
\end{itemize}

\subsection{Computation of conditional expectation and derivative}
In this part, we present the methodology adopted to compute conditional expectations and their derivatives, which appear in $L_i(\cdot,\cdot)$ and $F_i(\cdot,\cdot)$

\begin{adjustwidth}{-1cm}{0cm}
\begin{itemize}
    \item  Estimation of conditional expectation using the kernel method
    $$
    g(z)=\mathbb{E}[X \mid Z=z]\approx\frac{\sum_{l=1}^L x_l K\left(\frac{z-z_l}{h}\right)}{\sum_{l=1}^L K\left(\frac{z-z_l}{h}\right)}
    $$
    \item Estimation of the derivative of the conditional expectation using the kernel method
    $$g^\prime(z)\approx\frac{\sum_{l=1}^L x_l K'\left(\frac{z - z_l}{h}\right) \sum_{m=1}^L K\left(\frac{z - z_m}{h}\right) - \sum_{l=1}^L K'\left(\frac{z - z_l}{h}\right) \sum_{m=1}^L x_m K\left(\frac{z - z_m}{h}\right)}{h \left( \sum_{l=1}^L K\left(\frac{z - z_l}{h}\right) \right)^2}$$

$$
\begin{aligned}
g^{\prime\prime}(z) &
\approx\frac{ \left( \sum_{l=1}^L x_l K''\left(\frac{z - z_l}{h}\right) \sum_{m=1}^L K\left(\frac{z - z_m}{h}\right) + \sum_{l=1}^L x_l K'\left(\frac{z - z_l}{h}\right) \sum_{m=1}^L K'\left(\frac{z - z_m}{h}\right) \right) }{h^2 \left( \sum_{l=1}^L K\left(\frac{z - z_l}{h}\right) \right)^2} \\
& - \frac{ \left( \sum_{l=1}^L K''\left(\frac{z - z_l}{h}\right) \sum_{m=1}^L x_m K\left(\frac{z - z_m}{h}\right) + \sum_{l=1}^L K'\left(\frac{z - z_l}{h}\right) \sum_{m=1}^L x_m K'\left(\frac{z - z_m}{h}\right) \right) }{h^2 \left( \sum_{l=1}^L K\left(\frac{z - z_l}{h}\right) \right)^2} \\
& - 2 \frac{ \left( \sum_{l=1}^L K'\left(\frac{z - z_l}{h}\right) \right) \left( \sum_{l=1}^L x_l K'\left(\frac{z - z_l}{h}\right) \sum_{m=1}^L K\left(\frac{z - z_m}{h}\right) - \sum_{l=1}^L K'\left(\frac{z - z_l}{h}\right) \sum_{m=1}^L x_m K\left(\frac{z - z_m}{h}\right) \right) }{h^2 \left( \sum_{l=1}^L K\left(\frac{z - z_l}{h}\right) \right)^3}.
\end{aligned}
$$
\end{itemize}
\end{adjustwidth}

\subsection{Simulation method}
To evaluate the functions $L_i(\cdot,\cdot)$ and $F_i(\cdot,\cdot)$ on real data, a lot of observations are needed. In the following, we present the simulation method adopted to obtain more data.
\begin{itemize}
    \item For each weather variable $Y_j$, we calibrate a marginal distribution using the package univariateML in the software R. This choice of marginal distribution will be later used to calibrate a copula for the multivariate weather variable $\boldsymbol{Y}$. 
    \item Once the univariate distribution for each weather variable is calibrated, we use vine copula (see \cite{thesisKernelMethodsVine}) to calibrate the joint distribution $\boldsymbol{Y}$ of the weather variables.
\end{itemize}

\section{Some additional results \label{app4}}

\begin{figure}[H]
        \centering
                \includegraphics[width=0.9\linewidth]{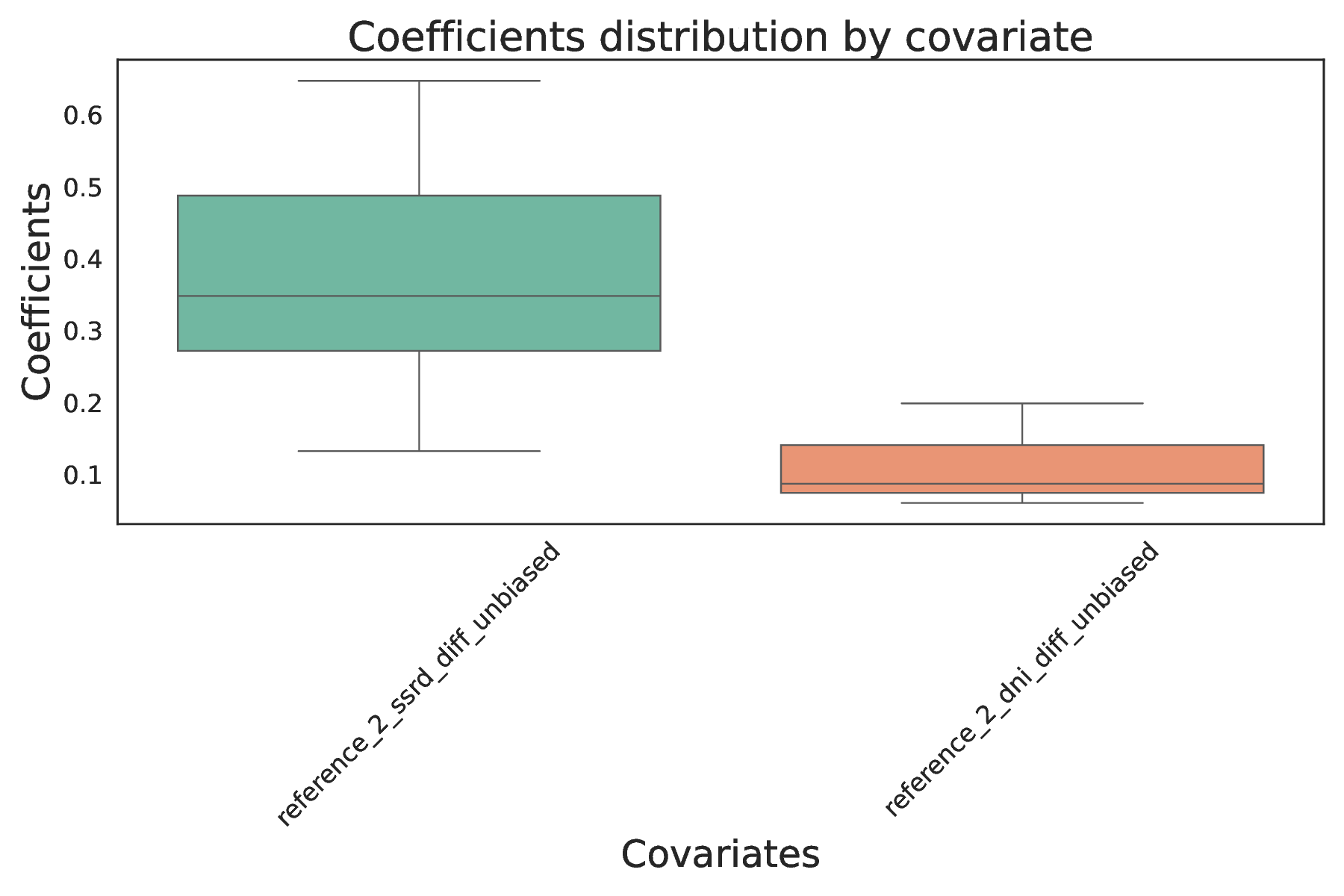}
                \caption{Gaussian GLM coefficients boxplots by covariate.}
                 \label{fig:illustration8}
\end{figure}

\begin{figure}[H]
\begin{subfigure}{.5\textwidth}
  \centering
  \includegraphics[width=.9\linewidth]{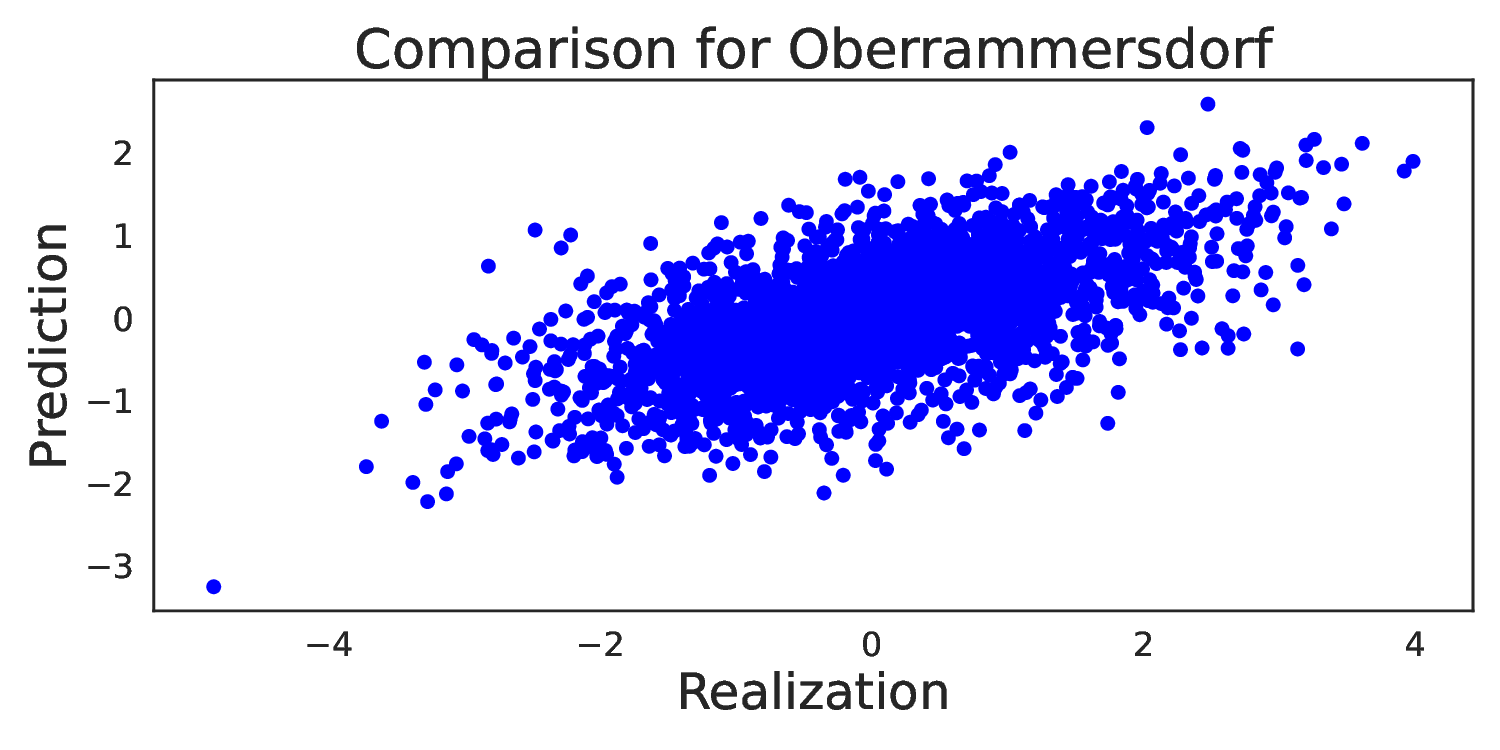}
  \caption{}
  \label{fig:sfigadd1}
\end{subfigure}%
\begin{subfigure}{.5\textwidth}
  \centering
  \includegraphics[width=.9\linewidth]{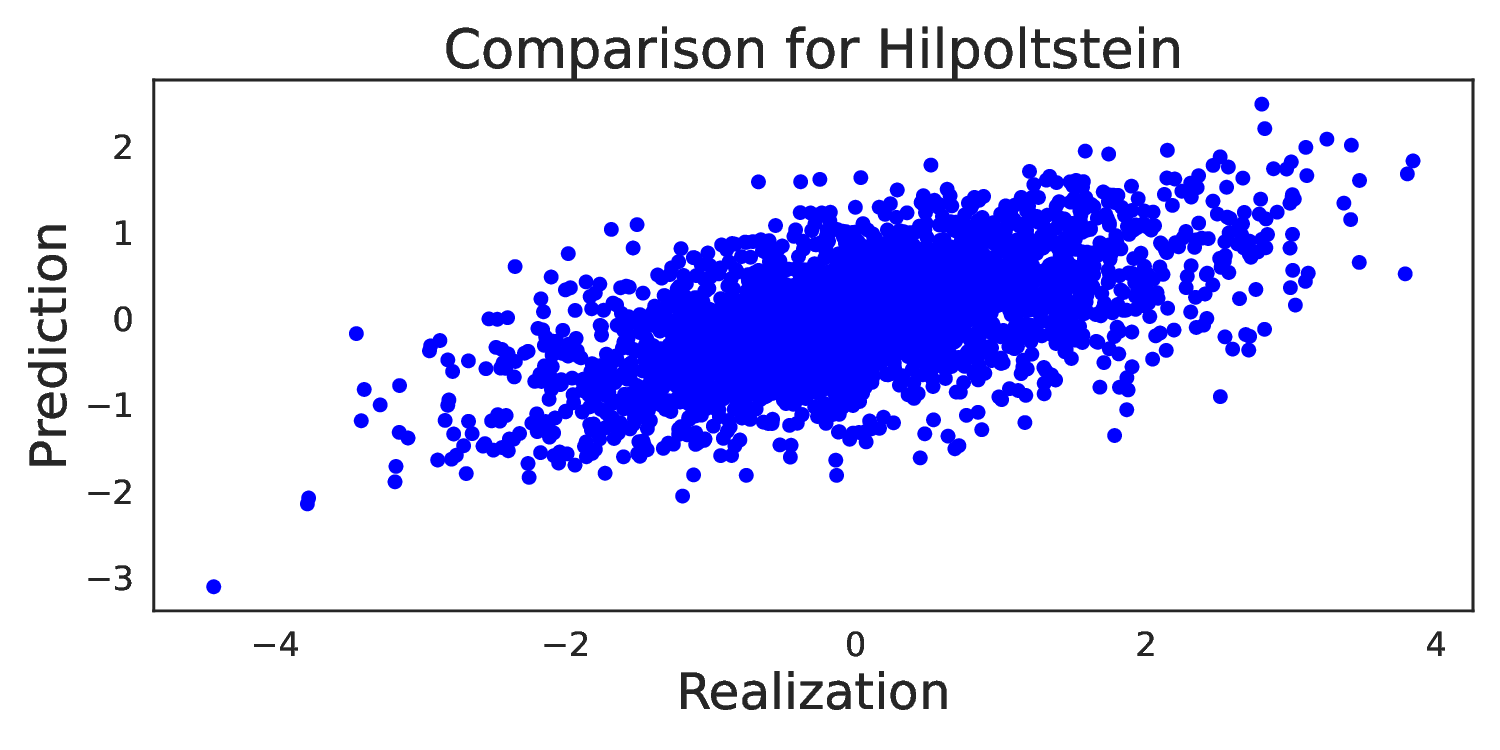}
  \caption{}
  \label{fig:sfigadd2}
\end{subfigure}
\begin{subfigure}{.5\textwidth}
  \centering
  \includegraphics[width=.9\linewidth]{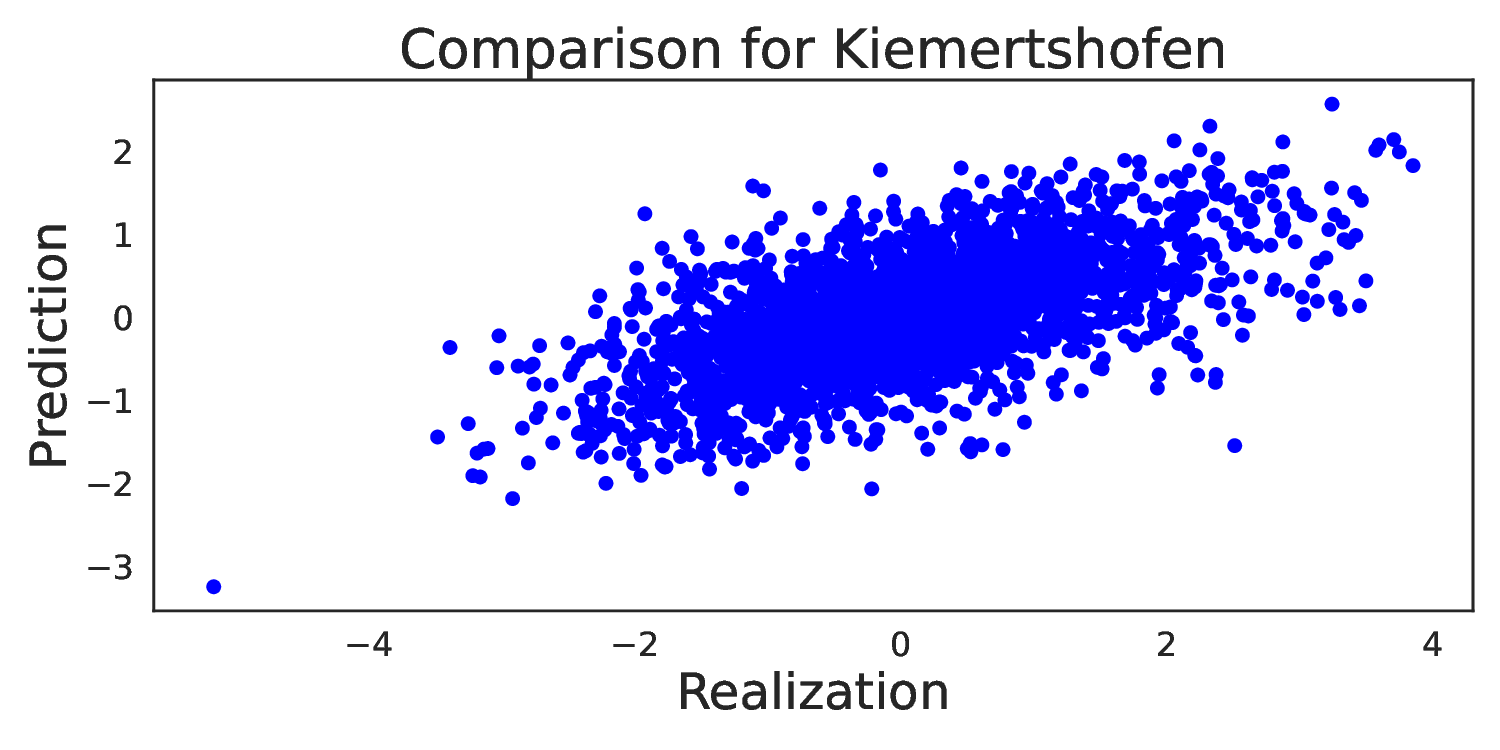}
  \caption{}
  \label{fig:sfigadd3}
\end{subfigure}
\begin{subfigure}{.5\textwidth}
  \centering
  \includegraphics[width=.9\linewidth]{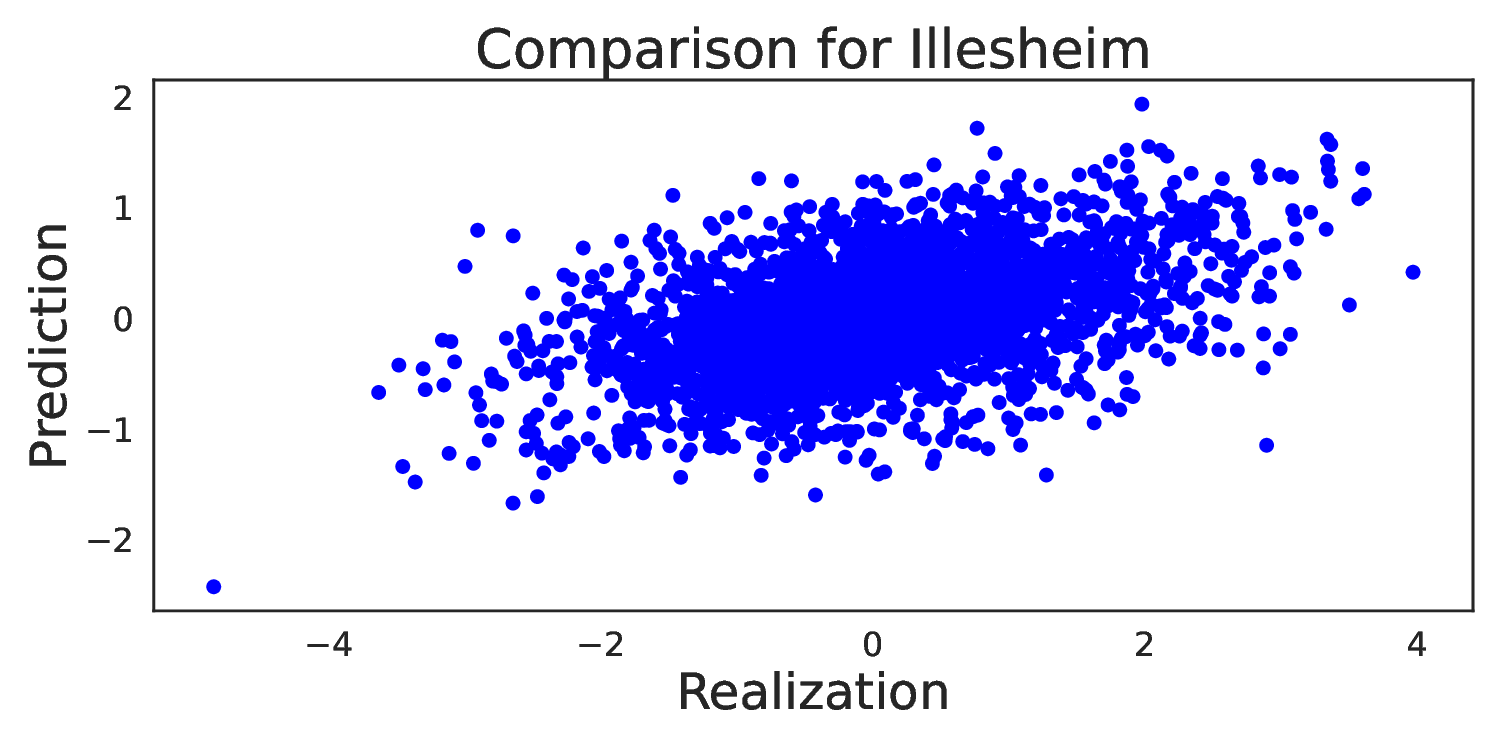}
  \caption{}
  \label{fig:sfigadd4}
\end{subfigure}
\caption{Scatter plot of the realization losses and the Gaussian GLM prediction.}
\label{fig:illustrationadd9}
\end{figure}

\begin{figure}[H]
\begin{subfigure}{.5\textwidth}
  \centering
  \includegraphics[width=.9\linewidth]{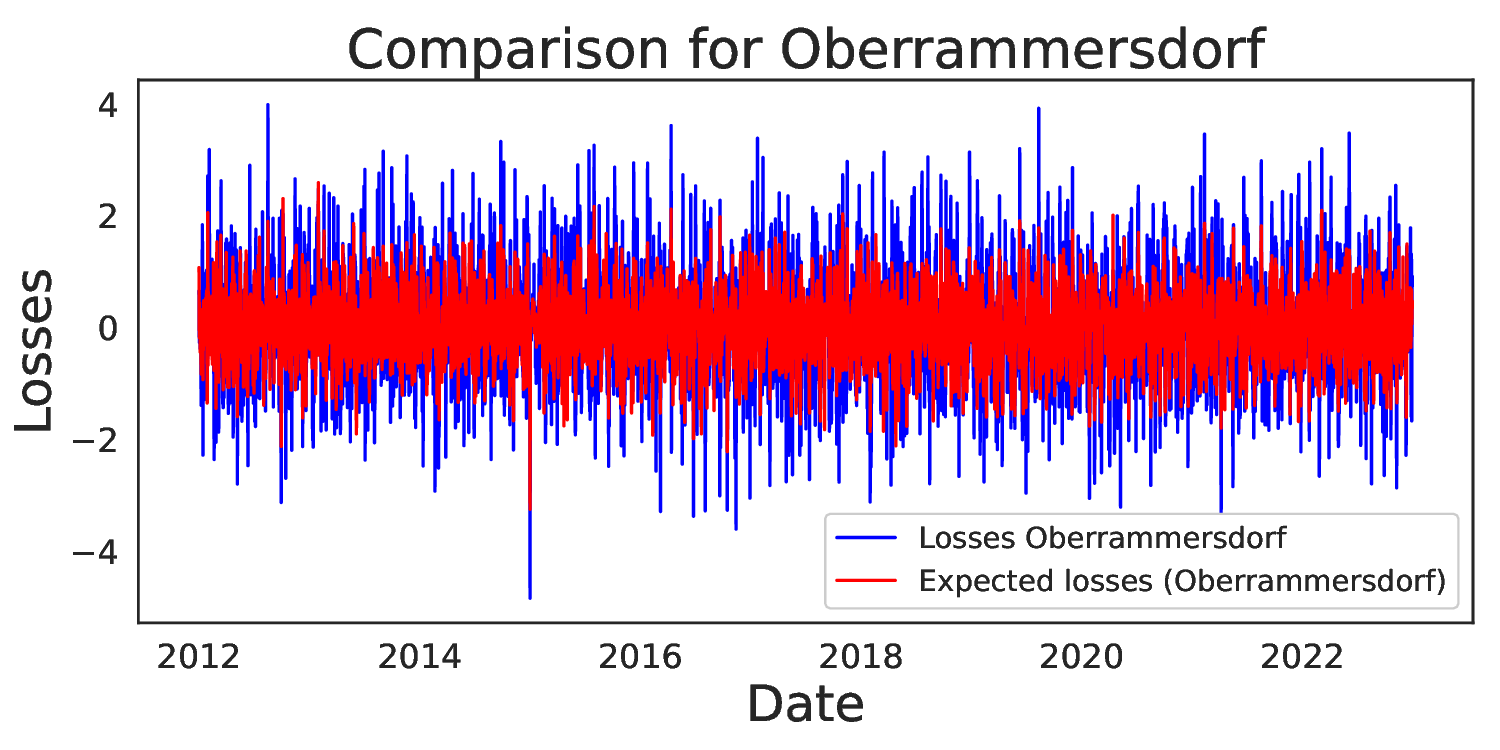}
  \caption{}
  \label{fig:sfig1}
\end{subfigure}%
\begin{subfigure}{.5\textwidth}
  \centering
  \includegraphics[width=.9\linewidth]{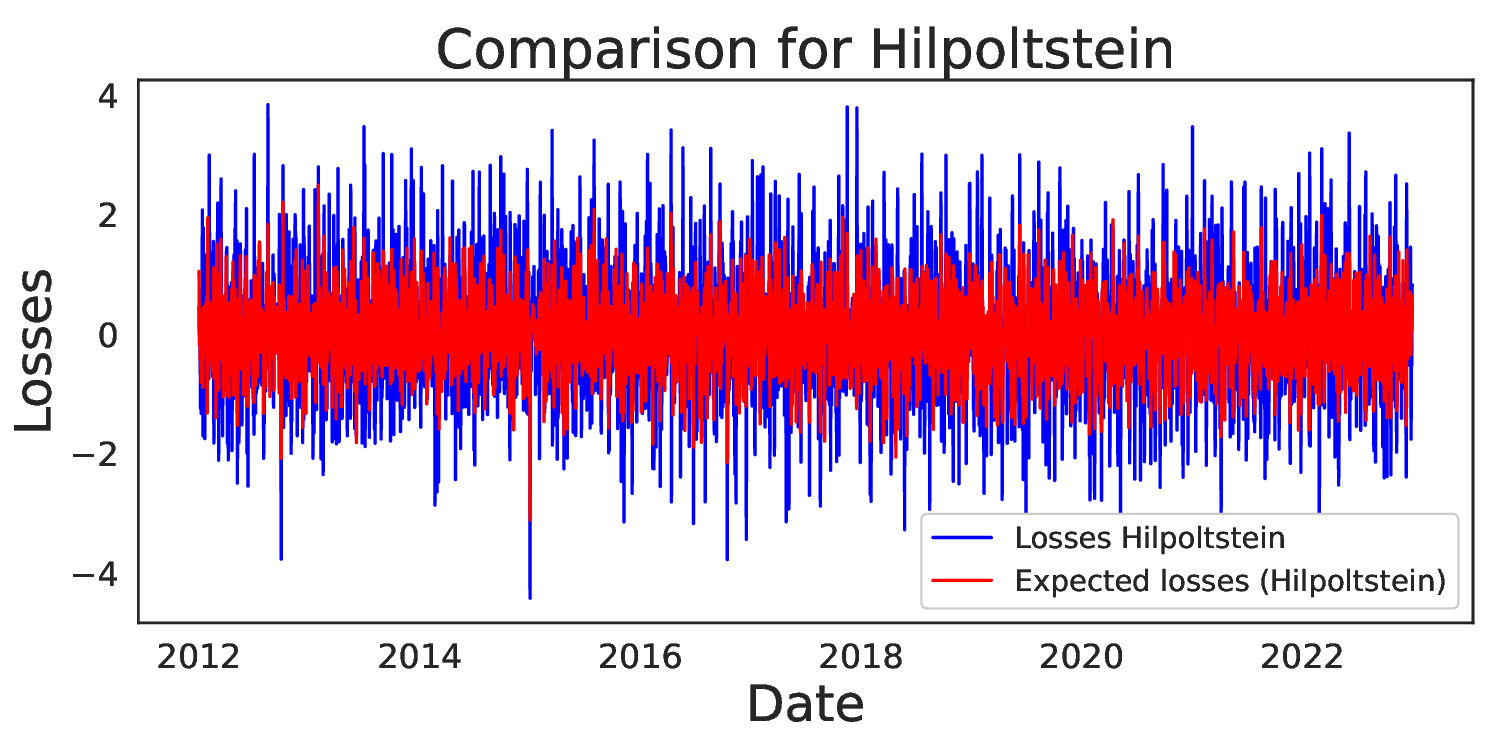}
  \caption{}
  \label{fig:sfig2}
\end{subfigure}
\begin{subfigure}{.5\textwidth}
  \centering
  \includegraphics[width=.9\linewidth]{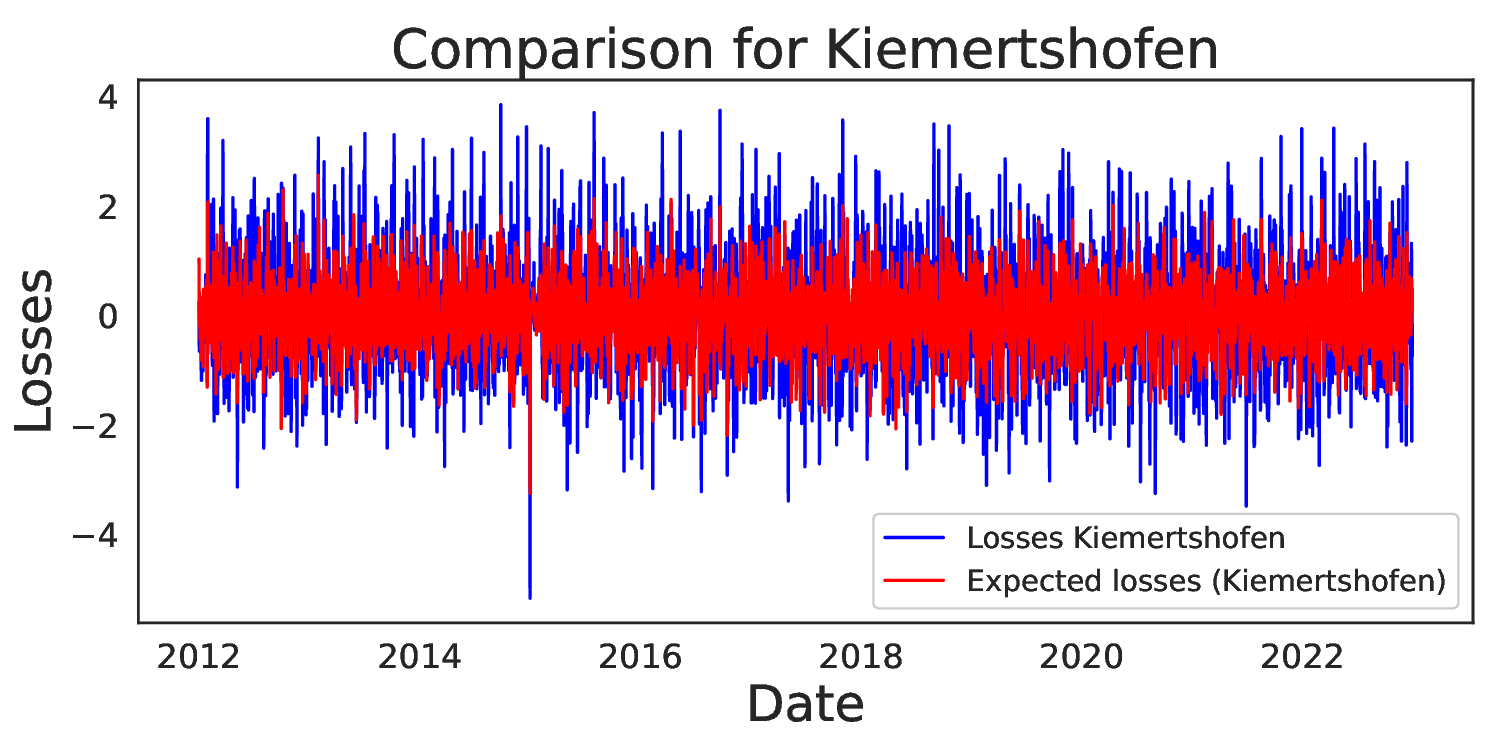}
  \caption{}
  \label{fig:sfig3}
\end{subfigure}
\begin{subfigure}{.5\textwidth}
  \centering
  \includegraphics[width=.9\linewidth]{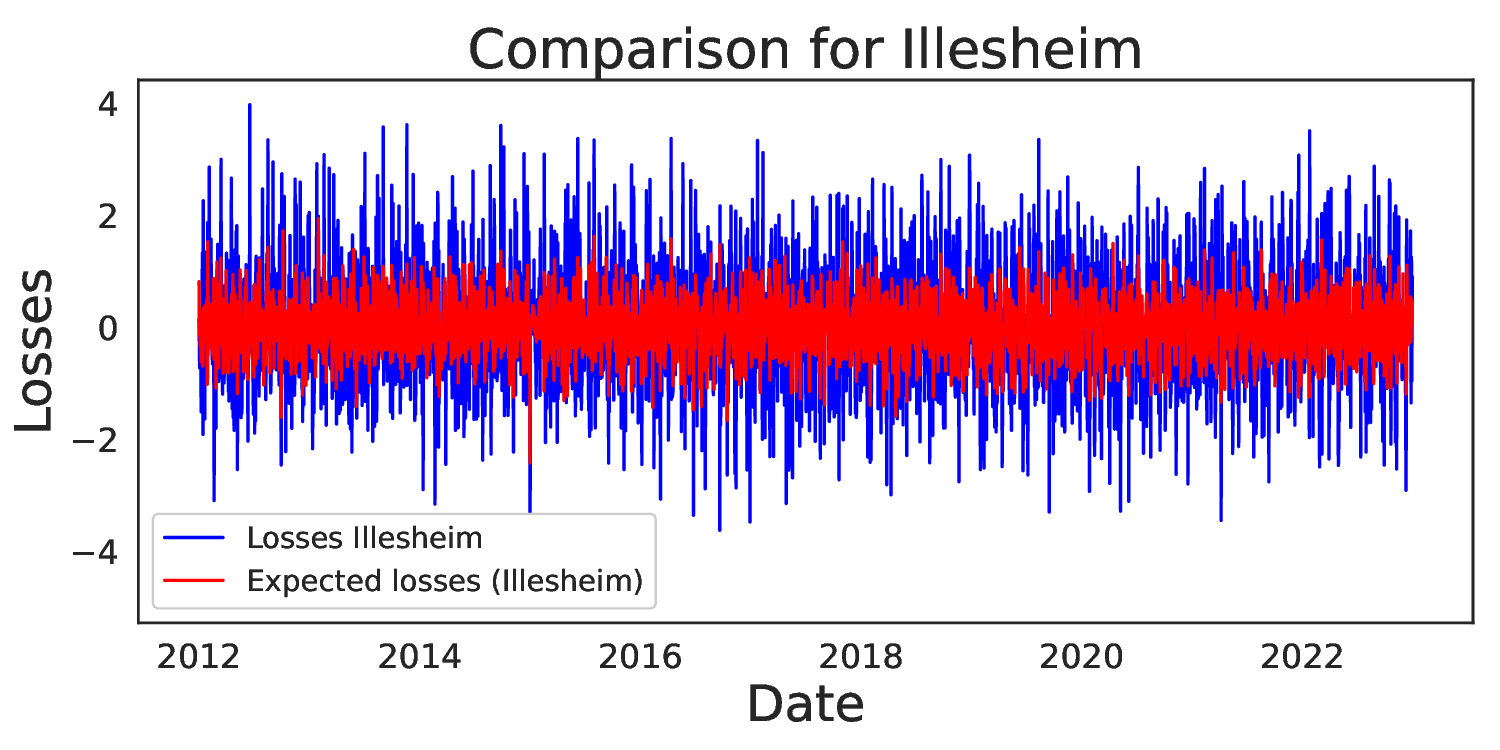}
  \caption{}
  \label{fig:sfig4}
\end{subfigure}
\caption{Prediction of the Gaussian GLM.}
\label{fig:illustration9}
\end{figure}

\begin{figure}[H]
        \centering
                \includegraphics[width=0.9\linewidth]{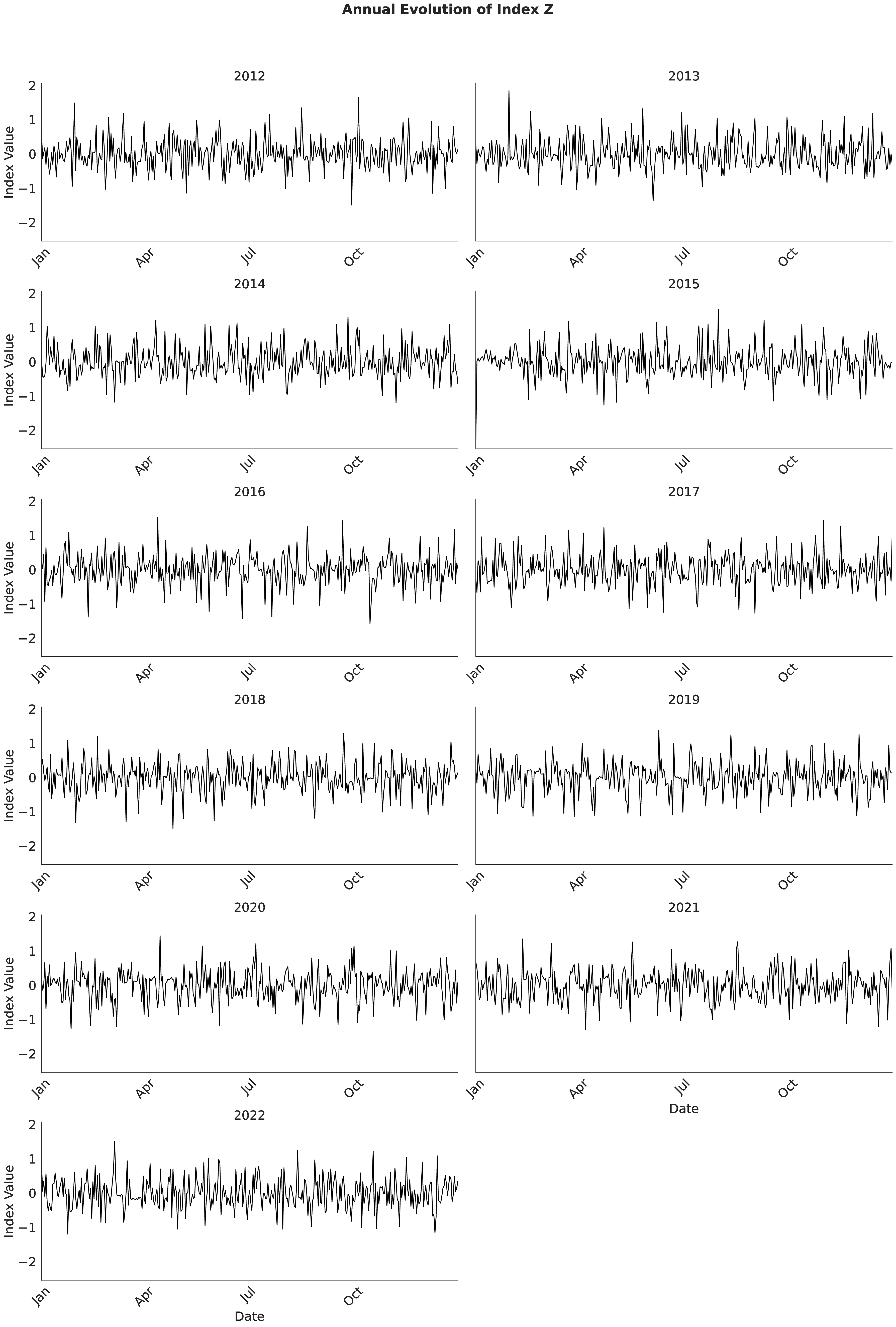}
                \caption{Initial weather index $\tilde{Z}$.}
                 \label{fig:illustration10}
\end{figure}

\begin{figure}[h!]
\begin{subfigure}{.5\textwidth}
  \centering
  \includegraphics[width=.9\linewidth]{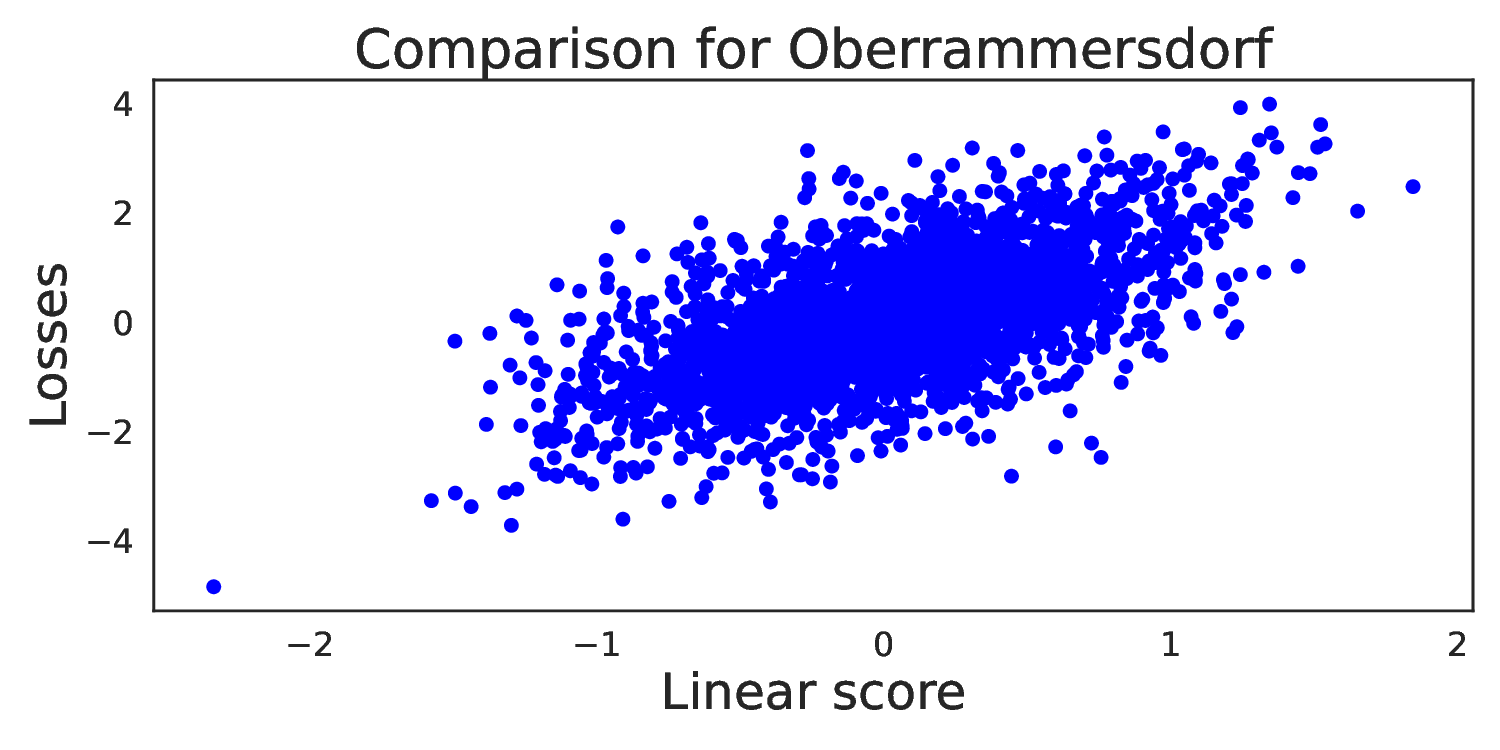}
  \caption{}
  \label{fig:sfig5}
\end{subfigure}%
\begin{subfigure}{.5\textwidth}
  \centering
  \includegraphics[width=.9\linewidth]{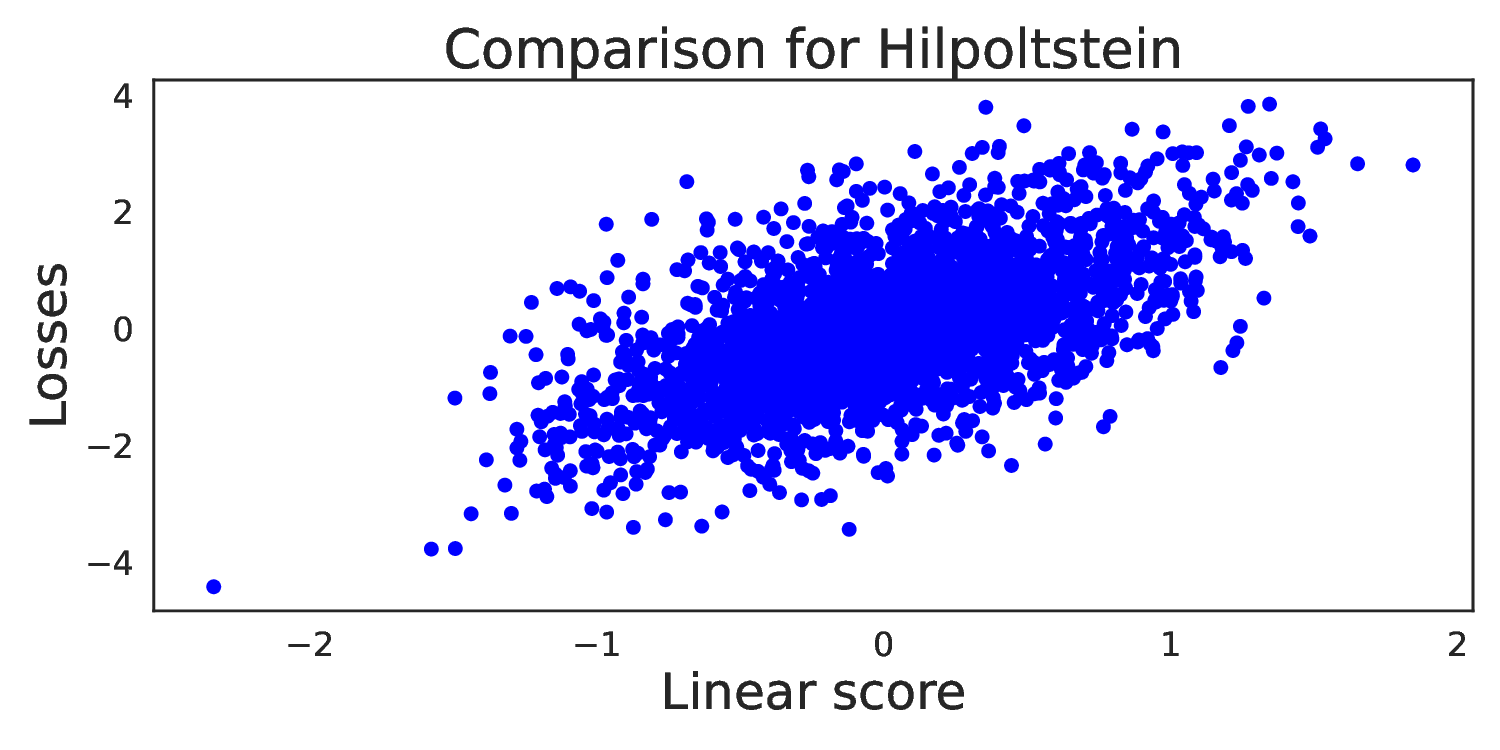}
  \caption{}
  \label{fig:sfig6}
\end{subfigure}
\begin{subfigure}{.5\textwidth}
  \centering
  \includegraphics[width=.9\linewidth]{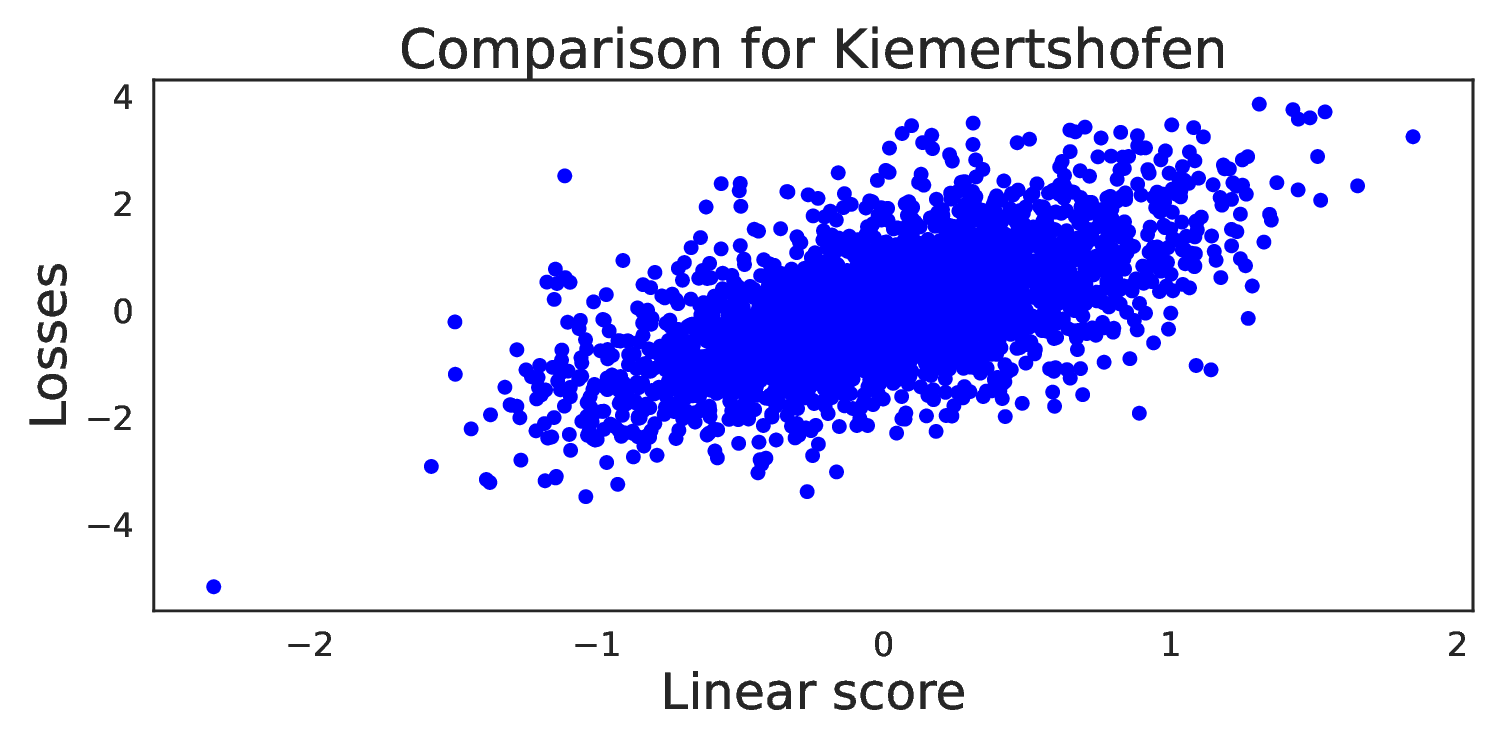}
  \caption{}
  \label{fig:sfig7}
\end{subfigure}
\begin{subfigure}{.5\textwidth}
  \centering
  \includegraphics[width=.9\linewidth]{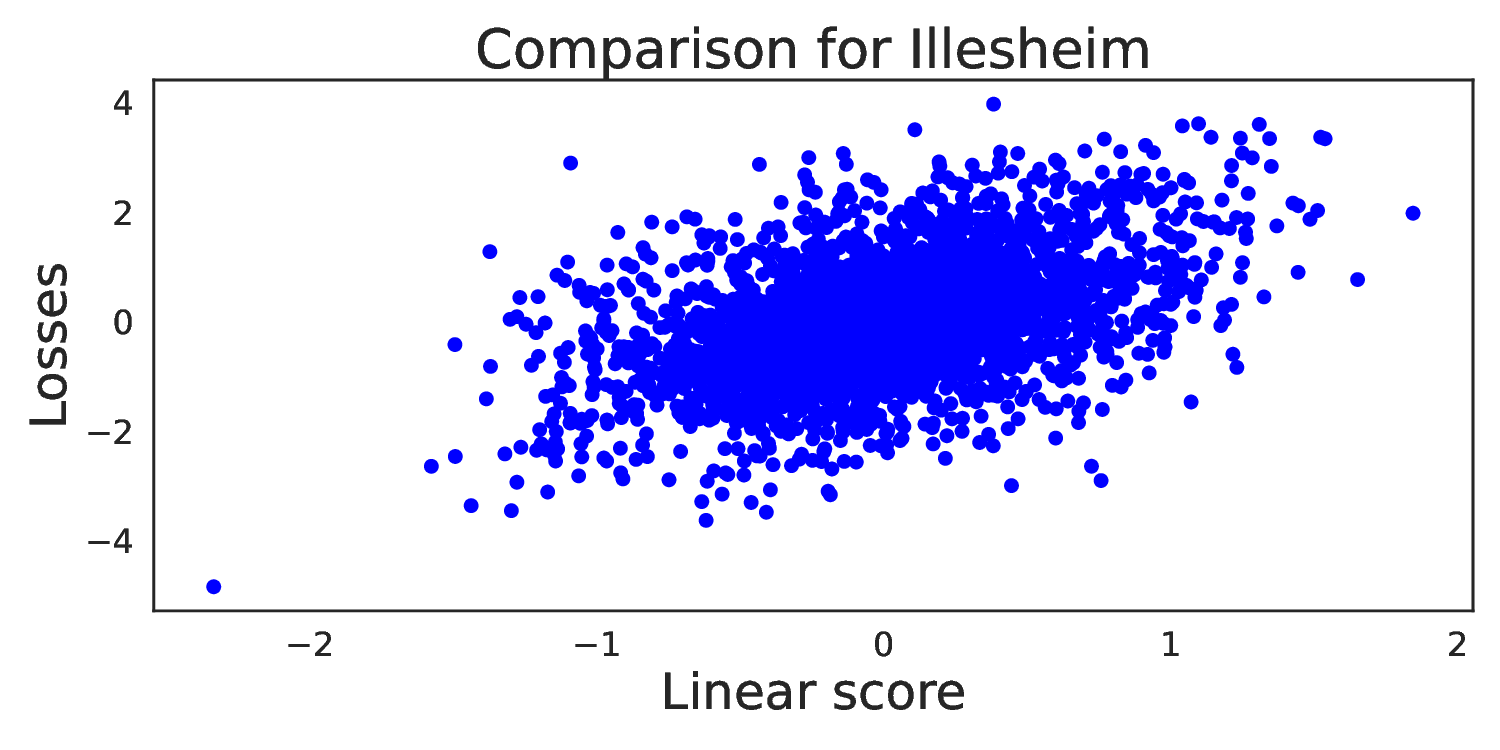}
  \caption{}
  \label{fig:sfig8}
\end{subfigure}
\caption{Scatter plot of the linear score $\tilde{Z}$ with respect to solar production losses.}
\label{fig:illustration11}
\end{figure}

\begin{figure}[H]
        \centering
                \includegraphics[width=0.9\linewidth]{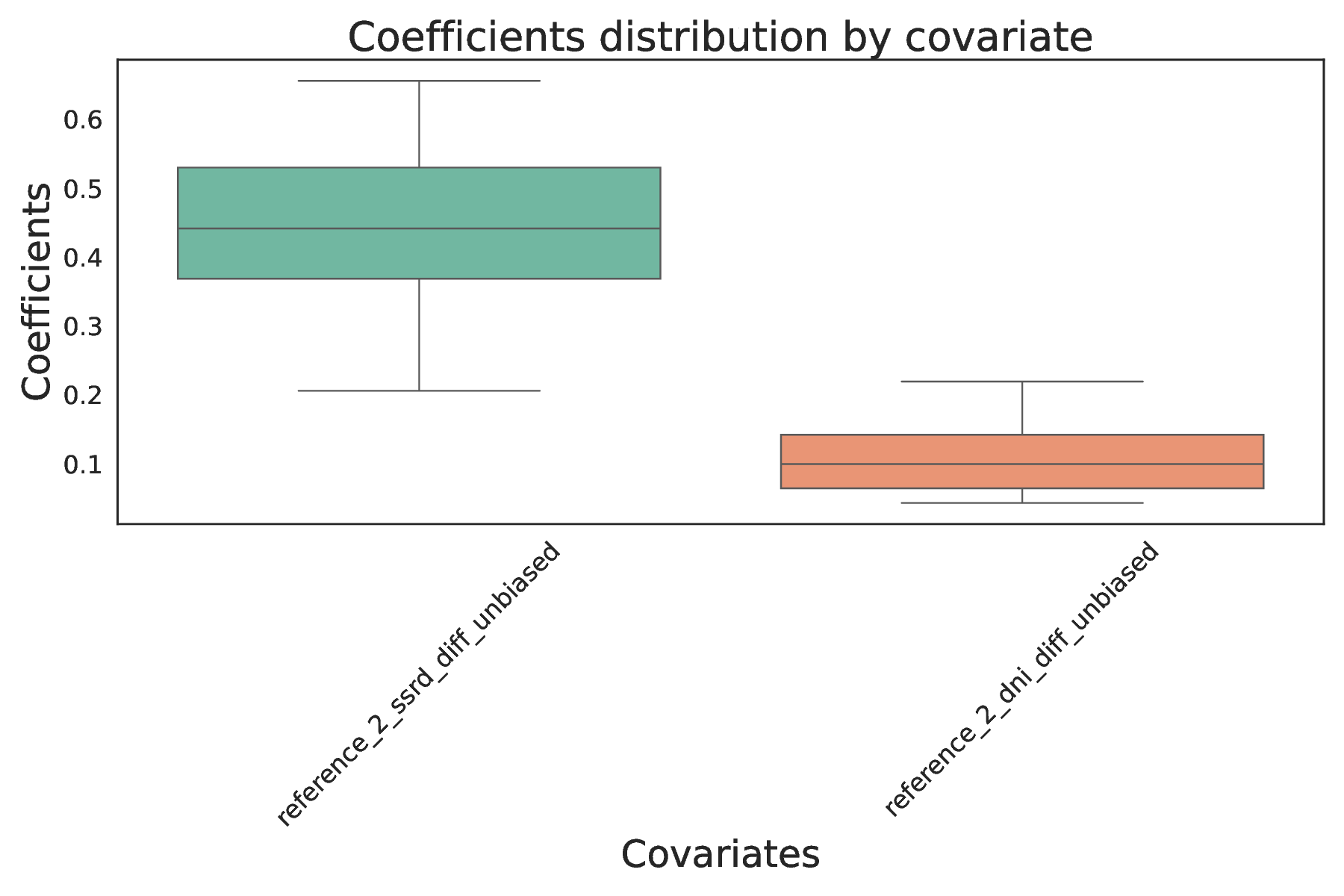}
                \caption{Tweedie GLM coefficients boxplots by covariate.}
                 \label{fig:illustration12}
\end{figure}

\begin{figure}[h!]
\begin{subfigure}{.5\textwidth}
  \centering
  \includegraphics[width=.9\linewidth]{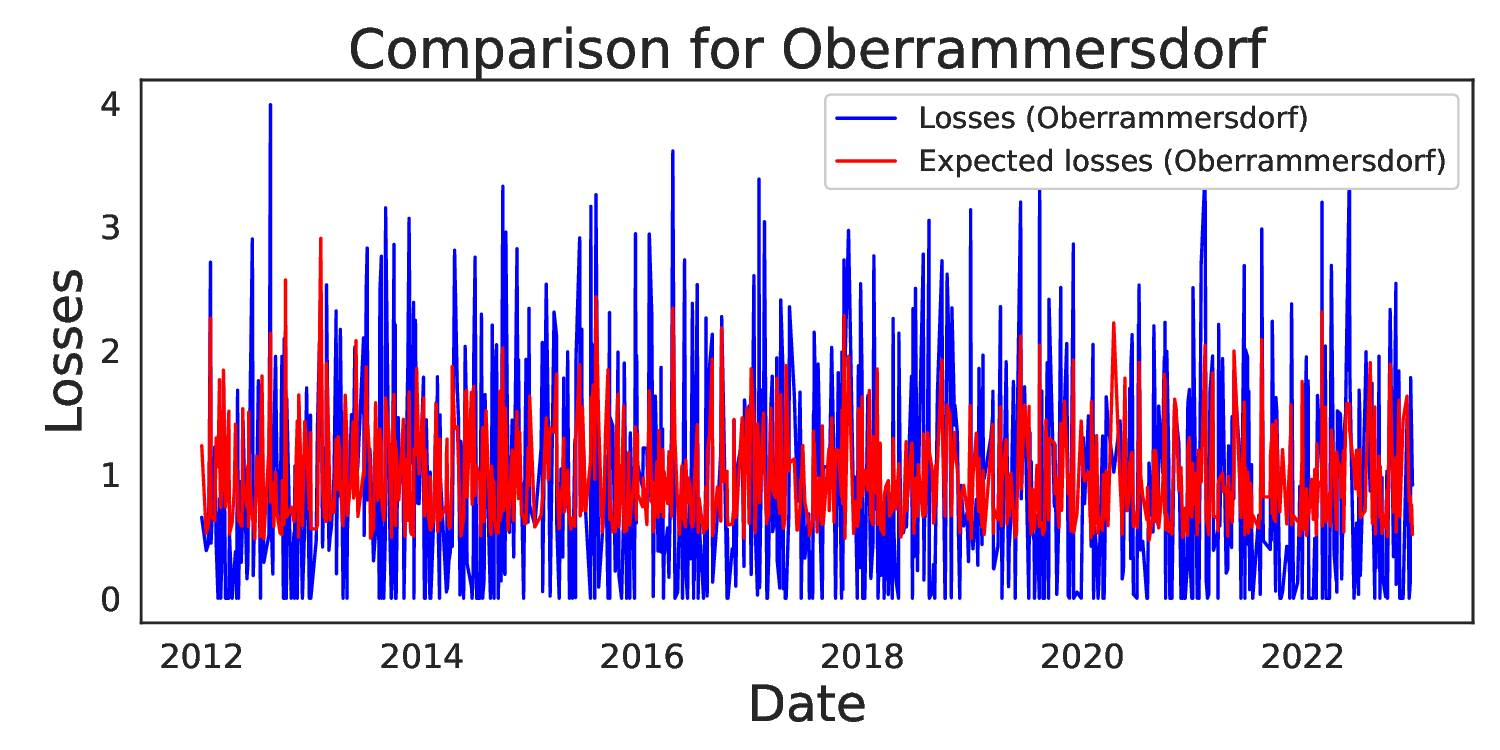}
  \caption{}
  \label{fig:sfig9}
\end{subfigure}%
\begin{subfigure}{.5\textwidth}
  \centering
  \includegraphics[width=.9\linewidth]{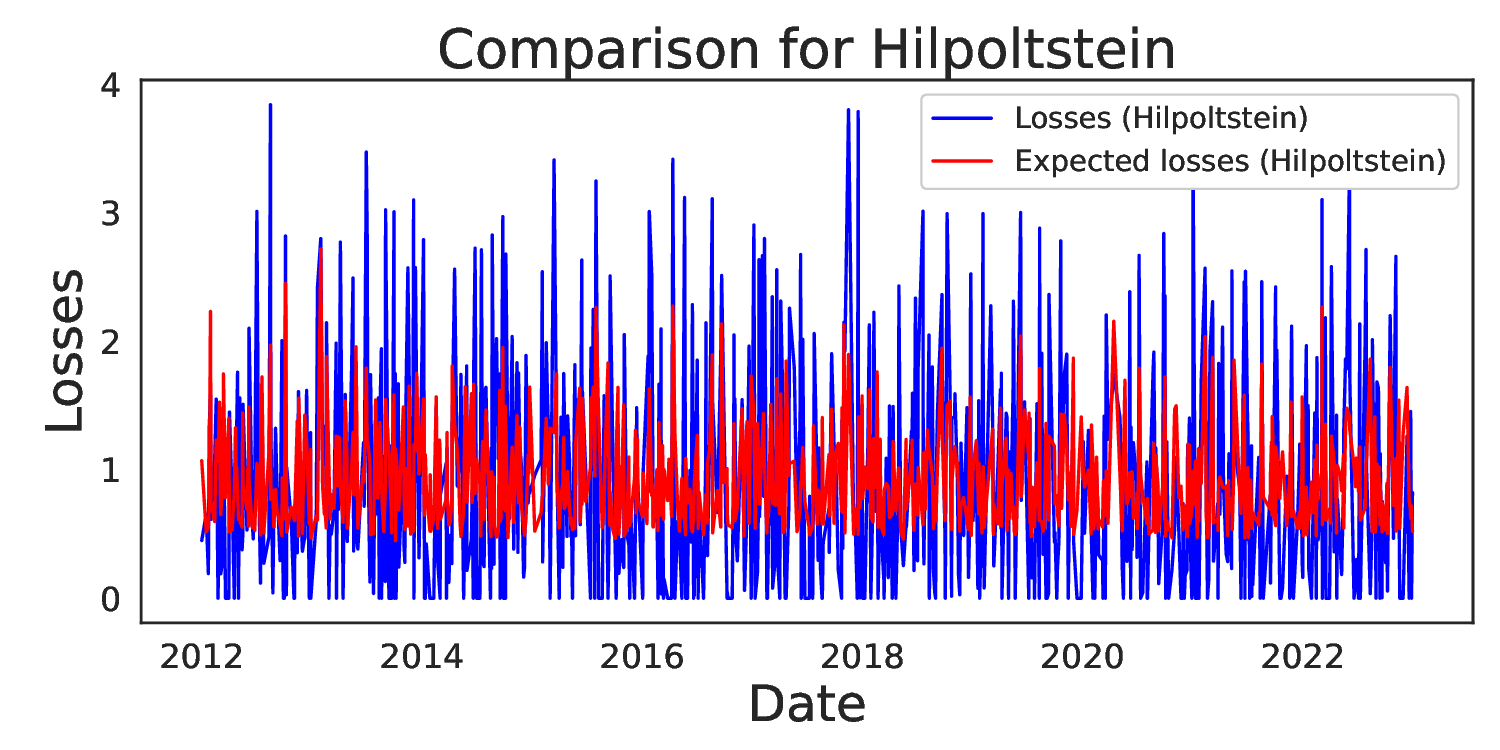}
  \caption{}
  \label{fig:sfig10}
\end{subfigure}
\begin{subfigure}{.5\textwidth}
  \centering
  \includegraphics[width=.9\linewidth]{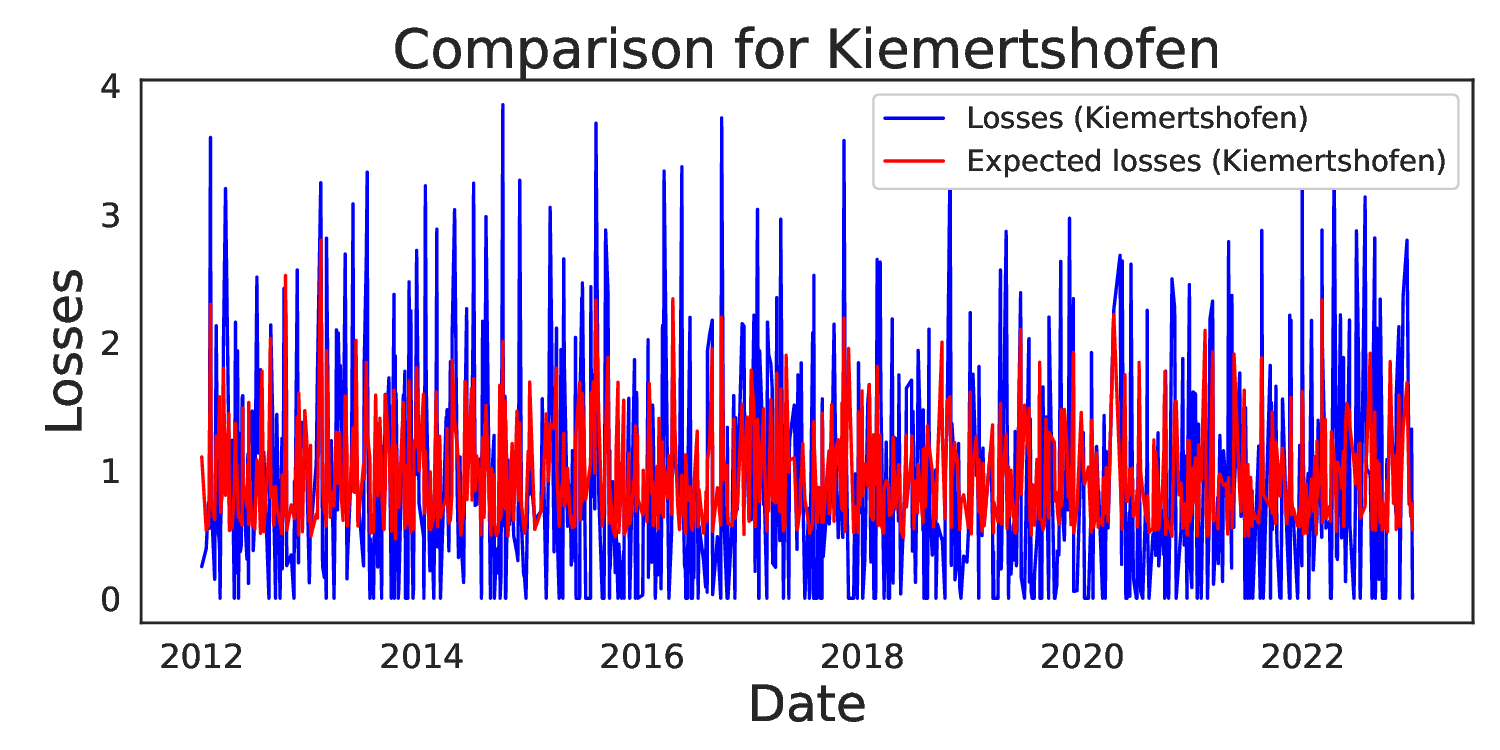}
  \caption{}
  \label{fig:sfig11}
\end{subfigure}
\begin{subfigure}{.5\textwidth}
  \centering
  \includegraphics[width=.9\linewidth]{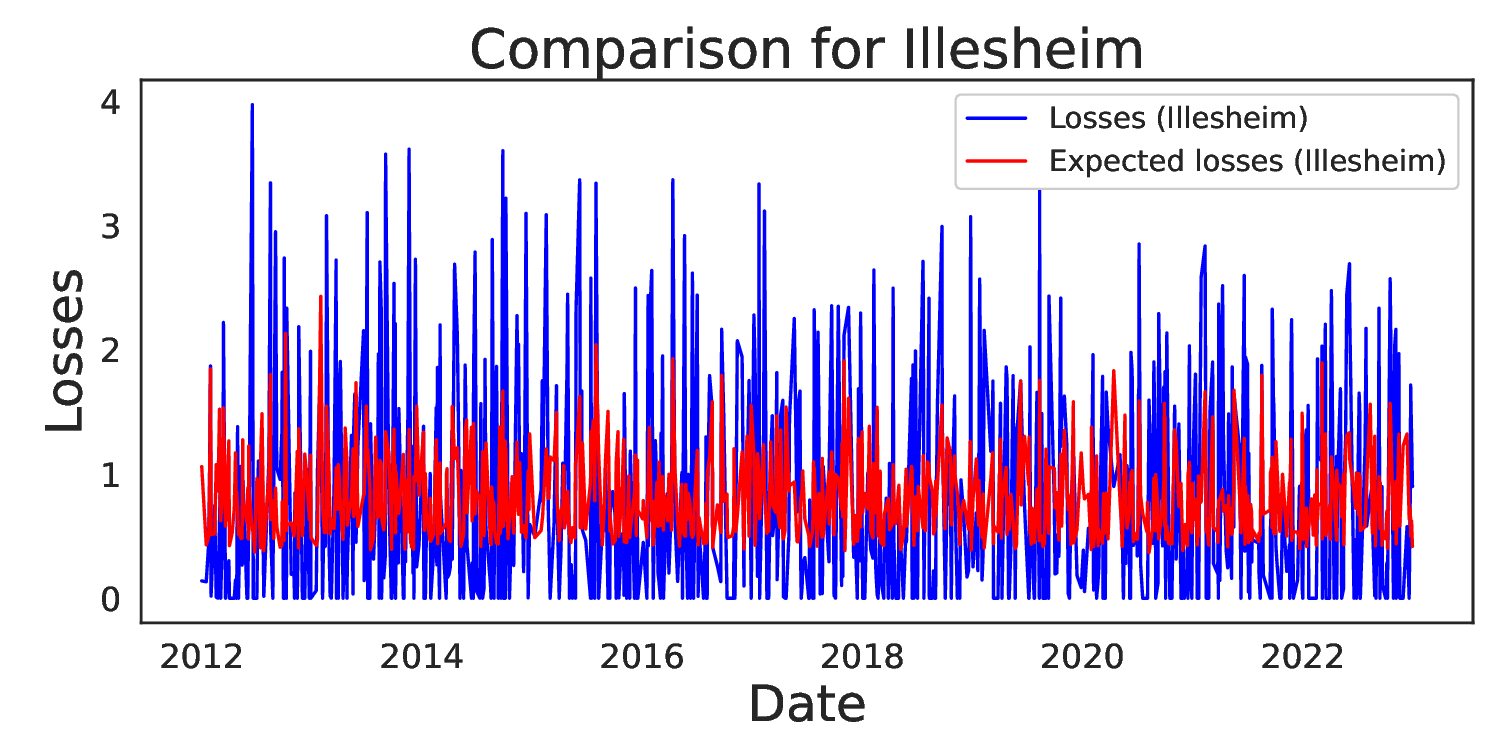}
  \caption{}
  \label{fig:sfig12}
\end{subfigure}
\caption{Prediction of the Tweedie GLM.}
\label{fig:illustration13}
\end{figure}

\begin{figure}[H]
        \centering
                \includegraphics[width=0.7\linewidth]{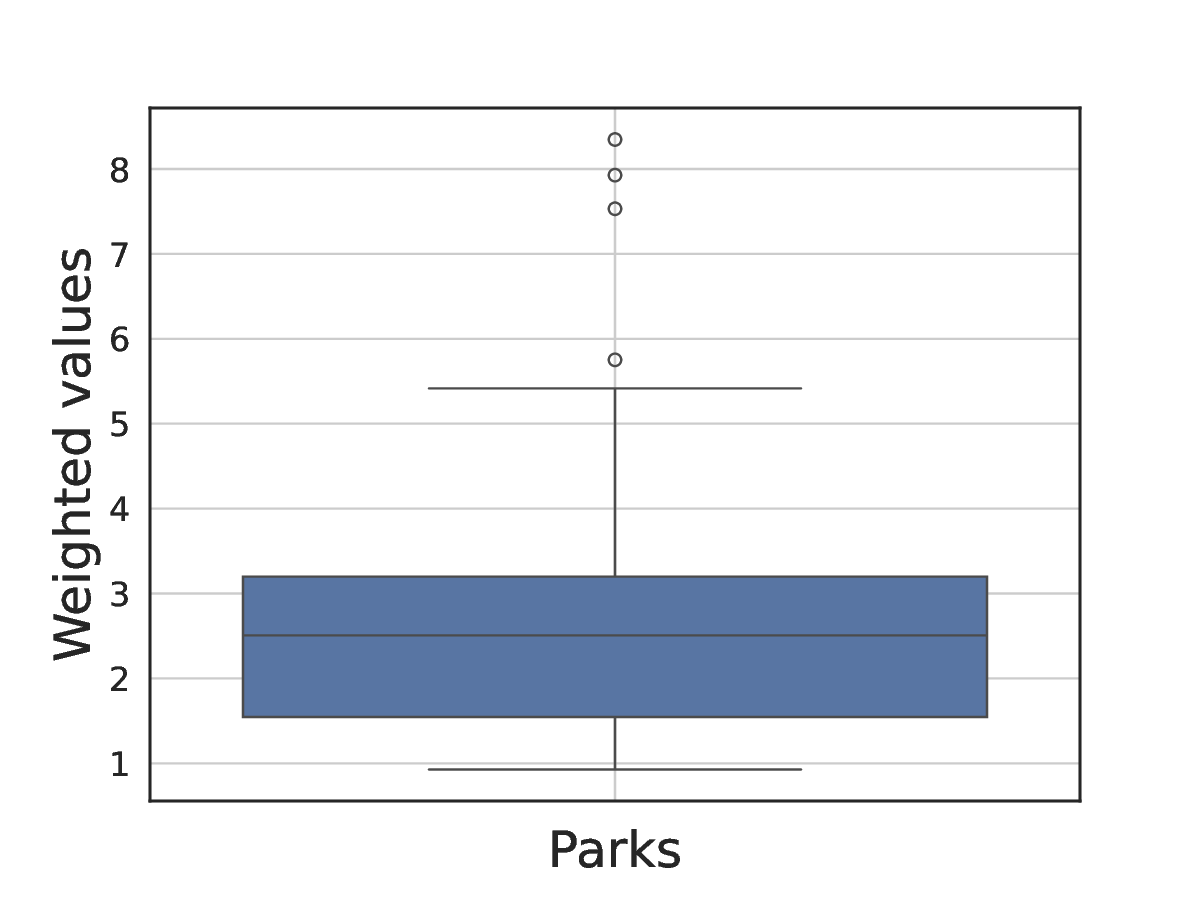}
                \caption{Box plot of the weighted values $\frac{\sigma_{i,d}^2(Z_d)(\sigma_{m, y}^{\Delta P_{i,d}})^2}{\sum^{n}_{j=1}\sigma_{j,d}^2(Z_d)(\sigma_{m, y}^{\Delta P_{j,d}})^2} \times \sigma(S^{*}_{\varepsilon, d})$}
                 \label{fig:illustrationa1}
\end{figure}

\begin{figure}[H]
        \centering
                \includegraphics[width=0.9\linewidth]{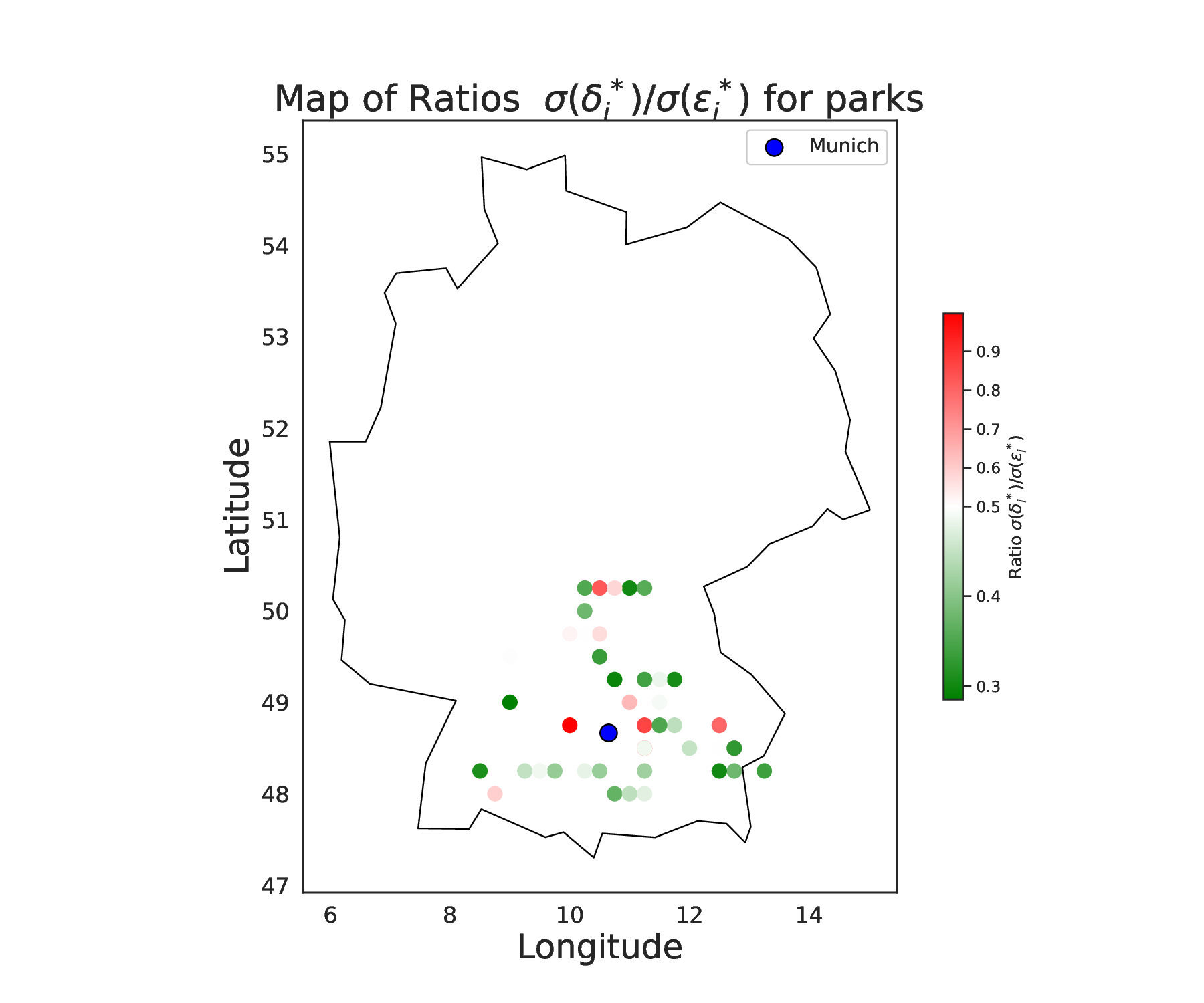}
                \caption{Risk map and reference locations}
                 \label{fig:illustrationa3}
\end{figure}

\end{appendices}

\end{document}